\documentclass[12pt]{article}
\usepackage[symbol]{footmisc}
\usepackage{cancel}
\usepackage{color}
\usepackage{ams}
\usepackage{amsmath,amssymb}
\usepackage{mathtools}
\usepackage{ulem}
\usepackage{float}
\usepackage{lmodern}
\usepackage{graphicx}
\usepackage[symbol]{footmisc}

\renewcommand{\thefootnote}{\arabic{footnote}}

\def\ds{\displaystyle}
\def\bls{\baselineskip}
\def\beq{\begin{equation}}
\def\eeq{\end{equation}}
\def\bi{\begin{itemize}}
\def\be{\begin{enumerate}}
\def\bc{\begin{center}}
\def\bt{\begin{tabular}}
\def\bd{\begin{description}}
\def\bq{\begin{quotation}}
\def\ba{\begin{array}}
\def\beas{\begin{eqnarray*}}
\def\bea{\begin{eqnarray}}
\def\bv{\begin{verbatim}}
\def\ev{\end{verbatim}}
\def\eea{\end{eqnarray}}
\def\eeas{\end{eqnarray*}}
\def\ea{\end{array}}
\def\eq{\end{quotation}}
\def\ed{\end{description}}
\def\ec{\end{center}}
\def\et{\end{tabular}}
\def\ee{\end{enumerate}}
\def\ei{\end{itemize}}
\def\sss{\subsubsection}

\def\d{\cdot}
\def\j{ & }
\def\ada{ \j \dfn \j}
\def\aea{ \j = \j}

\def\aba{ \j   \j}

\def\rdot{\right.}
\def\ldot{\left.}

\def\nn{\nonumber \\}

\def\a{\h{\alpha}}
\def\b{\h{\beta}}
\def\g{\h{\gamma}}
\def\th{\h{\theta}}
\def\de{\h{\delta}}
\def\De{\h{\Delta}}
\def\k{\h{\kappa}}
\def\om{\h{\omega}}
\def\m{\h{\mu}}
\def\n{\h{\nu}}
\def\sg{\h{\sigma}}

\def\e{\h{\epsilon}}
\def\l{\h{\lambda}}
\def\L{\h{\Lambda}}
\def\vp{\h{\varphi}}

\def\s{\section}
\def\ss{\subsection}

\def\sz{\h{\uu 0}}
\def\so{\h{\uu 1}}
\def\sw{\h{\uu 2}}
\def\sr{\h{\uu 3}}

\def\sn{\h{\uu n}}
\def\si{\h{\uu i}}
\def\sj{\h{\uu j}}
\def\sm{\h{\uu m}}

\def\sb{\h{\uu b}}
\def\uz{\h{^ 0}}
\def\uo{\h{^ 1}}
\def\sq{\h{^ 2}}
\def\cu{\h{^ 3}}

\def\upn{\h{^ n}}

\def\ua{\h{^ a}}
\def\ub{\h{^ b}}

\def\str{\h{^*}}
\def\upm{\h{^\mu}}
\def\dnm{\h{_\mu}}
\def\upa{\h{^\a}}
\def\dna{\h{_\a}}
\def\upb{\h{^\b}}
\def\dnb{\h{_\b}}
\def\upn{\h{^\nu}}
\def\dnn{\h{_\nu}}

\def\dni{\h{_{_I}}}

\def\inv{\h{^{-1}}}
\def\dmn{\h{_{\m\n}}}
\def\umn{\h{^{\m\n}}}
\def\dab{\h{_{\a\b}}}
\def\uab{\h{^{\a\b}}}

\def\sk{\h{\dn k}}

\def\tpc{\h{(2\pi)\cu}}

\def\tpf{\h{(2\pi)\tu 4}}
\def\tok{\h{2\ok}}

\def\dvk{\h{d\, \vec k}}

\def\dvx{\h{d\, \vec x}}
\def\dvy{\h{d\, \vec y}}
\def\vk{\h{\vec k}}

\def\vq{\h{\vec q}}

\def\vx{\h{\vec x}}
\def\vy{\h{\vec y}}

\def\vr{\h{\vec r}}

\def\etal{{\it et al.}}

\def\cuks{\h{{ \hat{\bf c}(\uvk,s)}}}

\def\cdukpsp{\h{{ \hat{\bf c}^\dagger(\uvk\,',s')}}}
\def\cduks{\h{{ \hat{\bf c}^\dagger(\uvk,s)}}}
\def\cukpsp{\h{{ \hat{\bf c}(\uvk\,',s')}}}

\def\gmu{\ensuremath{\gamma^\mu}}
\def\gnu{\ensuremath{\gamma^\nu}}
\def\gru{\ensuremath{\gamma^\rho}}
\def\gsu{\ensuremath{\gamma^\sg}}

\def\gau{\ensuremath{\gamma^\a}}

\def\gad{\ensuremath{\gamma_\a}}

\def\ok{\ensuremath{\omega(\vk\,)}}
\def\pmd{\ensuremath{\pl_\mu}}
\def\pmu{\ensuremath{\pl^\mu}}
\def\pld{\ensuremath{\pl_\l}}

\def\pnd{\ensuremath{\pl_\nu}}

\def\plt{\h{\pl\dn t}}
\def\izi{\int_{0}^\infty}
\def\imp{\int_{-\infty}^\infty}
\def\ra{\h{\rightarrow}\, }

\def\pl{\partial}
\def\plt{\partial\dn t}

\def\dfx{d\tu 4 x}

\def\dfy{d\tu 4 y}

\def\df{d\tu 4}
\def\dnc{\dn{\, c}}

\def\uk{\un{k_{}}}
\def\uvk{\un{\vk}}

\def\sl{\!\!\!/}
\def\psl{\pl\sl}
\def\asl{A\sl}

\def\bu{\bar u(q,s')}
\def\ju{u(p,s)}

\def\cam{(A\dn{c})\dn\m(p-q)}
\def\nlb{\\ \aba}

\def\hc{\h{\tu\dagger}}
\def\psl{\h{\cancel{\pl}}}

\def\emiuvkx{\h{\eto{-i\uvk x}}}
\def\eiuvkx{\h{\eto{i\uvk x}\ }}

\def\emiuvkx{\h{\eto{-i\uvk x}\ }}
\def\eiuvkx{\h{\eto{i\uvk x}\ }}
\def\emiuvkx{\h{\eto{-i\uvk x}\ }}
\def\eiuvkx{\h{\eto{i\uvk x}\ }}

\def\cth{\cos(\th)}
\def\sth{\sin(\th)}

\def\vac{\ket 0}

\def\unm{\h{\tu{\n\m}}}
\def\vacm{|\, 0\dn M\, \rangle}

\def\hii{\h{H\dn{int}\tu I}}
\def\kpi{\ket{\psi_{_I}(t)}}

 \def\bkpi{\ket{\Psi\dni (t)}}
 \def\zm{0_{_M}}

 \def\sep{\vspace{12pt} \hrule width \hsize \kern 1mm \hrule width \hsize height 2pt \vspace{12pt} }
 
 \def\cls{\h{{\cal S}}}
 
  \def\clfp{\h{{\cal F}\uu{\cal P}}}
 \def\clr{\h{{\cal R}}}
 \def\clm{\h{{\cal M}}}
\def\cla{\h{{\cal A}}}
\def\clb{\h{{\cal B}}}
\def\clf{\h{{\cal F}}}
\def\clh{\h{{\cal H}}}
\def\cli{\h{{\cal I}}}
\def\clj{\h{{\cal J}}}
\def\cll{\h{{\cal L}}}

\def\clq{\h{{\cal Q}}}
\def\clo{\h{{\cal O}}}

\def\clp{\h{{\cal P}}}
\def\clt{\h{{\cal T}}}
\def\clu{\h{{\cal U}}}

\def\deq{\h{\ = \ }}

\def\beass{\begin{small}\beas}
\def\eeass{\eeas\end{small}}
 \def\beasm{\begin{small}\bea}
\def\eeasm{\eea\end{small}}
 \def\beafns{\begin{footnotesize}\bea}
\def\eeafns{\eea\end{footnotesize}}
\def\vcp{\vec p}

\def\rtd{\right.}
\def\ltd{\left.}

 \def\clr{\h{{\cal R}}}
 \def\clm{\h{{\cal M}}}
\def\cla{\h{{\cal A}}}
\def\clb{\h{{\cal B}}}
\def\clf{\h{{\cal F}}}
\def\clh{\h{{\cal H}}}
\def\cll{\h{{\cal L}}}

\def\clq{\h{{\cal Q}}}
\def\clo{\h{{\cal O}}}

\def\clp{\h{{\cal P}}}
\def\cle{\h{{\cal E}}}
\def\cln{\h{{\cal N}}}

\def\opa{{\bf\hat a}}
\def\opad{{\bf\hat a}\hc}

\def\dnfree{\uu{\mbox{\scriptsize free}}}
\def\dnint{\uu{\mbox{\scriptsize int}}}

\newcommand\lc[3]{f\tu{#1#2#3}}

\def\dfn{\h{:=}}
\def\me{m\uu e}

\def\bu{\bar u}

\def\esl{\e\,\sl}

\def\oso{\om\so}
\def\osw{\om\sw}

\def\ko{k\so}
\def\kw{k\sw}
\def\qo{q\so}
\def\qw{q\sw}
\def\eo{\e\so}
\def\ew{\e\sw}

\def\ipsr{autoregularization\, }
\newcommand\h[1]{\ensuremath{#1}}

\newcommand\tu[1]{\h{^{#1}}}
\newcommand\dn[1]{\h{ _{#1}}}
\newcommand\f[2]{\h{\frac{#1}{#2}}}
\newcommand\brf[2]{\h{\lrr{\f{#1}{#2}}}}
\newcommand\bsf[2]{\h{\lrs{\f{#1}{#2}}}}
\newcommand\bcf[2]{\h{\lrc{\f{#1}{#2}}}}
\newcommand\dsf[2]{\h{\ds \f{#1}{#2}}}
\newcommand\colb[1]{\textcolor{black}{#1}}

\newcommand\eto[1]{\h{e^{\ds #1}}}
\newcommand\gu[2]{\h{g\tu{#1 #2}}}

\newcommand\sps[1]{\h{^{(#1)}}}

\newcommand\delf[1]{\h{\de^4(#1)}}

\newcommand\vev[1]{\h{\left\langle\, 0 \left| \, #1 \, \right|0\, \right\rangle}}

\newcommand\mtrx[2]{\h{\lrs{\ba{#1} #2 \ea}}}
\newcommand\lrr[1]{\h{\left( #1 \right)}}
\newcommand\lrs[1]{\h{\left[ \, #1 \, \right]}}
\newcommand\lrc[1]{\h{\left\{ #1 \right\}}}
\newcommand\lrb[1]{\h{\left| #1 \right|}}

\newcommand\lt[2]{\lim_{#1 \ra #2}}
\newcommand\ct[1]{(\ref{eqn#1})}
\newcommand\evl[1]{\rule{0in}{#1 pt}}
\newcommand\hs[1]{\h{\hspace{#1in}}}
\newcommand\vs[1]{  \ \vspace{#1in} }
\newcommand\nl[1]{\nn [0.#1 in]}

\newcommand\gef[1]{\h{\ g {_e}\, (\,{#1}\,; \clf, \clp)}}
\newcommand\gpf[1]{\h{\ g {_p}\, (\,{#1}\,; \clf, \clp)}}

\newcommand\gpfv[1]{\h{\ g {_p} (\,\uline{\vec{#1 {}_{}}}\,; \clf, \clp)}}

\newcommand\dotp[2]{\h{\left\langle \, #1\, | \, #2 \, \right\rangle}}
\newcommand\mtrxel[3]{\h{\left\langle #1\,
\left| \, #2 \ \right| #3 \, \right\rangle}}

\newcommand\spn[2]{\h{\sg\tu{#1 #2}}}

\newcommand\ti[1]{\h{\tilde{#1}}}
\newcommand\ket[1]{\h{  \left.\left| \,#1\, \right\rangle\right. }}
\newcommand\bra[1]{\h{\left.\left\langle\,#1\, \right.\right|}}

\newcommand\lrab[1]{\h{\langle \, #1 \,  \rangle}}

\newcommand\uu[1]{_{_{#1}}}

\newcommand\leqn[1]{\label{eqn#1}}
\newcommand\lii[1]{\int \dsf{d\vec{#1}}{ \tpc (2\omega(\vec #1))}\ }

\def\prd{\pl\dn\rho}

\newcommand\sump[1]{\ds\sum_{#1=0}^{3}}
\newcommand\crop[4]{}

\newcommand\un[1]{\uline{#1}}

\newcommand\msrf[1]{\int \dsf{d\tu 4  #1}{\tpf}\ }
\newcommand\msrt[1]{\int \dsf{d\tu 3  #1}{\tpc}\ }

\newcommand\wick[1]{\underbracket{#1}}
\newcommand\fs[3]{{\fontsize{#1}{#2}\selectfont #3 }}
\newcommand\upd[2]{\tu{#1}\,\dn{#2}}

\newcommand\scn[2]{\h{#1 \times 10^{#2}}}

\newcommand\gue[1]{g\uu e(#1)}
\newcommand\gues[1]{g\uu e\sq(#1)}
\newcommand\gup[1]{g\uu p(#1)}

\def\esl{\e\,\sl}
\newcommand\epsl[2]{\esl(\vec #1, #2)}

\newcommand\lmu[2]{\eta\tu{#1 #2}}

\newcommand\ubc[2]{\underbrace{#1}_{#2}}  
  
\def\nlasal{\\ [\bls]}

\def\nln{\\ \aba}

\newcommand\od[1]{\om\hc\ \tu #1}

\newcommand\mc[3]{\multicolumn{#1}{#2}{#3}}
\newcommand\igh[2]{\includegraphics[height=#1in]{#2}}
\newcommand\igw[2]{\includegraphics[width=#1in]{#2}}
\newcommand\gm[2]{(G\tu #1)\tu #2}
\newcommand\gk[2]{(G\tu #1)  \dn #2}

\newcommand\mm[2]{\tu{#1#2}} 
\newcommand\mk[2]{\tu{#1}\dn{#2}}
\newcommand\kk[2]{\dn{#1#2}}
\newcommand\dm[2]{\de\tu {#1#2}}

\def\zm{Z\uu m}

\usepackage{float}
\usepackage{graphicx}
\usepackage{lmodern}
\usepackage{titlesec}
\usepackage[title]{appendix}
\date{}

\def\esm{\cle\uu{SM}}
\def\rhoc{\rho\uu c}
\begin{document}
 
\noindent[This is the Accepted Manuscript version of an article accepted for publication in Journal of Physics Communications.  IOP Publishing Ltd is not responsible for any errors or omissions in this version of the manuscript or any version derived from it. This Accepted Manuscript is published under a CC BY license. The Version of Record is available online at https://doi.org/10.1088/2399-6528/ad0649.]

\begin{center}
\fs{18}{23}{\hs{0.05}  
\colb{
A new method
that automatically regularizes
\\[0.2\bls] scattering 
amplitudes  
}
}\\[\bls]
Nagabhushana Prabhu\footnote{prabhu@purdue.edu}\\
Purdue University, West Lafayette, IN 47907 
\end{center}
\begin{abstract}
We present a new regularization procedure called {\it autoregularization}. {\color{black}
The new procedure regularizes the divergences, encountered previously in a scattering process,
using the intrinsic scale of the process.} We use autoregularization to calculate the amplitudes of several scattering processes in QED
and compare the calculations with experimental   
{\color{black} measurements} over a broad
range of  center-of-momentum energies\ \  \ \ \ \  ($\lesssim$ MeV to $\gtrsim$ 200 GeV). The calculated
amplitudes are found to be in good agreement with experimental data\footnote{\label{fn:results}
Specifically, the $O(\a)$ correction to electron's gyromagnetic ratio  predicted
by  autoregularization agrees with
experimental measurement to within 0.06\% (\colb{Section} 
\ref{amm}), which is to be 
compared to  Schwinger's $O(\a)$
correction which agrees with experimental measurement to within $0.15\%$; the $O(\a)$ estimate 
of the Lamb shift  
predicted by autoregularization 
agrees with the experimental measurements to within $0.33\%$ (see 
\colb{Section} 
\ref{ls}); the running fine structure constant 
predicted by autoregularization at $O(\a)$ agrees with the prediction of cutoff regularization to within 0.8\% over
one to four orders of magnitude above the electron's mass scale (\colb{Section} 
\ref{rcc});
the tree-level prediction of autoregularization for Compton scattering 
is in better agreement with experimental data than the prediction of the well-known
Klein-Nishina formula by about 4.02\%
 ({\color{black} 
 \colb{Section}  
 \ref{acs}}); the tree-level predicition of autoregularization for pair annihilation
 at center-of-momentum energy of 206.671 GeV agrees with the experimental data 
 about 0.67\,\% better than the prediction of the standard QED ({\color{black} \colb{Section}  
 \ref{apa}}).}.  
To test autoregularization in a non-Abelian gauge theory, we calculate the QCD coupling constant at 1-loop and show that, like the known regularization schemes, autoregularization also predicts 
asymptotic freedom in QCD.
{\color{black} Finally,  we show that the vacuum energy density of the 
free fields in the Standard Model, calculated using autoregularization, is
smaller than the current estimate of the cosmic critical density.
}
\end{abstract}

\s{Introduction}\label{intro}
Several regularization schemes have been used for renormalizing  scattering amplitudes in 
 quantum field theory.  A regularization scheme introduces an arbitrary energy scale into the renormalization
process---such as the cutoff scale in Wilson's renormalization,  the masses of  
fictitious particles in Pauli-Villars regularization or the energy scale  that is
introduced in dimensional regularization to consistently 
extend the action   to arbitrary
spacetime dimension \cite{hatf,huan,itzu,mand,pesc,ryde,sred,wein,zee}.   In all cases the 
regularization schemes, as well as the energy scales they introduce, are {\it independent
of the kinematics of the scattering process}  of interest.  

On the other hand  every scattering process or phenomenon has an intrinsic `energy' scale that  provides
a natural candidate for the energy scale needed in regularization.  For example, the 
Mandelstam variables provide Lorentz-invariant kinematic `energy' scales that are intrinsic
to the scattering process under consideration. One can construct  a customized 
regularization scheme for each scattering process or  phenomenon using the intrinsic 
`energy' scale of the scattering  process or the phenomenon itself, rather than an energy scale introduced 
by fiat.  
Regularization that uses the intrinsic `energy' scale of  the
very process or phenomenon being regularized  will be called
{\it autoregularization}.
 We present such an autoregularization scheme  in the following discussion.

A key new feature of autoregularization is that  in it the representations of free fields   depend on the 
scattering process or phenomenon of  interest.   Specifically, the creation and annihilation
operators of free fields are scaled by the so-called Gibbs factors\footnote{
The term is borrowed from the formalism of the Grand Canonical Ensemble,  in which the
Gibbs factor suppresses the fluctuation of any system  
that is in thermal and diffusive equilibrium with a reservoir. 
See Appendix \ref{agf}.}, which depend
on the Lorentz-invariant intrinsic scale of the scattering process or phenomenon
under consideration.  
As a result of the above scaling the scattering amplitudes are rendered naturally
divergence-free at all orders of perturbation theory.  Autoregularization is 
described in the next section.

 As  preliminary test  of autoregularization we compare its  
 predictions at 1-loop with the experimental
measurements of the anomalous magnetic moment of the electron\footnote{{\color{black}
The $O(\a)$ correction to electron's anomalous magnetic moment, $a\uu e$,  calculated using autoregularization,
 is closer to experimental
 data than Schwinger's $O(\a)$ prediction \cite{schw}.  $a\uu e$ has been calculated previously
up  to the tenth order \cite{akni}, and the theoretical calculations are found to agree with
 experimental measurements \cite{hann,hhga} up to the tenth decimal place.  Calculation of higher order corrections of $a\uu e$,
 using autoregularization, will be the focus of follow-up work.}} and Lamb shift. 
We also compare the  the running fine structure constant predicted by autoregularization
at 1-loop with the corresponding prediction of the cutoff regularization at 
momentum transfer scales up to $10^4$ times the electron mass scale. 
At the tree-level, we compare the  predictions of autoregularization with the experimental 
measurements of  Compton scattering (at $\sim$ MeV;
 Friedrich and Goldhaber \cite{frgo})  and $e^+-e^-$ pair annihilation (at $\sim$ 206 GeV;
 ALEPH collaboration \cite{alep}). 
The predictions of autoregularization are found to be in
 good agreement with the experimental data 
  and with the prediction of 
 the cutoff regularization.   {\color{black}{}}
 
 The above tests pertain to scattering processes in QED, an Abelian gauge
 theory. To test autoregularization in a non-Abelian gauge theory we consider
 the running of the QCD coupling constant.  Known regularization 
 schemes predict the remarkable phenomenon of asymptotic freedom in QCD.
 We calculate the QCD coupling constant at 1-loop and confirm that autoregularization also predicts asymptotic freedom in QCD. 
 
Following the  the above tests, we use 
autoregularization to compute the vacuum energy density of the  free fields in the Standard Model, and obtain a value smaller than the current estimate of the critical density.

The paper is organized as follows.  Following a description of  autoregularization  in the next section, 
we  use it to \colb{calculate the amplitudes of selected
scattering processes in Section \ref{var}.  We} compute the 1-loop correction to electron's anomalous
magnetic moment  in Section \ref{amm}  and the Lamb shift  in Section \ref{ls}.
In Section \ref{rcc}  we use  \ipsr to compute the running of the 
fine structure constant at 1-loop---for momentum transfer scales that are up to four orders of magnitude
above the electron's rest mass scale.  
Section \ref{amm} -- Section \ref{rcc} contain the summaries of the above calculations,
with the underlying details deferred to Appendices \ref{aamm}--\ref{afpp}.  
In Section \ref{QCD} we calculate the running of the QCD coupling constant 
and show that autoregularization \colb{also} predicts that QCD is an asymptotically free
theory. 
\colb{In Section \ref{tls} we calculate tree-level
amplitudes of two scattering processes.  In Section \ref{acs} we compare the tree-level 
prediction of autoregularization for Compton scattering with the experimental data and with the prediction of 
the Klein-Nishina
formula.  In Section \ref{apa} we compare the tree-level predictions of both
autoregularization and the standard QED with the experimental data for pair annihilation.}
In  Section \ref{ved},
we use  \ipsr to calculate the vacuum energy density of the free fields in the Standard Model.
{\color{black} 
Section \ref{remrks} contains the concluding remarks.}
Unless specified otherwise  
we work in   natural units.
\s{\colb{Description of and formalism for autoregularization}}\label{psr} 
\colb{
In a scattering process,  particles of a quantum field can be created or annihilated and momentum can be transferred to or from the particles of the field.  That is, particles and energy can flow into or out of a quantum field during scattering.  Hence, one
can view a quantum field as a system that is in thermal and
diffusive equilibrium with a reservoir made of the other quantum fields that participate in the scattering process of interest.}

\colb{The statistical behavior of a system that is in thermal and diffusive equilibrium with a reservoir is well-described, in statistical mechanics, by the  Grand Canonical Distribution (GCD), first derived by J.W. Gibbs \cite{gibb,kikr}.  GCD states that the probability of a fluctuation that puts the system in a state with a
certain particle number and 
energy $E$   is proportional to the {\it Gibbs factor}, which decreases exponentially with  $E$.  The stochastic description
of fluctuations of a system, given by GCD, provides a natural paradigm for describing
stochastic  fluctuations of a quantum field, and by extension stochastic scattering processes.}

\colb{Modeled after the GCD, autoregularization constrains 
the probabilities of  fluctuations of quantum fields 
by including Lorentz-invariant Gibbs factors\footnote{\colb{
The Lorentz-invariant (frame-independent) Gibbs factors, described in Section \ref{gf}, differ slightly from the 
(frame-dependent) Gibbs factors in GCD.   With some abuse of notation we use the term
``Gibbs factors'' for the Lorentz-invariant factors, described in Section \ref{gf}, as well.
}} in the description of quantum fields.
Constraining the probabilities of fluctuations of quantum fields with Gibbs factors
has the immediate consequence of 
eliminating  divergences from  
scattering amplitudes, at all orders of perturbation theory.
The Lorentz-invariant
Gibbs factors are described in Section \ref{gf}
and Appendix \ref{agf}.  We begin the discussion by describing how Gibbs factors are included in autoregularization.} 

Consider a scattering process
$\clp$ seen by an observer who is at rest in frame $\clf$.  
We seek to calculate
the $S$-matrix element of the process\footnote{Thus we assume that 
the incoming and outgoing particles in the scattering process, the
asymptotic momenta of the incoming and outgoing particles
(in some Lorentz frame $\clf$), and the Lagrangian  
governing the particles' interactions are specified at 
the outset.   Further, we assume that the tree-level Feynman 
diagrams are connected; that is, we assume that $\clp$ does 
not describe two or more non-interacting scattering processes.}. 
Let $\cll$ be the Lagrangian that underlies $\clp$ and 
let $\varphi$ denote a generic quantum field in $\cll$.  
For simplicity, we assume $\vp$ is a scalar field of mass $m$. 
In autoregularization  we scale the creation
and annihilation operators in the free-field expansion of $\vp$
(in the interaction picture) 
with a Lorentz-invariant Gibbs\footnote{
The discussion in Appendix \ref{agf} clarifies the 
reason for calling it the Gibbs factor. 
} factor $g\uu\vp$ as follows.
\bea
\vp\dnint(x; \clf, \clp) \aea \msrt k g\uu{\vp}(\uvk; \clf,\clp) \lrc{\opa(k) \emiuvkx + \opad(\vk) \eiuvkx}
\leqn{2p1}
\eea
where $\opa$ and $\opad$ denote the annihilation and creation operators.
We follow the convention of underlining a 4-vector to indicate that it is
on mass shell.   Specifically, $\uvk \dfn (\sqrt{|\vk|\sq + m\sq}, \vk)$.
$\vp\dnint$ remains a solution of the Klein-Gordon equation\footnote{We
drop the subscript in $\vp\dnint$ hereafter.}. The creation and annihilation
operators of a generic free quantum field--for example, Dirac field or Maxwell
field-- are similarly scaled by a Gibbs factor 
 and a generic free field satisfies the corresponding Euler-Lagrange equation.

The Gibbs factor, and hence $\vp$, depends on the scattering process $\clp$.
 The Gibbs factor is constructed using the Lorentz invariant  described below.   
\ss{\colb{Lorentz-invariant extension of center-of-momentum energy for construction of Gibbs factor}}\label{li}
 Let $\clfp$ 
be a center-of-momentum frame of the scattering process $\clp$.  $\clfp$ is determined up to 
spatial rotations. Let $\clu(\clfp; \clf)$ denote the 4-velocity of an observer who is at rest in frame $\clfp$,
as seen by an observer who is at rest in frame $\clf$.
Let $\pi\uu\vp$ denote the projection of a momentum 4-vector onto  the  mass shell of $\vp$.  
Specifically,  for 4-momentum $k\dfn (k\uz,\vk)$, 
\bea
\pi\uu\vp(k) \ada (\om(k), \vk), \qquad \om(k) \dfn \lrs{|\vk\,|\sq + m\sq}\tu{1/2}
\leqn{pro}
\eea
where $m$ is the mass corresponding to the free field $\vp$.
Let  $\Lambda(\clfp, \clf)$ denote the Lorentz transformation that maps a 4-momentum in 
$\clf$ to a corresponding 4-momentum in $\clfp$.  Given a 4-momentum $k$ in $\clf$, $\xi$, defined
as 
\bea
\xi \uu\vp(k; \clf,  \clp) \ada \Lambda\inv(\clfp, \clf) \lrr{\pi\uu\vp\lrs{ \rule{0in}{14pt}\Lambda(\clfp,\clf) k}}
\leqn{2p2}
\eea
maps $k$ to an on-mass-shell 4-momentum.   Although, $\clfp$ appears on the right hand side
above, it is easy to see that the left hand side is independent of the particular   center-of-momentum
frame $\clfp$ that one chooses, and depends only on the process $\clp$.  We observe that 
\bea
E\uu\vp(k;\clf,\clp)\ada \xi\upa\uu\vp(k; \clf, \clp) \,\clu\dna(\clfp; \clf)
\leqn{2p3}
\eea
is a Lorentz invariant.   Setting $\clf \deq \clfp$ in \ct{2p2} we see that in a center-of-momentum
frame $\clfp$, $E\uu\vp$ reduces to 
\bea
E\uu\vp(k;\clfp, \clp) \aea \om(k)
\leqn{2p4}
\eea
\ss{\colb{Gibbs factor formulas for Fermions and Bosons}}\label{gf}
Deferring a derivation of the Gibbs factor as well as a discussion of the form of the Gibbs factor
to Appendix \ref{agf}, we state the explicit form of the Gibbs factor 
below. 
Using the abbreviation $E\uu\vp(k) \dfn E\uu\vp(k;\clf,\clp)$, the Gibbs factor for a field $\vp$ 
in frame $\clf$, for process  $\clp$, is defined as 
\bea
g\uu\vp(k; \clf,\clp) \aea \left\{ 
\ba{ll}
\bsf{\ds {d\uu\vp}_{_{\ }}}{\ds e\tu{(E\uu\vp(k)-\mu\uu\vp)/\tau}-1}\tu{1/4}, \j \vp: \mbox{ massive boson}\\ 
\ \\
\bsf{\ds {d\uu\vp}_{_{\ }}}{\ds e\tu{(E\uu\vp(k)-\mu\uu\vp)/\tau}+1}\tu{1/4}, \j \vp: \mbox{ massive fermion}\\
\ \\
\bsf{\ds {d\uu\vp}_{_{\ }}}{\ds e\tu{ E\uu\vp(k)/\tau + \tau/E\uu\vp(k)}-1}\tu{1/4}, \j \vp: \mbox{ massless boson}
\ea
\rtd \leqn{2p5}
\eea
$d\uu\vp$, the degeneracy factor, is the number of distinct creation operators in the free field
expansion of $\vp$ and its conjugate. For example, the degeneracy factor for an electron is $d\uu e = 4$
and for a photon it is $d\uu p = 2$.  $\m\uu\vp$, the chemical potential of field $\vp$
corresponding to a particle of mass $m$ and an electric charge\footnote{{\color{black} 
Given that we do not know if neutrinos are Majorana fermions,
electric charge is the only conserved charge of interest for the 
particles involved in the scattering processes that we consider in this paper.  However,
in scattering processes  involving particles that carry other conserved 
charges,  such as the color
charge,  the definition of chemical potential will  likely need to be 
extended to encompass other conserved charges. }}   
of magnitude $q\,|e|$ is defined as
\bea
\m\uu\vp \aea  
\sqrt{\dsf{9 \a  m\sq q\sq}{16\pi}}
\leqn{chp}
\eea
in units in which $\hbar = c = \me = \e\sz = 1$.  $\a$ denotes the fine structure
constant, $\me$, the rest mass of the electron and the other symbols have
the standard meanings. A massless boson is assumed to have
vanishing electric charge and thus a vanishing chemical potential.  As described
after Equation \ct{chpo},  a particle and its antiparticle are stipulated to have equal
chemical potentials.

$\tau\uu\clp$, the intrinsic `energy' scale of a scattering process $\clp$ with $m$ incoming particles of 
momenta $p\so, \ldots, p\sm$ and $n$ outgoing particles of momenta $q\so, \ldots, q\sn$,
is defined as
\bea
\tau\uu\clp \aea \max_{\cli, \clj} \lrc{\sqrt{|k\dnm k\upm|} \ \left|   \  k \deq \sum_{i\in \cli} p\si - \sum_{j\in \clj} q\sj, 
\rtd}
\leqn{tau}
\eea
where $\cli \subseteq \lrc{1, \ldots, m}, \clj \subseteq\lrc{1, \ldots, n},  
0<  | \cli| +| \clj | < m+n$.
$\tau\uu\clp$ is abbreviated to $\tau$ in \ct{2p5} and in the rest of this section.   

It should not be surprising that the Gibbs factor for massless boson is not the 
limit of the Gibbs factor of a massive boson where the mass $\ra 0$.  Even  in
canonical quantization the quantization of  the massless Maxwell field
cannot be implemented as the massless limit of the quantization of  a 
massive boson field;  the quantization of the massless field is a more delicate 
procedure than the quantization of a massive field.  Further a continuous limit
is precluded by the fact that massive and massless bosons have different numbers 
of physical degrees of freedom.

In the other regularization schemes   the
regularization parameter is driven to a limit (for eg., $\e = 4 - d \ra 0$ in dimensional
regularization) and eliminated after regularization is completed.
On the other hand, the
regularization parameter $\tau$ in autoregularization, derived from 
the kinematics of $\clp$, is not driven to a limit and 
persists in the calculation.

From the 
structure of the Gibbs factors we see that, in autoregularization, more energetic processes---that is
processes with higher $\tau$---probe deeper into the UV regime (in the sense that 
they receive greater contributions from high `energy' modes) than less energetic processes.
Thus the cutoff is neither sharp nor pre-specified.  

The Lorentz invariance of the Gibbs factor follows from the Lorentz invariance of $\tau, \m\uu\vp$ and
$E\uu\vp(k)$. We note that $g\uu\vp\tu 4$ resembles the Bose-Einstein and Fermi-Dirac distributions
for massive boson and fermion fields.  Gibbs factor for massless bosons, such as photons,  
includes  a term in the exponent that suppresses infrared divergence.  

\ss{\colb{\fs{14}{14}{Derivation of Feynman propagators for electrons and photons}}}\label{fq}

\vs{-0.35}

\noindent
If the Gibbs factor is set to $g\uu\vp = 1$ in \ct{2p1} one recovers the process-independent 
representation of the free field, denoted $\Phi(x)$. It is well known that imposing equal-time 
commutation relations on $\Phi(x)$ and its conjugate momentum  is equivalent to 
imposing commutation relations on the creation and annihilation operators $\opad$ and $\opa$. 
The insertion of the Gibbs factor into the free field expansion, as shown in \ct{2p1}, breaks the
above equivalence.  Thus we can postulate either the equal-time commutation relations on the field
$\vp$ and its conjugate momentum or we can impose commutation relations on $\opa$ and 
$\opad$, but not both.  We impose the standard commutation\footnote{
Commutation relations for the creation and annihilation operators of boson fields and 
anticommutation relations for the creation and 
annihilation operators of fermion fields.
} relations on the creation and annihilation
operators, and omit the equal-time commutation relations on the field and its conjugate 
momentum.   

Using the standard commutation/anticommutation relations of the creation and annihilation
operators, a straightforward calculation yields the following Feynman propagators for 
electrons and photons. 
\bea
D\uu{\clf,\clp}(x,y) \aea i \msrf k \bsf{ e\tu{-ik(x-y)}}{k\sl - \me + i\e} g\uu e\sq (k; \clf,\clp)
\leqn{2p6}
\eea
The only change in the propagator in \ct{2p6} is the extra factor comprising the square of the electron's
Gibbs factor $g\uu e$.  Similarly, in Lorenz gauge the photon propagator is
\bea
M\uu{\clf,\clp}\umn(x,y) \aea i \msrf k \bsf{ (-\eta\umn)\, e\tu{-ik(x-y)}}{k\sq   + i\e} g\uu p\sq (k; \clf,\clp)
\leqn{2p7}
\eea
where $g\uu p$ denotes the photon's Gibbs factor.  When the frame $\clf$ and the process $\clp$
are evident from the context, we omit the subscripts of $D$ and $M\umn$. We also  write
the Gibbs factors as $g\uu e(k)$ and $g\uu p(k)$ in the above propagators.

\ss{\colb{Derivation of  the LSZ (Lehmann-Symanzik-Zimmermann) scattering amplitude in terms of the Gibbs factors}}\label{lsz}
\colb{As is well known, the
LSZ reduction formula expresses 
the amplitude of a scattering process ($S$-matrix element) in terms of time-ordered 
correlation functions of the participating quantum fields. The 
formula is described in most textbooks on quantum field theory
including \cite{hatf,huan,pesc,sred,wein} to which the reader is referred for  
details.}

\colb{Previous derivations of the LSZ reduction formula (for eg., see
\cite{sred}), use creation and annihilation operators that 
occur in process-independent representations of quantum
fields.  Autoregularization, on the other hand, uses process-dependent representations of quantum fields.  
}   
A
straightforward calculation  shows that 
for a 
scattering process $\clp$ involving $m$ incoming particles of momenta $p\so, \ldots, p\sm$ and 
$n$ outgoing particles of momenta $q\so, \ldots, q\sn$
\bea
\cla\uu{AR}(\clp) \aea \lrs{\prod_{i=1}^m \dsf 1 {g\uu{\vp\si}(p\si)} \prod_{j=1}^n \dsf 1 {g\uu{\vp\sj} (q\sj)}}
\cla(\clp) 
\leqn{2p8}
\eea
where $\cla\uu{AR}(\clp) $ represents the LSZ scattering amplitude in autoregularization and $\cla(\clp)$  
denotes the   LSZ scattering amplitude  in the standard theory.   
As \ct{2p8} shows, the Gibbs factors impact  the scattering amplitude $\cla\uu{AR}(\clp) $  even at the 
tree level. Therefore, we check the prediction of \ct{2p8} against experimental data by computing
the tree-level differential cross-section for Compton scattering  in Section \ref{acs}
and pair annihilation in Section \ref{apa}.   As shown in the appendices, the tree-level predictions of autoregularization 
are  in good agreement with experimental data as well.

\colb{The main advantage of autoregularization {\it vis-\`a-vis} standard theory  becomes evident by 
observing that $\cla(\clp)$ in \ct{2p8} can be expanded perturbatively 
in terms of Feynman propagators; see \cite{huan,sred}.  As shown in \ct{2p6} and \ct{2p7}, the Feynman propagators in autoregularization contain (squares of)
Gibbs factors, which exponentially suppress contributions from 
high-energy modes\footnote{\colb{As seen in the center-of-momentum frame.}}.  
As a 
result  
 all scattering amplitudes,
including those that diverge in standard theory, become 
finite in autoregularization.
}
\s{\colb{Validity of the Autoregularization method}}\label{var}
\colb{In this section we demonstrate how autoregularization avoids 
divergences in scattering amplitudes   
by incorporating the  
Gibbs factors,  described in Section \ref{gf}.  
We also test the validity of the
autoregularization method by applying it to  the 
following calculations: 
{\it   
1-loop correction to the anomalous magnetic
moment of the electron}  (Section \ref{amm});    
 {\it Lamb shift}  at 1-loop (Section \ref{ls});   the 
{\it running fine structure constant} at 1-loop (Section \ref{rcc}); 
demonstration of {\it asymptotic
freedom in QCD} at 1-loop (Section \ref{QCD});
tree-level {\it Compton scattering} (Section \ref{acs}); 
tree-level {\it pair annihilation} (Section \ref{apa}). 
In Sections \ref{amm}--\ref{tls},  the relevant
formulas     
are derived,  
 in terms of the Gibbs factors,  
and   then validated using 
 experimental data and previous theoretical 
 predictions.}
\ss{\colb{Anomalous magnetic moment of the electron}}\label{amm}
Two landmark calculations  played a significant role in the early
development of quantum field theory---the calculation of
the  $O(\a)$ correction to electron's gyromagnetic ratio  by 
Schwinger \cite{schw}, and  the calculation of the
 Lamb shift  by Bethe \cite{bethe}.  The remarkable agreement 
between the theoretical prediction and the experimentally measured
value of electron's gyromagnetic ratio is regarded as one of the 
major triumphs of quantum electrodynamics.   We use autoregularization to 
calculate the $O(\a)$ correction to electron's gyromagnetic 
ratio in this section, and the $O(\a)$ correction to the Lamb shift
in the next section.  The summaries of the calculations are presented in this section
and the next, and the underlying detailed calculations are presented in 
Appendices \ref{aamm} and \ref{als}. 

Following Schwinger \cite{schw} we
consider the  process $\clp$ in which an electron that is initially at rest in frame $\clf$
scatters off a weak background 
classical electromagnetic field $\de A\uu c$.  
\colb{
\bea
\clp: \qquad {e\tu -}\str\dn s  
\cdots   
\de A\dnc {\longrightarrow}  \  {e\tu-}\str \dn{s'}  
\leqn{3p1}
\eea}
\hs{-0.1}The dimensionless parameter $\de\ll 1$ determines the strength of the 
 background field.  \colb{$p$ and $q$ are the 4-momenta of the incoming and outgoing electrons.} In frame $\clf$,  $\vec p = 0$   and we work in the limit $|\vq\,|/\me \ll 1$.  $s, s'$ 
 represent the $z$-components of the spins of the incoming and outgoing electrons.  Since both the
 incoming and outgoing electrons are on mass shell, from
 \ct{tau}, we see that in the weak field limit the intrinsic energy scale of the process 
 is $\tau = \me$.
 
\begin{figure}[hbt]
\centering
\includegraphics[height=1.5in]{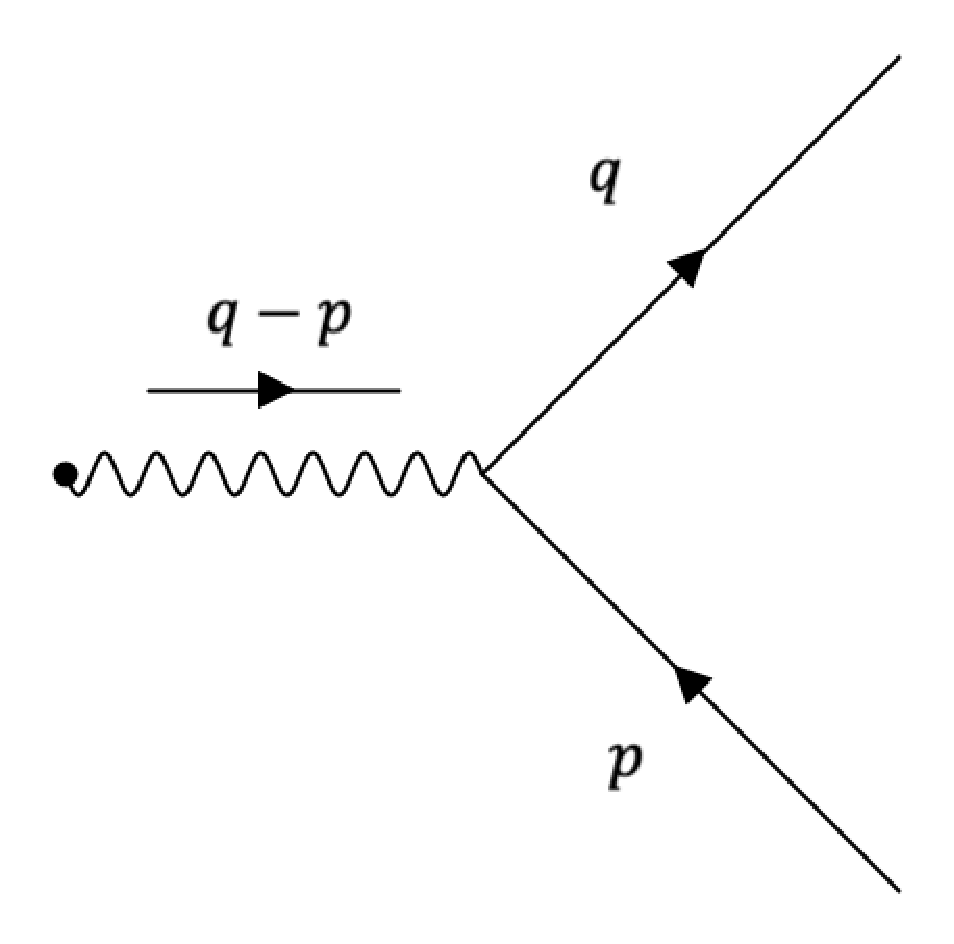}
\hs{1}
\includegraphics[height=1.5in]{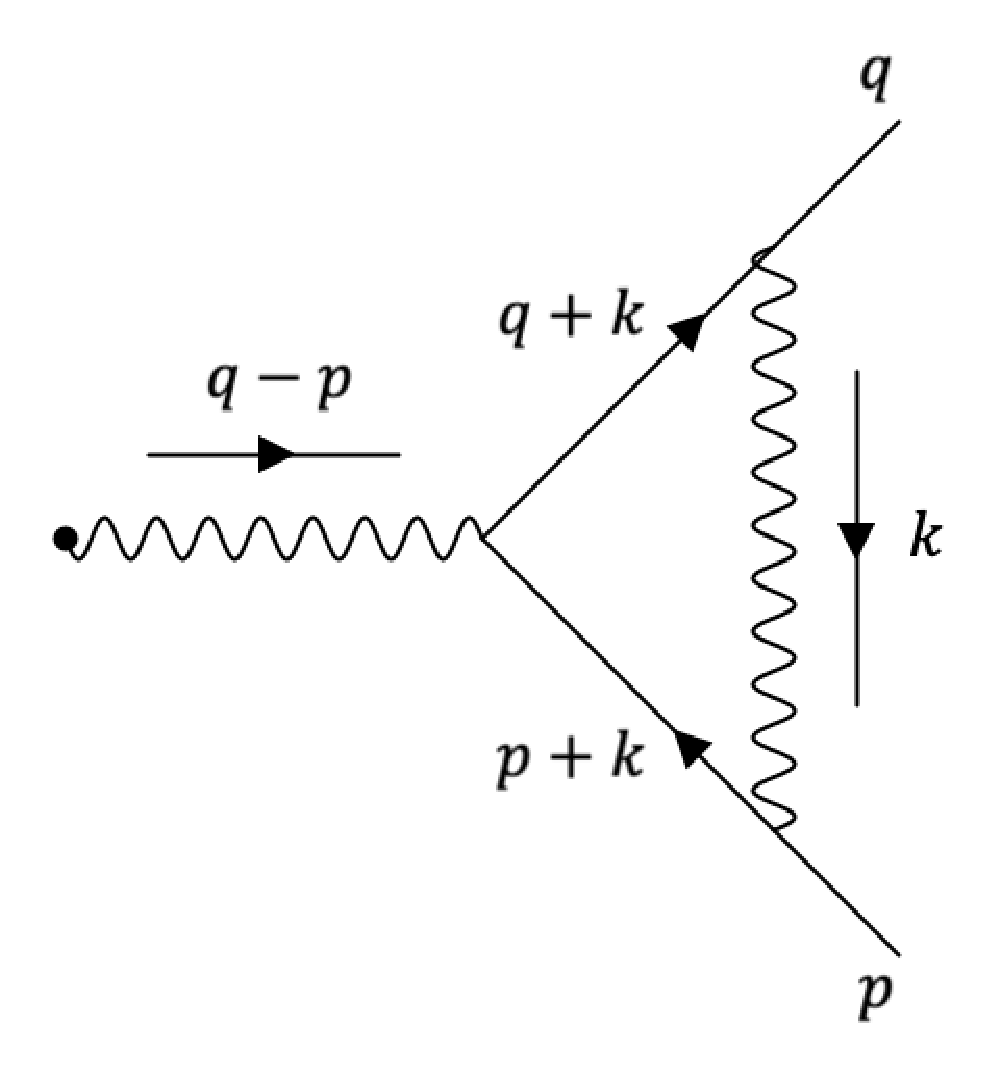}
\caption{\it Tree-level and 1-loop 1-particle irreducible vertex diagrams for an electron
scattering off a classical background electromagnetic field.  The solid circles
represent the classical source.}
\label{figure1}
\end{figure}

 The tree-level and 1-loop 1-particle irreducible Feynman diagrams describing the above
scattering process are shown in Figure \ref{figure1}.
At $O(\de)$ we can expand the total scattering amplitude $\cla(p,s; q,s')$ in powers of $e$.
For brevity, we omit the arguments of $\cla$ and write
\bea
\cla \aea \sum_{j=1}\tu\infty \cla\sj \, e\tu j
\leqn{for1}
\eea
where $\cla\sj$ is the sum of the Feynman amplitudes with $j$ interaction vertices.

The general form of $\cla\sj$ is 
\bea
\cla\sj \aea \lrr{\mathfrak{ g}\uu{tree}} \lrc{\chi\sj(\me)}  
\underbrace{\lrs{\bu{(q,s')} \lrc{ \sigma\umn (p-q)\dnm (A_c)\dnn(p-q)} \ju}}_{
{\cal M}
} + \clb\sj
\leqn{3p2}
\eea
where $\clb\sj$ is the sum of terms that do not contain the factor  ${\cal M}$ 
shown in \ct{3p2}, and hence are not relevant to the calculation of the magnetic moment.
$\mathfrak{g}\uu{tree} = 2$ is the tree-level gyromagnetic ratio of the electron.  

Noting that $A\sw =0$, we have
\beas
\cla \aea \lrs{\mathfrak{g}\uu{tree} \lrc{1 + e\sq \brf{\chi\sr }{\chi\so }+ \ldots }} \lrs{e \,\chi\so \, \clm} +
\clb\so + \clb\sr + \ldots
\eeas
Thus the 1-loop correction to the gyromagnetic ratio is 
\bea
\k\uu{1{\text -}loop} \aea 4\pi \a \brf{\chi\sr}{\chi\so}
\leqn{3p3}
\eea
where $\a$ is the fine structure constant. 

As shown in Appendix \ref{aamm}, 
\bea
\chi\so \aea \dsf{g\uu e(p) g\uu e(q)}{(2 \me) Z \cln}
\leqn{chi1}
\eea
where $\sqrt{Z}$ is the wavefunction {\color{black} normalization} constant for the 
electron field and $\cln$ is a 
normalizing factor.  The Gibbs factors  $g\uu e(p; \clf, \clp)$ and $g\uu e(q; \clf, \clp)$
have been abbreviated to $g\uu e(p)$ and $g\uu e(q)$ for brevity.
 
In Appendix \ref{aamm},  it is also shown that
\beasm
\chi\sr
\aea
-(2i) \d \chi\so
\int
\dsf{d\tu 4 k}{\tpf}
\bsf{ \lrs{g\uu e (p+k)\, g\uu e (q+k) g\uu p (k)}\sq \lrc{(p+k)\upa (q+k)\dna - \me\sq}}{
\lrr{(p+k)\sq - \me\sq + i\e}\lrr{(q+k)\sq - \me\sq + i\e}
\lrr{k\sq  + i\e}}
\leqn{chit}
\eeasm 
From \ct{3p3}, \ct{chi1} and \ct{chit} we have
\beass
\k\uu{1{\text -}loop} \aea - (8\pi \a i) \int
\dsf{d\tu 4 k}{\tpf}
\bsf{ \lrs{g\uu e (p+k)\, g\uu e (q+k) g\uu p (k)}\sq \lrc{(p+k)\upa (q+k)\dna - \me\sq}}{
\lrr{(p+k)\sq - \me\sq + i\e}\lrr{(q+k)\sq - \me\sq + i\e}
\lrr{k\sq  + i\e}}
\eeass 
We note that if set the Gibbs factors $g_e$ and $g_p$ to 1, the integral diverges.
The inclusion of the Gibbs factors renders the integral finite.
The above integral is evaluated numerically.  
First the $k\uz$ integral is calculated using contour integration. The 
subsequent integration is done in spherical coordinates.  Numeric evaluation of the 
integral  yields
\beas
\k\uu{1{\text -}loop} \aea 0.001158990105
\eeas
which differs from  the
experimentally measured value of 0.001159652180
\cite{hann,hhga} by less than 0.06\%.  The above 
prediction is to be compared with Schwinger's $O(\a)$ 
prediction \cite{schw}, which differs from the experimentally
measured value by about $0.15$\%.

\ss{\colb{Lamb shift}}\label{ls}
Lamb and Retherford's measurements \cite{lamb} showed that the 
energy of the $^2S_{\frac 1 2}$ level in Hydrogen atom was 
 higher than 
the energy of the $^2P_{\frac 1 2}$ level by about 1 GHz or 
4.14 $\m$eV \cite{bethe,hind}; also see \cite{anne,eide,kiye,krla,lambnl}.
The measured difference, called the Lamb shift, conflicted with Dirac's
theory, which predicted equal energies for the two levels \cite{hind,grif}. 
The subsequent theoretical explanation of the Lamb shift by Bethe, using
a renormalization argument, has 
been hailed as one of the pivotal advances in the early development
of quantum field theory.  We present a calculation of  the  Lamb shift
based on autoregularization.  
The summary of the calculation is presented in this section, and the
detailed calculation in Appendix \ref{als}.    

Following Bethe, we regard the motion of the electron in Hydrogen atom 
to be non-relativistic  because the relativistic correction to the energy
levels in Hydrogen is known to be small and much of the shift can be
explained using a non-relativistic correction.  In fact, Baranger,
Bethe and Feynman \cite{bbfe} showed that the relativistic correction, which is of
$O( \a^6 )$, contributes  about 7.13 MHz, while the experimentally measured
value of Lamb shift is about 1057 MHz\footnote{\label{fnLambShift} 
The experimental measurements of the Lamb shift are 
1057.77(6) $MHz$ \cite{tpla,tdla}, 1057.90(6) $MHz$ \cite{tpla,rosh},
, 1057.8576 (2.1) $MHz$ \cite{psya,soya}
1057.862(20) $MHz$,  \cite{naun},
1057.845(9) $MHz$ \cite{lupi},
1057.852(15) $MHz$ \cite{vhdr} and
1057.842(12) $MHz$ \cite{hapi}.}.  
The photon field with  which the electron interacts is regarded as a quantum field,
which is regularized using autoregularization. 

At $O(\a)$, the energy of both a free electron as well as a bound electron are shifted on account
of radiative correction---the emission and subsequent re-absorption of a photon by the electron.
Thus the scattering process can be described as 
\colb{
\bea
\clp: {e\tu -}\str  \ra e\tu -   + \g  \ra {e\tu -}\str
\leqn{lsp}
\eea}
where both the incoming and outgoing electrons are on mass shell. 
From \ct{tau}, the intrinsic `energy' scale of the process is $\tau = \me$.  We 
work in the reference frame that is  momentarily comoving with the electron. The 
calculation presented in Appendix \ref{als} shows that the shift in the energy of an electron in
a Hydrogen orbital labeled by  quantum number triple\footnote{
$n\uu s, l\uu s$ and $m\uu s$ denote the principal, orbital and magnetic 
quantum numbers. } $s = (n\uu s, l\uu s, m\uu s)$  is
\beass
\Delta E\uu s \aea  
-\dsf {e\sq} {3\pi\, \me\sq} \int_0^\infty d\kappa\,  \kappa\,
\ti g\uu p\sq (\kappa) 
\, \sum_r \dsf{\left| \mtrxel s {\vec {\bf P}} r \right|\sq}{E\dn r - E\uu s + \kappa}; \ \mbox{ where } 
 \ti g\uu p(\kappa) \dfn \bsf{2}{e^{\kappa/\tau + \tau/\kappa} - 1}\tu{1/4}
\eeass
 The sum above is over all the orbitals of the Hydrogen atom.  The orbitals are labeled with
 triplets of quantum numbers as 
 $r = (n\uu r, l\uu r, m\uu r)$.   $\vec {\bf P}$ is the 3-momentum operator.
 $E\uu r$ and $E\uu s$ denote the energies of the orbitals labeled by  quantum number triples
 $r$ and $s$.  We note that the energy shift $\Delta E\uu s$ diverges if we set $\ti g\uu p(\kappa)=1$,
 but is rendered finite by the Gibbs factor.
   
The self-energy of a free electron also receives a radiative correction
$\Delta E\sps{free}$ at $O(\a)$.   Following Bethe, the observable shift in the energy of the orbital with quantum number $s$ is therefore  
 \beas
 \Delta E\sps{observed}\uu s \aea \Delta E\uu s - \Delta E\sps{free}  
 \eeas
 In Appendix \ref{als} we show that 
 \beasm
 \Delta E\sps{observed}\uu s 
 \aea \dsf {e\sq} {3\pi\, \e\sz\, \me\sq\, c\cu\, \hbar}
\sum_r  \left| \mtrxel s {\vec {\bf P}} r \right|\sq\d (E\dn r-E\uu s) \lrs{\int_0^\infty dz  \d
\dsf{ \ti g\sq (z)}{E\dn r - E\uu s + z}}\qquad 
\leqn{4p1}
 \eeasm
 where we have explicitly shown the fundamental constants $\e\sz, \me, c$ and $\hbar$.  
 
 A straightforward calculation shows that (see Appendix \ref{als}) 
\bea
\sum_r  \left| \mtrxel s {\vec {\bf P}} r \right|\sq\d (E\dn r-E\uu s) \aea 
\dsf{e\sq \hbar\sq}{2\e\sz} |\psi\uu s(0)|\sq
\leqn{4p3}
\eea
where $\psi\uu s(x)$ is the wavefunction of an electron in the orbital that is
labeled by quantum number $s$. 

Since the wavefunction of an electron in the $^2 P\dn{1/2}$ state vanishes at the origin,
from \ct{4p1} and \ct{4p3} we note that  an electron
in the $^2 P\dn{1/2}$ orbital experiences no energy shift due to radiative
correction at $O(\a)$.   The Lamb shift is therefore entirely due to the energy
shift of the $^2 S\dn{1/2}$ level.

Using the known energies of the Hydrogen orbitals   a straightforward
calculation shows that (see Appendix \ref{als})
\bea
1.244289 < \int_0^\infty dz  \d
\dsf{ \ti g\sq_p (z)}{E\dn r - E\uu s + z} < 1.244294
\leqn{4p2}
\eea
We note that the integral in \ct{4p2} diverges logarithmically if $\ti g(z) = 1$. 
Bethe used a 
 UV cutoff to  handle the logarithmic divergence.  Specifically, as
Kroll and Lamb \cite[Footnote, page 388]{krla} observed,
if the retardation and recoil effects are included in Bethe's non-relativistic approximation then
the cutoff $K=2m_e c^2$, which leads to a Lamb shift of 1131.9 $MHz$.
Bethe chose a value of $K=m_e c^2$ to obtain a prediction of  1038 $MHz$ for the shift.

The Gibbs factor $\ti g\uu p\sq(z)$ appearing in the integral in \ct{4p2}
regularizes the logarithmic divergence.  The upper and
lower bounds in \ct{4p2} are   numerical predictions of autoregularization
  that can be compared
with experimental measurements. 
Using \ct{4p1}, \ct{4p3} and \ct{4p2} and 
$\Delta E\sps{observed}\uu {(2,0,0)} = h \Delta \n\uu{LS}$ we 
see that at $O(\a)$ autoregularization predicts a Lamb shift  
of
\beas
1060.476 \,MHz \ <\  \Delta \nu_{_{LS}} \ <\  1060.480\, MHz 
\eeas
which differs from the experimentally measured value of about 1057 MHz
by about 0.33\%.

\ss{\colb{Running fine structure constant}}\label{rcc}
The  calculations  in
Sections \ref{amm} and \ref{ls} pertain to
`{\it soft}' scattering processes in that they involve
momentum transfers that are well below the electron mass
scale. In this section  we present a
 1-loop calculation
that involves momentum transfers that
are considerably above the electron mass scale.
Specifically, we
calculate, at 1-loop, the running of the fine structure `constant' at scales that
are from one to four orders of magnitude above the electron mass threshold.
The prediction of autoregularization, denoted $\a_{_{AR}}(k)$,
agrees with the previously known result, derived using cutoff 
regularization scheme, denoted $\a_{_{CR}}(k)$,
to within 0.8\% as shown in Figure 4.

\begin{figure}[h!]
\centering
\includegraphics[height=1.5in]{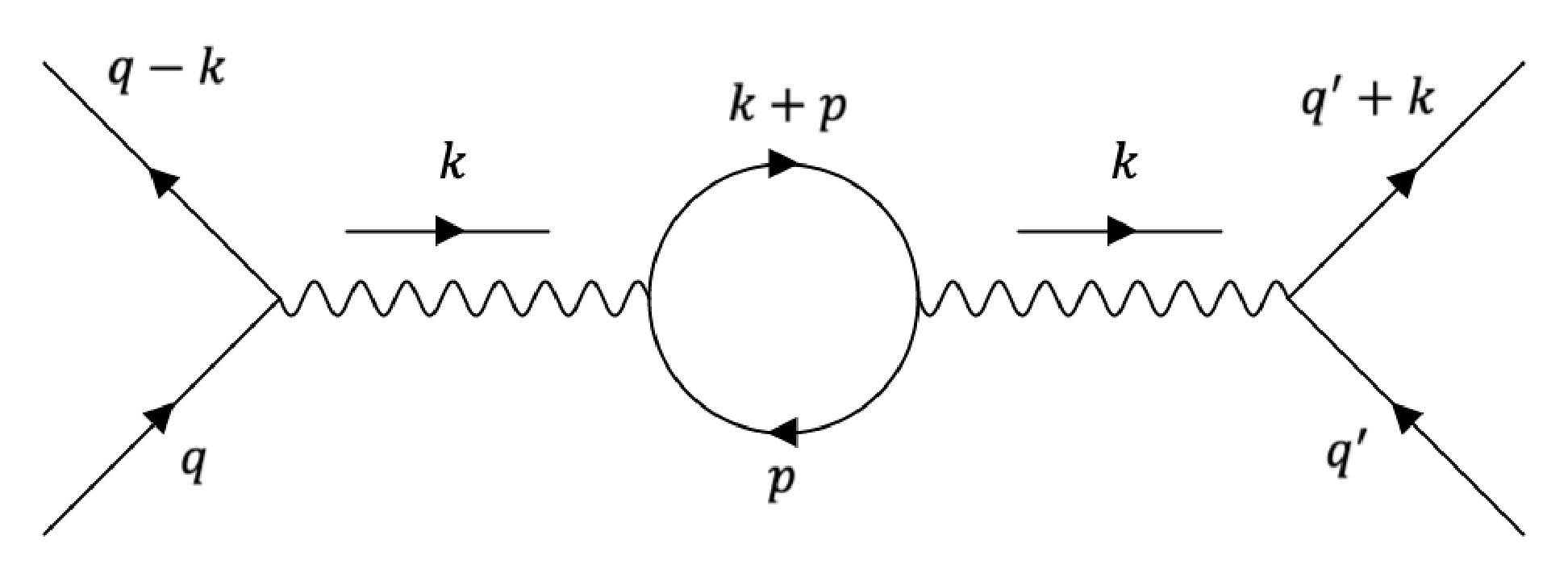}
\caption{\it Electron-electron scattering with 1-loop correction to photon propagator.}
\label{figees}
\end{figure}

Consider the electron-electron scattering process $\clp$\ 
\beas
\clp: \qquad e\tu - + e\tu - \ra e\tu - + e\tu -
\eeas
shown in Figure \ref{figees}. The 4-momentum transferred between the scattering electrons
is denoted $k$.  
We consider scattering in which 
 the incident electrons are nonrelativistic in the 
center-of-momentum frame $\clf$, in which we work, but the transferred momentum $k$ has large 
invariant norm, $\sqrt{|k\upm k\dnm|} \gg \me$.   Thus, from 
\ct{tau}, the intrinsic  scale of the process is 
$\tau \deq \sqrt{|k\upm k\dnm|}$.

The full photon propagator of the interacting photon field is (see Appendix \ref{afpp})
\bea
\ti M\umn\uu{full}(k) \aea \dsf {\ti M\umn(k)}{1- e \sq \d g\uu p\sq(k) \d \chi(k)}
\leqn{phprop}
\eea
where $\ti M\umn(k)$ is the propagator of the free photon field, $g\uu p(k;\clf, \clp)$ is 
abbreviated to $g\uu p(k)$  and 
\bea
\Pi\uu{\a\b}(k) \aea i \, e \sq\, (k\sq \eta\uu{\a\b} - k\dna k\dnb) \,\chi(k), 
\qquad \chi(k) \dfn \dsf{\Pi\upd\m\m(k)
}{3\, i\, e \sq\, k\sq}
\leqn{doch}
\eea 
Following \cite{huan}, we define a scale-dependent coupling
 as
\bea
e\sq(k) \ada \dsf{ Z e \sq}{1-  e \sq \d g\uu p\sq(k) \d \chi(k)}
\leqn{rec}
\eea
where $e$ is the QED coupling constant that appears in the unrenormalized 
QED lagrangian, and $Z$ is a constant.    Since
\beas
\lt k 0 \ g\uu p\sq(k) \, \chi(k) \aea 0
\eeas
we can set $Z=1$ by imposing the boundary condition $e\sq(0) = e\sq = \sqrt{\a/(4\pi)}$,
where $\a$ is the fine structure constant.
That is, we set the coupling constant $e$ in the 
QED lagrangian to be the scale-dependent coupling at zero momentum
transfer, also called the `physical charge' of an electron.   
Again, with some abuse of notation, we define 
$4\pi\a(k) \dfn 
 e\sq(k)$.   $\a(k)$ is the {\it running fine structure
 constant.}   With some abuse of notation
we have denoted $\a(0)$ as $\a$.  From \ct{rec} we have
\beasm
\dsf{\a(k)}{\a} \aea  \dsf{1}{1 - 4\pi \,\a\, g\uu p\sq(k)  \, \chi(k)} \deq 1 + 4\pi \, \a \,g\uu p\sq (k)  \, \chi(k) + O(e^4)
\leqn{npp}
\eeasm
The running of $\a(k)$ with $k$ at $O(\a)$, shown in \ct{npp}, is a prediction of autoregularization.  We compare the
prediction \ct{npp} with previously known results below.

From Figure \ref{figees}, at $O(\a)$ we have
\beass
\Pi\umn(k) \aea e\sq \msrf q \, Tr\lrc{\g\upm D(k+q) \g\upn D(q)},  \qquad D(q) \deq i\d \dsf{g\uu e\sq(q)  \d (q\sl + \me)}{q\sq - \me\sq + i\e}
\eeass
where we have abbreviated $g\uu e(q; \clf, \clp)$ as $g\uu e(q)$.
Therefore,
\beasm
\Pi\upd\m\m(k) \aea e\sq \msrf q \dsf{\gues{k+q}  \d \gues q  \d \lrs{8(k+q)\d q - 16\me\sq}}{((k+q)\sq -\me\sq + i\e)(q\sq - \me\sq + i\e)}
\leqn{pim}
\eeasm
The integral is calculated numerically. The $q\uz$ integral is calculated first using contour
integration and the subsequent integration done using spherical coordinates.  
Substituting \ct{pim} into \ct{doch}, we can evaluate   $\chi(k)$,
and thus the ratio $ \a(k)/\a $ shown in \ct{npp}.

It is well-known \cite[Equation 13.63]{huan} that for large, negative $k\sq/\me\sq$,
cutoff regularization scheme (CR)  yields
\bea
\dsf{\a(k)}{\a } \aea 1 + \dsf \a{3\pi} \, \ln \dsf{|k\sq|}{\me\sq} + O(\a \sq)
\leqn{hpa}
\eea
Denoting the $\a(k)$ in \ct{hpa}, obtained using cutoff regularization,
as $\a\uu{CR}(k)$ and the $\a(k)$ in \ct{npp}, obtained using
autoregularization, as $\a\uu{AR}(k)$, at O($\a $), we obtain, using \ct{npp} and \ct{hpa}, the ratio
\bea
\dsf{\a\uu{AR}(k)}{\a\uu{CR}(k)} \aea \dsf{1 + 4\pi \, \a \, \gues k \d \chi(k)}{1 + \dsf \a {3\pi} \, \ln \dsf{|k\sq|}{\me\sq}}, \qquad \mbox{ at } O(\a )
\leqn{npvshuang}
\eea 

\begin{figure}[h!]  
\hs{-0.5}\includegraphics[width=1.1\textwidth]{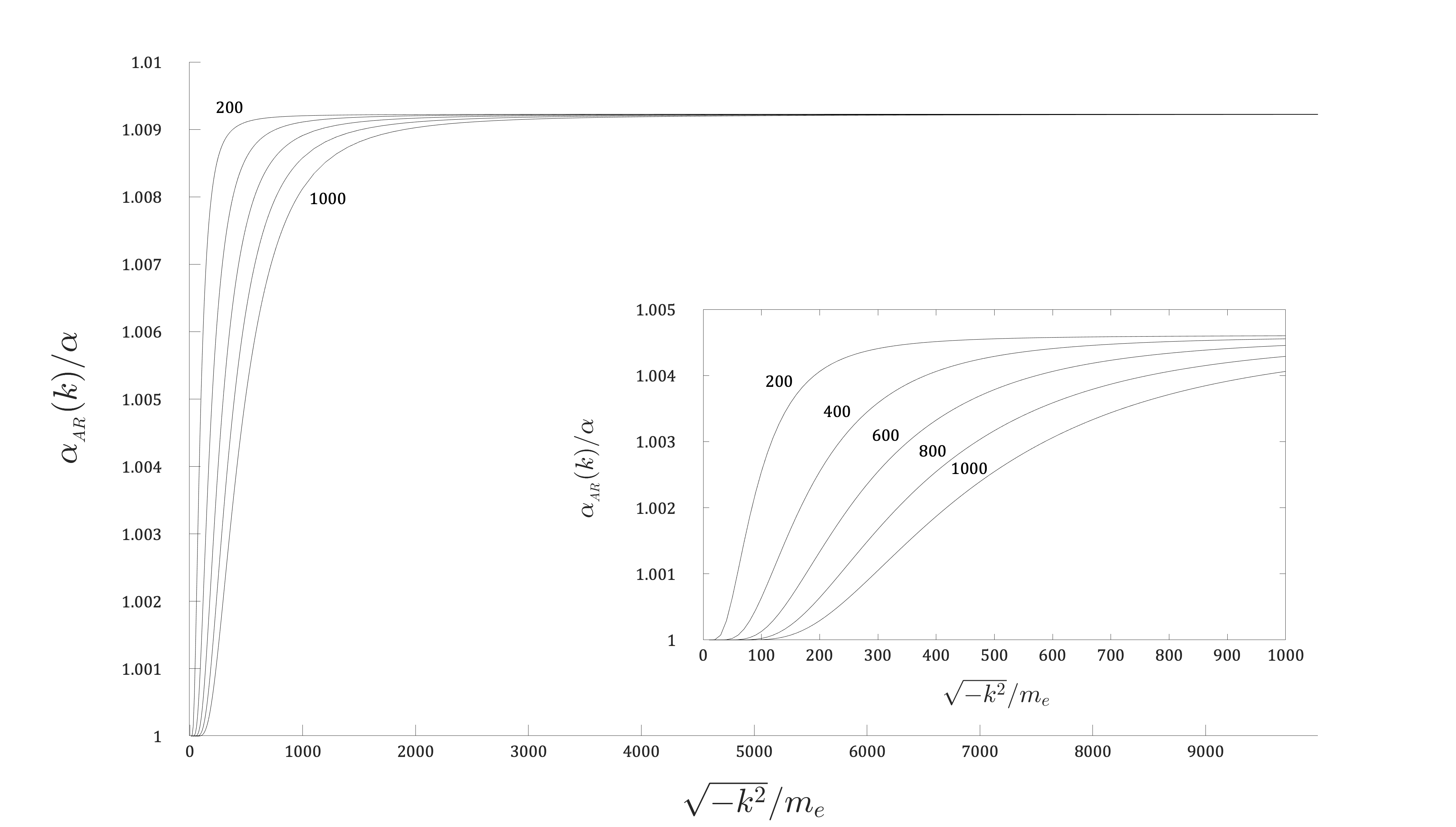} 
\caption{\it Growth of $\alpha_{_{AR}}$ with $\sqrt{-k\sq}/\me $ over the range $10 \leq \sqrt{-k\sq}/\me \leq 10^4$, and for
$k\dnm \clu\upm(\clf;\clf) = 200, 400, 600, 800$ and $1000$. The inset shows the plot for $10 \leq \sqrt{-k\sq}/\me \leq 10^3$.}
\label{figGibbs}
\end{figure}

Figure \ref{figGibbs} shows the plot of $\a_{AR}(k)/\a$, for $10  \leq \sqrt{-k\sq}/\me \leq 10^4 $.
The graphs confirm that the QED coupling constant $\a\uu{AR}(k)$, predicted by autoregularization,
increases with invariant momentum transfer $\sqrt{ -k\sq}$, as expected \cite{huan}.  Figure 4 provides a  
comparison with the previously known results.  The plot of $\a_{AR}(k)/\a_{CR}(k)$
shows that the discrepancy between
$\a_{AR}(k)$, and
$\a_{CR}(k)$ is less than 0.8\% over the above range of $\evl{11}\sqrt{-k\sq}/\me$ and
less than 0.5\% at
momentum transfer scale of 10$^4\,\me$. 
 
\stepcounter{figure}
\bc
\hspace{-0.0in} \includegraphics[width=\textwidth]{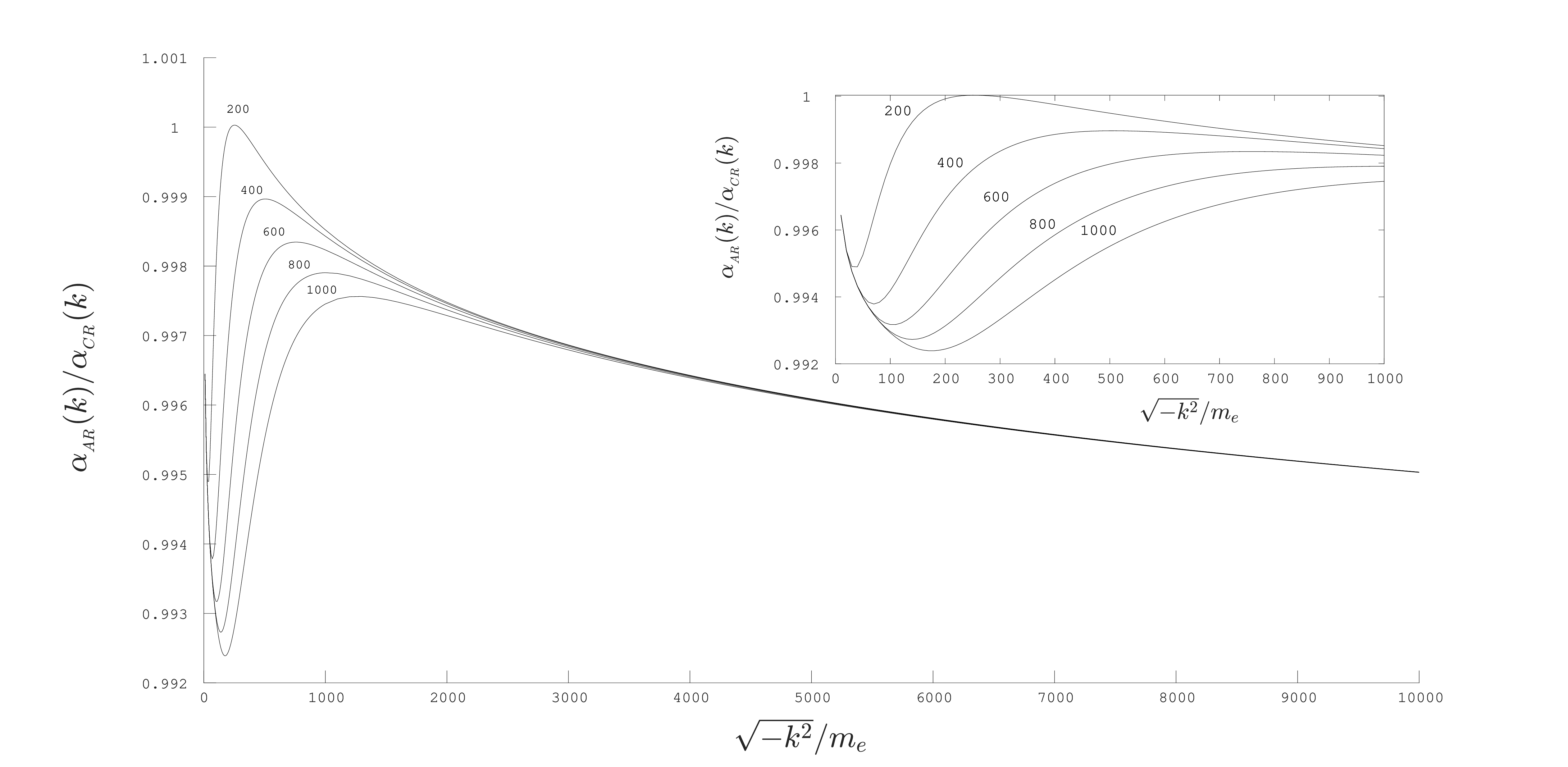} 
\ec
Figure \thefigure: 
{\it Comparison of the prediction of autoregularization ($\alpha_{_{AR}}$) with the prediction of 
cutoff regularization 
($\alpha_{_{CR}}$),  for  $10 \leq \sqrt{ -k\sq }/\me  \leq 10^4 $ and $k\dnm   \clu\upm(\clf;\clf) = 200, 400, 600, 800$ and $1000$.
The inset shows the plot for $10 \leq \sqrt{-k\sq}/\me \leq 10^3$.}

\ss{\colb{Asymptotic freedom in QCD}}\label{QCD}
We compute the running of the QCD coupling constant  at 1-loop and show that autoregularization, like other known regularization schemes, also
predicts the well-known
asymptotic freedom in QCD.

The lagrangian of the QCD sector, denoted $\cll\dn{QCD}$, is 
\bea
\cll\dn{QCD} \ada \bar q\tu f\dn c \lrr{i\psl - g G\!\sl\tu a (T\tu a)\dn{cd}-m\dn f} q\tu f\dn d - \f14 G\dmn\ua G\ua\, \umn  - \dsf 1{2\xi} \lrr{\pmu
G\dnm\tu a}\sq  + \pmu \od a D\dnm \om\ua \nn
G\dmn\ua \ada \pmd G\dnn\ua -\pnd G\dnm\ua - g \lc ars G\dnm\tu r G\dnn \tu s\nn
D\dnm\om\ua \ada \pmd \om\ua - g \lc ars \om\tu r G\dnm\tu s \leqn{lqcd}
\eea 
where $q, G, \om, \om\hc$ denote the quark, gluon, ghost and antighost fields respectively.  The sum over  repeated flavor index $f$ (quark flavors), the color indices $c,d$ and group indices $a, r, s$ is implied. $T\ua = \l\ua/2$ are the generators of $SU(3)$ where $\l\ua$ 
are the Gellmann matrices. $\lc ars$ are the structure constants of $SU(3)$ and $g$   the 
QCD coupling constant.

At $O(g\sq)$  the gluon propagator receives corrections from the four loops shown in 
Figure 5.  The quark loop is shown in solid line, the ghost loop in dashed line and the gluon loops in curly lines.  The gluon loops at $O(g\sq)$ arise from 3-gluon and 4-gluon self-interaction terms in $\cll\dn{QCD}$.  The labels represent momenta.

The full gluon propagator can be written as 
\beass
K\umn(x-y) \aea \msrf k \eto{-ik(x-y)} \ti K\umn(k)\\
\ti K\umn(k) \aea \ti K\sz\tu{\m\a\so}(k) \, 
\sum_{n=1}^\infty \Sigma\uu{\a\so \b\so}(k) 
\ti K\sz\tu{\b\so\a\sw}(k) \, \ldots \Sigma\uu{\a\uu n\b\uu n}(k) \ti K\sz\tu{\b\uu n\n}(k),  \quad
\ti K\uab \sz(k) \deq \dsf{-i \eta\umn \, g\sq_g(k)}{k\sq + i\e}  
\eeass
where $g_g$ denotes the Gibbs factor\footnote{We consider the gluon propagator for a $q + q \ra q + q$ scattering process and work in the center-of-momentum frame
of the scattering process.} of the gluon and $\ti K\sz$ the propagator\footnote{As we show below,
$k\upm \Sigma\dmn(k) \deq \Sigma\dmn(k) k\upn \deq 0$.} of the free gluon.    
 As with the photon propagator we use the ansatz that 
 \bea
 \Sigma\dab(k) \aea i g\sq (g\dab k\sq - k\dna k\dnb) \chi\uu g(k),  \qquad \chi\uu g(k) \deq\dsf{\Sigma\upa\dna(k)}{3 \, i\, g\sq \, k\sq}
 \ \leqn{ansa}
 \eea
 The ansatz implies 
 \bea
 k\upa \Sigma\dab(k) \aea 0 
 \leqn{just}
 \eea
For the photon propagator  $k\upa \Pi\dab(k) \deq 0$  followed from current conservation at the vertex to which the photon line attaches. On the other hand,
the gluon line 
attaches to four types of vertices, as shown in Figure 5.  In order to use the ansatz 
\ct{ansa}, therefore, we need to establish the transversality condition \ct{just} 
for all the four diagrams in Figure 5.  

\stepcounter{figure}
\bc
\igw 3 {quarkLoop} \igw 3 {ghostLoop}\\
\igw 3 {gluonLoop3} \igw {3} {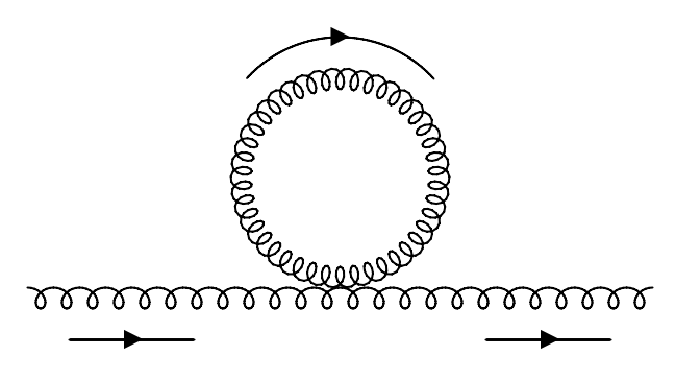}
\ec
Figure \thefigure: {\it 1-loop corrections to the gluon propagator. The quark loop (left) and ghost loop (right) are shown above and the gluon loops below.}\\

Consider the  quark loop at $O(g\sq)$ 
shown in Figure 5.
\beass
I\dn {(q)}\sps{ab}\,\uab(x-y)   \aea \dsf 1{2!}\langle 0 \mid
T \gm a \a(x) \lrs{-i g \int \gm r \m(w)  J\tu r\dnm(w) \, dw}  \nln
\hs {1.2} \lrs{-i g \int \gm s \n(z)  J\tu s\dnn(z) \, dz}   \gm b \b(y)\mid 0\rangle\\
\aea -g\sq \msrf k \eto{-ik(x-y)}  \dm  a b \ \ti K\tu{\a\m}\sz (k) \ \Sigma\sps q\dmn(k) \ \ti K\tu{\n\b}\sz(k)
\eeass
where
\beas
J\tu r\dnm(w) \aea \lrs{ \bar q\tu f \dn c \g\dnm q\tu f \dn d} T\tu r \dn{cd}
\eeas
Since $q\tu f\dn c$ and $q\tu f\dn d$ have the same mass, 
\bea
\pmu J\tu r\dnm  \aea 0 \leqn{cc}
\eea
Owing to the current conservation  \ct{cc}, $k\upm \Sigma\dmn\sps q (k) \deq 0$.

Next, consider the ghost loop.  The ghost lagrangian is
\beas
\cll_{gh} \deq \pmu \od\ua D\dnm \om\ua  \deq 
\pmu\lrc{\od\ua D\dnm \om\ua}- \od\ua \pmu D\dnm \om\ua
\equiv - \od\ua \pmu D\dnm \om\ua 
\eeas
where we have disregarded the divergence term. 
Hence the ghost lagrangian can be written equivalently as
\beas
\cll_{gh}
\aea 
\f12 \lrc{\pmu \od\ua D\dnm \om\ua-\od\ua \pmu D\dnm \om\ua}
\eeas
The ghost-gluon interaction term in the lagrangian is 
\beas
V_{gh} \aea - g \lc abc \lrr{\pmu \od\ua} \om\ub \gk c\m  + 
 g\lc abc \od\ua 
\lrr{\pmu\om\ub}  \gk c \m   +  g\lc abc \od\ua \om\ub \pmu\gk c \m 
\eeas
The third term vanishes due to Lorenz gauge condition $\pmu \gk c\m  \deq 0$.
Therefore
\beas
V_{gh}\aea  g \lc abc \ubc{\lrr{- (\pmu \od\ua) \om\ub  + \od\ua \pmu \om\ub }}{J\upm\, \tu{ab}} \gk c \m
\eeas
Note that 
\bea
\pmd J\upm\, \tu{ab} \aea -\lrr{\Box\,\od\ua } \om\ub + \od\ua \Box\, \om\ub \deq 0
\leqn{cogc}
\eea
since  $\Box\, \od\ua \deq \Box\, \om\ub \deq 0$ for fields $\od\ua$ and $\om\ub$ in the interaction
picture.
The current conservation \ct{cogc} implies that 
$k\upm \Sigma\dab\sps{gh}(k) \deq 0$, establishing the transversality condition for the ghost loop.

The $O(g\sq)$ correction due to the gluon loop containing 3-gluon vertices (bottom left in Figure 5) is\footnote{
since
$
\lc r qs (\pmd \gk r \n - \pnd \gk r\m)   \gm q \m \gm s \n \deq 2 \lc rqs \pmd \gk r \n \gm q \m \gm s \n
$}
\beass
I\tu{ab}_{3g}\, \uab(x-y) 
\aea  g\sq \lc jkl \lc ers \int d\tu 4 w \int d\tu 4z  \, 
\left[-\dm aj   \lrr{K\sz }\mm\a\m(x-w) \rtd\nln
\vev{T \lrs{ \pl\dn\l\lrr{\gm k \l \gk l \m}}_w\lrs{\prd \gk e \sg \gm r \rho \gm s \sg}_z \gm b\b (y)}_c\nln
+ \dm a k \, (K\sz)\tu{\a\m} (x-w) 
\vev{T \lrs{\pmd \gk j \l   \gm l \l}_w\lrs{\prd \gk e \sg \gm r \rho \gm s \sg}_z \gm b\b (y)}_c\nln
\ltd +\dm a l \, (K\sz)\tu{\a\m} (x-w) 
\vev{T \lrs{\pl\dn \l \gk j \m   \gm k \l}_w\lrs{\prd \gk e \sg \gm r \rho \gm s \sg}_z \gm b\b (y)}_c\right]\\
\aea  g\sq \, \lc jkl \, \lc ers\ \dm a b\, \msrf k \eto{-ik(x-y)} \ti K\sz\tu{\a\m}(k) \ \Sigma\sps{3g}\dmn(k) \ \ti K\sz\tu{\n\b}(k) 
\eeass
where $\vev{T \ldots}_c$ denotes the connected diagram, in which the $G$ fields at one vertex Wick-contract
necessarily with the $G$ fields at the other vertex.

In order to show that $k\upm \Sigma\sps{3g}\uu{\m\n}(k) \deq 0$, it is sufficient to show that
\bea
\lc jkl \lc ers\pmu\lrs{- \pld\lrr{\gm k \l \gk l\m } \dm aj  + \dm ak \pmd \gk j \l \gm k \l + \dm al 
\pld \gk j \m   \gm k \l} \aea 0\leqn{tc}
\eea
The first term in \ct{tc} vanishes due to Lorenz gauge condition $\pmu \gk {{(.)}} \m \deq 0$.  The 
second term vanishes since $\Box \gk j \l \deq 0$ in the interaction picture and $\lc jkl$ is totally
antisymmetric.  The third term vanishes due to Lorenz gauge condition and the antisymmetry of 
$\lc jkl$.

Finally, consider the gluon loop comprising the 4-gluon interaction vertex.
\beas
I_{4g}\tu{ab}\, \uab(x-y) \aea -i g\sq \int \lc ajk \lc ars d\tu 4 z \, \vev{T\gm a\a(x) \lrs{\gk j \m \gk k \n
\gm r \m \gm s \n }_z \gm b \b(y)}\\
\aea - i g\sq \, \dm a b \msrf k \eto{-ik(x-y)} \ti K\sz\tu{\a\m} (k) \ \Sigma\sps{4g}\uu{\m\n} (k) \ 
\ti K\sz\tu{\n\b} (k)
\eeas
We note that two of the four gluon fields $\gk j \m,  \,\gk k \n, \,
\gm r \m, \,  \gm s \n$ Wick-contract to form the loop shown (at the bottom right) in Figure 5 leaving 
two terms of the form $\gk u\l \gm v \l$ that Wick-contract with $\gm a\a$ and $\gm b \b$.   Further,
due to the antisymmetry of $\lc ajk$ and $\lc ars$, $u$ and $v$ cannot both belong to either $\lrc{j,k}$ 
or $\lrc{r,s}$.    

Without loss of generality assume that $\gm a\a$ Wick-contracts with $\gm r \m$.  Then to show that
$k\upm \Sigma\sps{4g}\dmn(k) \deq 0$ it suffices to show that 
\beas
\pmu \lrs{\gk j\m \wick{\gk k \n \gm s \n} + \gk k\n \wick{\gk j \m \gm s \n}} \aea 0
\eeas
where $\wick{AB}$ represents Wick-contraction.
The first term vanishes due to Lorenz gauge condition. The Wick-contraction in the second term 
makes the remaining term $\gk k\m$ and hence the second term also vanishes due to Lorenz
gauge condition. 

The above argument shows that 
\beas
k\upm \Sigma\dmn (k) \aea 0
\eeas
where
\bea
\Sigma\dmn(k) \dfn \Sigma\dmn\sps{q}(k) + 
\Sigma\dmn\sps{gh}(k) +\Sigma\dmn\sps{3g}(k) +\Sigma\dmn\sps{4g}(k) \leqn{sg}
\eea
The argument can be adapted to also show that
\beas
k\upn \Sigma\dmn(k) \aea 0 
\eeas
 
Once again, defining 
\beas
P\kk\m\n(k) \ada \eta\kk\m\n - \dsf{k\dnm k\dnn}{k\sq}
\eeas 
and noting $P\upm\dna(k) P\upa\dnn(k) \deq P\mk\m\n(k)$  we see that
\beas
\Sigma\dn{\a\so \b\so}(k) \ti K\sz\tu{\b\so \a\sw}(k) \ldots \Sigma\dn{\a\dn{n-1} \b\dn{n-1}}(k) 
\ti K\sz\tu{\b\dn{n-1} \a\sn}(k)  \aea \l\tu{n-1} P\mk{\a\sn}{\a\so}(k) , \qquad \l \deq g\sq \, g\dn g\sq(k) \, \chi\dn g(k) 
\eeas
Therefore,
\beas
\ti K\umn(k) \aea \ti K\sz \tu{\m\n} (k)  +  \ti K\sz\tu{\m\a} (k)
\lrs{\sum_{n=1}^\infty \l\tu{n}} P\mk{\n }{\a}(k) 
\deq \ti K\sz\umn(k) + 
\ti K\sz\umn(k) \bsf \l {1-\l} P\mk\n\a(k) \\
\ada   \ti K\sz\umn(k) + 
\ti K\sz\umn(k) \lrs{\ti g\sq(k)  g\dn g\sq(k) \chi\dn g(k)} P\mk\n\a(k) \\
\eeas
where the scale-dependent coupling constant 
is 
\beas
\ti g\sq(k) \ada \dsf{g\sq}{1 - g\sq g\dn g\sq(k) \chi\dn g(k)}
\eeas
Since $g\dn g\sq \chi\dn g(k) \ra 0$ as $k\ra 0$, 
$g \deq \ti g(0)$.  Defining 
$\a\uu{QCD}(k) \dfn 4\pi \ti g\sq(k)$ to be the QCD coupling
constant at momentum $k$, we obtain
\bea
\a\uu{QCD}(k) \aea \dsf{\a\uu{QCD}(0)}{1- 4\pi \a\uu{QCD}(0) \, g\dn g\sq(k) \chi\dn g(k)}, 
\leqn{rqcd}
\eea
If $\chi\uu g(k) < 0$ and $|g\uu g\sq(k) \chi\uu g(k)|$ increases as $-k\sq$ increases then the QCD coupling
constant $\a\uu{QCD}(k)$ decreases with 
the gluon propagator scale
$\sqrt{-k\sq}$, signaling asymptotic freedom.

From \ct{ansa} and \ct{sg}, the calculation of $\chi\dn g(k)$ reduces to the calculation of
$\Sigma\dab\sps{q}(k), \Sigma\dab\sps{gh}(k),$ $\Sigma\dab\sps{3g}(k)$ and $ \Sigma\dab\sps{4g}(k)$.
A lengthy calculation using $\cll\uu{QCD}$  in \ct{lqcd} yields
\beasm
(\Sigma\sps q)\tu \a\dn \a  \aea  \brf{1} 2 g\sq \sum_{f} \msrf p Tr\lrc{\gau \ti Q\tu f(k+p) \gad \ti Q\tu f(p)}, \hs{0.7} \ti Q\tu f(p) \dfn \dsf{i (g\dn q\tu f(k))\sq}{k\sl - m\tu f + i\e} \nn
(\Sigma\sps{gh})\tu \a\dn \a  \aea 3\, g\sq \msrf p p\upa (k+p)\dna \ti H(k+p) \ti H(p), \hs{1.4}
\ti H(p) \dfn \dsf{i g\dn {gh}\sq (p)}{p\sq + i\e}
\nn
(\Sigma\sps{3g})\tu \a\dn \a  
\aea -3 g \sq
\msrf p \ti G(p) \ti G(k-p) \lrs{ 12 k\upa k\dna + 9 p\upa p\dna - 15 k\upa  p\dna  
  }, \hs{0.2} \ti G(p) \dfn \dsf{-i \, g\dn g\sq (p)}{p\sq + i\e}\nn
(\Sigma\sps{4g})\tu \a\dn \a  \aea   -36 \, i \, g\sq   \msrf p \ti G(p)  \leqn{hp}
\eeasm  

\stepcounter{figure}
\bc
\igh 2 {qqScattering}
\ec
Figure \thefigure: {\it Scattering of two quarks through gluon
exchange.  The labels represent 4-momenta.  The figure shows
only one of the possible gluon loops.}

\vs{0.1}

Consider  a scattering process $\clp$ in which two quarks 
scatter through a gluon exchange.   Since we seek to study the variation of $\a\uu{QCD}(k)$
with increasing $-k\sq$, we assume that 
 the  momentum
transfer scale $\sqrt{-k\sq}$  is the dominant scale
of the scattering process and hence $\tau\uu\clp\deq \sqrt{-k\sq}$.
The Gibbs factors of gluons and ghosts are given by the 
expressions for the Gibbs factors of massless bosons\footnote{Although ghosts and antighosts
are strange in that they
are described by Grassmann fields.}, both with two degrees
of freedom ($d\uu\vp \deq 2$ in \ct{2p5}).  
Numerical calculation yields the following plot for 
$\a\uu{QCD}(k) /\a\uu  Q(0)$, confirming that the QCD coupling
constant $\a\uu{QCD}(k)$ does 
decrease with increasing $\sqrt{-k\sq}$ as shown.  Thus 
autoregularization, like other known regularization schemes,
also predicts the asymptotic freedom in QCD. 

\stepcounter{figure}
\bc
\igh 3 {asymptoticFreedom}

Figure \thefigure: {\it Running of the QCD coupling constant $\a\uu{QCD}(k)$ with 
$\sqrt{-k\sq}$.}
\ec
\ss{\colb{Scattering cross sections at tree-level}}\label{tls}
In the following subsections we use autoregularization to 
calculate the tree-level cross sections  of  two processes---Compton scattering and $e^+$-$e^-$ pair 
annihilation---and compare the theoretical calculations with 
experimental data. While the energy scale of the experimental data for Compton scattering is   $\sim $ MeV, 
the center-of-momentum energy of the pair annihilation experiment is $\gtrsim 206 $ GeV.  For both processes,
the predictions of autoregularization are found to be in good agreement with experimental data. 
\sss{\colb{Compton scattering}}\label{acs}
 In
the following discussion
we  
compare a tree-level prediction
of autoregularization  with experimental
data.  Specifically, we calculate the angular
distribution of the differential
cross section for Compton scattering  at tree level using autoregularization and compare
the prediction with that of the well-known Klein-Nishina formula \cite{klni}
as well as the experimental measurements of Friedrich and Goldhaber
\cite{frgo}.
We perform the calculation  in the laboratory frame.

\begin{figure}[hbt]
\centering
\includegraphics[height=2.1in]{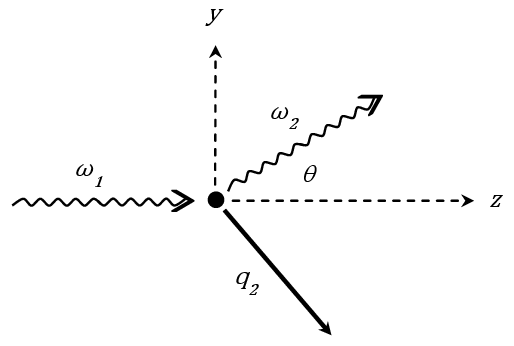}
\caption{\it The  kinematics of Compton scattering. }
\label{figurey}
\end{figure}

In Compton scattering
\bea
\clp: \quad \gamma + e^- \ra \gamma + e^- \leqn{cscat}
\eea
  incoming photon and electron scatter to outgoing photon and electron as
 shown in Figure \ref{figurey}.
We assume the
scattering occurs in the $y-z$ plane in the laboratory frame.  An
incoming photon of energy $E_\g=\oso \me c\sq = \oso$ (
$\me=c=1$ in natural units) collides with an electron, which
is initially at rest in the laboratory frame, and scatters to an outgoing
photon of energy $E_\g' = \osw\me c\sq = \osw$   at an angle $\th$ 
relative to the $z$-axis. The
4-momentum of the outgoing electron in laboratory frame
is denoted $\qw$.

Momentum conservation and the requirement that the outgoing
 electron is on mass shell determines $\osw$ given
 $\oso$ and $\th$.  Specifically, $\osw$ is given by
 \bea
 \osw \aea \dsf{\me\oso}{\me+\oso(1-\cth)} \leqn{osw}
 \eea

We choose $\e(\ko,1) = (0, 1, 0, 0)$ and $\e(\ko,2) = (0,0,1,0)$ to be the basis transverse
polarizations corresponding to the incoming photon and  $\e(\kw,1) = (0,1,0,0)$
and $\e(\kw,2) = (0,0,\cth,-\sth)$ to be the basis transverse polarizations corresponding
to the outgoing photon.  We label the polarizations of the incoming and
outgoing photons, $\e(\ko,r\so)$ and $\e(\kw,r\sw)$ respectively, with $r\so, r\sw = 1,2$.

\begin{figure}[hbt]
\centering
\includegraphics[height=2.1in]{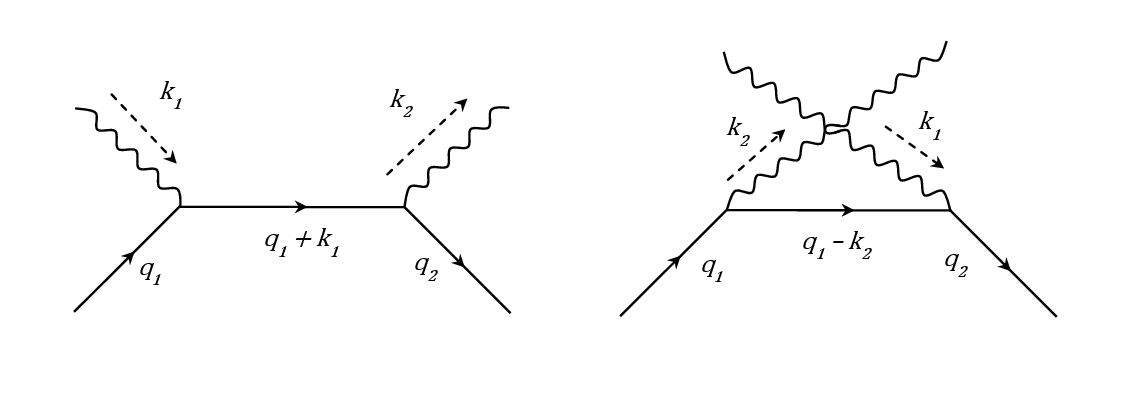}
\caption{\it The  Feynman diagrams that contribute to the tree-level amplitude
of Compton scattering.}
\label{figurex}
\end{figure}

At the tree level the scattering amplitude with incoming photon (electron) 
polarization\footnote{With slight abuse of notation we use the term
polarization to denote the helicity of an electron.}
labeled $r\so$ ($s\so$) and outgoing photon (electron) polarization $r\sw$ ($s\sw$),
denoted $\cla\sw$,
receives contributions from the two Feynman diagrams shown in Figure \ref{figurex},
and is given by
\beass
\cla\sw(\lrc{\ko,r\so}, \lrc{\qo,s\so}; \lrc{\kw,r\sw}, \lrc{\qw,s\sw}) \aea  -i \tpf \delf{\qw+\kw-\qo-\ko} \ \clm\sw
\eeass
where at $O(e\sz\sq)$ we have
\beasm
\ko \aea (\oso, 0, 0, \oso), \ \kw\deq (\osw,0,\osw\sth,\osw\cth),\nl 1
\qo \aea(\me,0,0,0), \
 \qw \deq (E\sw\,',0,-\osw\sth,\oso-\osw\cth),
\quad
 E\sw\,' := \me+\oso - \osw\nl 1
 \clm\sw \aea - e\sz\sq  \  \chi(\om\so) \  \lrs{ \bar u(\qw,s\sw) \ \Gamma\uu{AR} \ u(q\so,s\so)} \deq - 4 \pi \a
 \ \chi(\om\so) \lrs{ \bar u(\qw,s\sw) \ \Gamma\uu{AR} \ u(q\so,s\so)},\nl 1
 \chi(\om\so) \ada
 \dsf{\gpf{\ko} \gpf{\kw} \gef{\qo}\gef{\qw}}{Z\uu e\, Z\uu\gamma\, \cln} \ \nl 1
 \aea \gpf{\ko} \gpf{\kw} \gef{\qo}\gef{\qw}   \nl 1
 \aea  \gpf{\ko}\sq \gef{\qo}\sq \nl 1
\Gamma\uu{AR} \aea
\gef{\qo+\ko}\sq \bsf{\e\,\sl(k\sw, r\sw) \, (q\sl \so + k\sl\so + \me) \e\,\sl(k\so,r\so)}{(q\so + k\so)\sq - \me\sq + i\e} \nn
\aba
+
\gef{\qo-\kw}\sq \bsf{\e\,\sl(k\sw, r\sw) \, (q\sl \so - k\sl\sw + \me) \e\,\sl(k\so,r\so)}{(q\so -  k\sw)\sq - \me\sq + i\e}
\leqn{Gamm}
\eeasm
where $\sqrt{Z\uu e}$ and $\sqrt{Z\uu \gamma}$ are the wavefunction {\color{black} normalization} constants for electron and photon fields;
$\cln$ is a {\color{black} normalizing factor}.  
  At the lowest order\footnote{
 As $e  \ra 0$, $\psi \ra \psi\dnfree,$ $H\dni \ra 0$ and $U(\infty,-\infty) \ra 1,$ 
 $Z\uu e, Z\uu \g \ra 1$,
 and $\cln \ra 1$.  
} in Compton scattering $Z\uu e=Z\uu\g=\cln = 1$.  To show the last equality for
$\chi(\om\so)$,   we show below that $\gpf{\kw} \deq \gpf{\ko}$
and $\gpf{\qw} = \gpf{\qo}$.

The Lorentz tranformation $\Lambda(\clfp,\clf): \clf \ra \clfp$ from the laboratory frame $\clf$ to the
center-of-momentum frame $\clfp$ of the scattering process is
\bea
\Lambda(\clfp,\clf) \aea \mtrx{cccc}{\g \j \j \j -v\g\\ \j 1 \j \j \\ \j \j 1 \j \\ -v\g \j \j \j \g}, \quad v \deq \dsf{\om\so}{\me + \om\so}, \quad
\g\inv \dfn \sqrt{1-v\sq} \quad
\leqn{lort}
\eea
Recalling the definition in  
\ct{2p2} and noting that 
\bea
\xi\uu A\upa(q; \clf, \clp) \clu\dna(\clfp; \clf) \aea \xi\uu A\upa(\Lambda(\clfp, \clf) q; \clf, \clp) \clu\dna(\clfp; \clfp) 
\leqn{lig}
\eea
 we have
\beas
\xi\upa\uu A (\kw; \clf,\clp) \, \clu\dna(\clfp;\clf) \aea \pi\uu A\lrr {\Lambda(\clfp,\clf) \kw}\uz  \deq \lrs{ \Lambda(\clfp,\clf) \kw }\uz
\eeas
where the subscript $A$ denotes the Maxwell field.  As defined in \ct{pro}, the projection operator
$\pi\uu A$ projects a 4-momentum $k$ to a photon's mass shell.
The second equality above holds since $\kw$ is on mass shell.   Using the form of $\kw$, shown in \ct{Gamm},
and the expression for $v$, shown in \ct{lort}, and \ct{osw} we have
\bea
\xi\upa\uu A (\kw; \clf,\clp) \, \clu\dna(\clfp;\clf) \aea \om\sw \g (1- v\cth) \deq \dsf{\g\, \me\, \om\so }{\me+\om\so}\deq \me\, v\, \g \leqn{k2}
\eea
Again, using \ct{2p2},   
\ct{Gamm} and \ct{lort} we have
\bea
\xi\upa\uu A(\ko; \clf,\clp) \, \clu\dna(\clfp;\clf) \aea \g\, \om\so(1-v) \deq \me v \g \leqn{k1}
\eea
From \ct{k2}, \ct{k1} and \ct{2p5}  
it follows that
\bea
\gpf{\ko} \aea \gpf{\kw} \leqn{k12}
\eea
Similarly, we have, noting that $\om\so (1-v) \deq  \om\sw(1-v\cth) \deq \me v$, and that $\qo, \qw$ are on mass-shell
\bea
\xi\uu \psi\upa (\qo; \clf,\clp) \, \clu\dna(\clfp;\clf) \aea \me \, \g \deq 
\xi\uu \psi\upa (\qw; \clf,\clp) \, \clu\dna(\clfp;\clf)  \leqn{q12}
\eea
where the subscript $\psi$ denotes the electron field. 
From  
\ct{2p5} and \ct{q12} we have
\bea
\gef{\qo} \aea \gef{\qw} \leqn{qq12}
\eea
From \ct{k12} and \ct{qq12} it follows that $\chi$, defined in \ct{Gamm}, is
\bea
\chi  \aea \gpf{\ko} \, \gpf{\kw} \, \gef{\qo}\, \gef{\qw} \nl 1
\aea  \lrs{\gpf{\ko}\, \gef{\qo}}\sq  \leqn{chio}
\eea
as claimed in \ct{Gamm}.   \ct{chio} shows that $\chi$, depends only on $\om\so$ and is independent
of $\th$, although $\kw$ and $\qw$ depend on $\th$.  Hence we write $\chi$ as  $\chi(\om\so)$ in \ct{Gamm}.

The differential cross section in the laboratory frame $\clf$ for prespecified polarizations and helicities  of the incoming and outgoing photons
and electrons is given by
\cite[Equations 11.9, 11.12, 11.22, 11.23]{sred}
\beas
(d\sigma_{r\so, r\sw, s\so, s\sw})\uu{AR} \aea \lrs{\izi d|\vec k\sw| \, |\vec k\sw|\sq  \int d\vec\qw \lrc{\dsf {\delf{\ko+\qo-\kw-\qw}  \d |\clm\sw|\sq } {(16\, \oso\, \osw\, \me \, E\sw\,' )(2\pi)\sq }}}
 d\Omega
 \eeas
 where $d\Omega \dfn \sth \, d\th\, d\varphi
$.  Performing the integration with respect to $\vec \qw$, and setting
$\kappa \dfn |\vec\kw|$, we have
\beas
(d\sigma_{r\so, r\sw, s\so, s\sw})\uu{AR} \hs{5}\\
\deq \lrs{\izi d\kappa \, \kappa\sq \,
\bcf{\de\lrr{\oso + \me - \kappa - \sqrt{\me\sq +\oso\sq + \kappa\sq
-2\oso \kappa\cth}} |\clm\sw|\sq}{(16\, \oso\, \osw\, \me \, E\sw\,' )(2\pi)\sq} } d\Omega
\eeas
Using \ct{osw} and a property\footnote{
\beas
\de(f(x)) \aea \sum_{x\si} \dsf{\de(x-x\si)}{|f'(x\si)|}, \qquad \mbox{ where } f(x\si) = 0
\eeas
} of $\de$-function we note that
\bea
\de\lrr{\oso + \me - \kappa - \sqrt{\me\sq +\oso\sq + \kappa\sq
-2\oso \kappa\cth}} \aea  \de(\kappa - \osw)\d \brf{E\sw\,' \osw}{\me\oso},
\leqn{kint}
\eea
Performing the integration with respect to $\kappa$  using \ct{kint},
averaging over the
possible polarizations of the incoming photon, and summing over
the possible spins of the incoming and outgoing electrons and the possible polarizations
of the outgoing photon we have,
\bea
\brf{d\sigma}{d\Omega}\uu{AR} \aea \dsf{1}{\lrr{16 \,\me\sq}(2\pi)\sq } \brf{\osw}{\oso}\sq
\lrs{\brf 1 2 \sum_{r\so, r\sw=1}^2 \sum_{s\so, s\sw=1}^2 |\clm\sw|\sq
}
\leqn{trph1}
\eea
At the tree level, using \ct{Gamm}, we have
\bea
\brf{d\sigma}{d\Omega}\uu{AR} \aea \dsf{\a\sq}{8 \me\sq} \brf{\osw}{\oso}\sq \, \clt\uu{AR}, \quad \mbox{ where } \nl 1\quad
\clt\uu{AR} \ada
(\chi(\om\so))\sq \sum_{r\so, r\sw=1}^2 \sum_{s\so, s\sw=1}^2
|\bar u(\qw,s\sw) \ \Gamma\uu{AR} \ u(q\so,s\so)|\sq
\leqn{npdcr}
\eea
Using the trace identities we have
\beasm
\clt\uu{AR}
\aea
\sum_{r\so, r\sw = 1}^2  \left[
\lrc{\gef{\ko+\qo}^4} \bsf{T_{11}}{4(\qo\d\ko)\sq} +  \lrc{\gef{\qo-\kw}^4}\bsf{T_{22}}{4(\qo\d\kw)\sq}  \rdot \nl 1
\aba \ldot + \lrc{2 \gef{\ko+\qo}\sq \gef{\qo-\kw}\sq} \lrs{\dsf{T_{12}}{4 (\qo\d\ko)(\qo\d\kw)}}\right] (\chi(\om\so))\sq
\leqn{tgibbs}
\eeasm
where, using the abbreviations $\ew \dfn \e(\kw,r\sw),$ and $\eo \dfn\e(\ko,r\so)$,
\beas
T_{11} \aea 16 (\ko\d\qo) (\ew\d\ko)\sq + 8  (\qo\d\ko)( \qo\d\kw)\\
T_{12}
\aea 16 (\eo\d\ew)\sq  (\qo\d\ko)(\qo\d\kw) +  8(\ko\d\qo) (\eo\d\kw)\sq - 8 (\qo\d\kw) (\ew\d\ko)\sq \\
\aba -4  (\qo\d\kw)\sq + 4(\ko\d\kw)\sq   - 4(\ko\d\qo)\sq  \\
T_{22} \aea - 16 (\eo\d\kw)\sq (\qo\d\kw) + 8 (\qo\d\ko)(\qo\d\kw)
\eeas
The Gibbs factors  in \ct{tgibbs} are calculated using \ct{2p5}.  

If we set all the Gibbs factors to 1  in \ct{Gamm}
and in \ct{tgibbs} we get 
\beasm
\brf{d\sigma}{d\Omega} 
\aea \dsf{\a\sq}{8 \me\sq} \brf{\osw}{\oso}\sq \clt\uu{HP}
\deq \dsf{\a\sq}{8 \me\sq} \brf{\osw}{\oso}\sq \lrs{\dsf{2 (\ko\d\kw)\sq}{(\qo\d\ko)(\qo\d\kw)} + 8 (\eo\d\ew)\sq}
\leqn{hpdcr}
\eeasm
which is the well-known Klein-Nishina formula\footnote{
In the laboratory frame, using \ct{osw}, we have $\ko\d\kw \deq \oso\osw (1-\cth) \deq m\,\oso\,\osw\lrr{\dsf 1 {\osw} - \dsf 1 \oso}$,
$\qo\d\ko \deq \me \oso$, and $\qo\d \kw \deq \me\osw$. Therefore,
\ct{hpdcr} can be rewritten as
\beas
\brf{d\sigma}{d\Omega} 
\aea
\dsf{\a\sq}{4 \me\sq} \brf{\osw}{\oso}\sq \lrs{\dsf\oso\osw +
\dsf\osw\oso - 2 + 4 (\eo\d\ew)\sq}
\eeas
which is the familiar form of the Klein-Nishina formula \cite[Equation 8.7.39]{wein}.
} \cite{huan,klni}.

\begin{figure}[hbt]
\centering
\includegraphics[width=\textwidth]{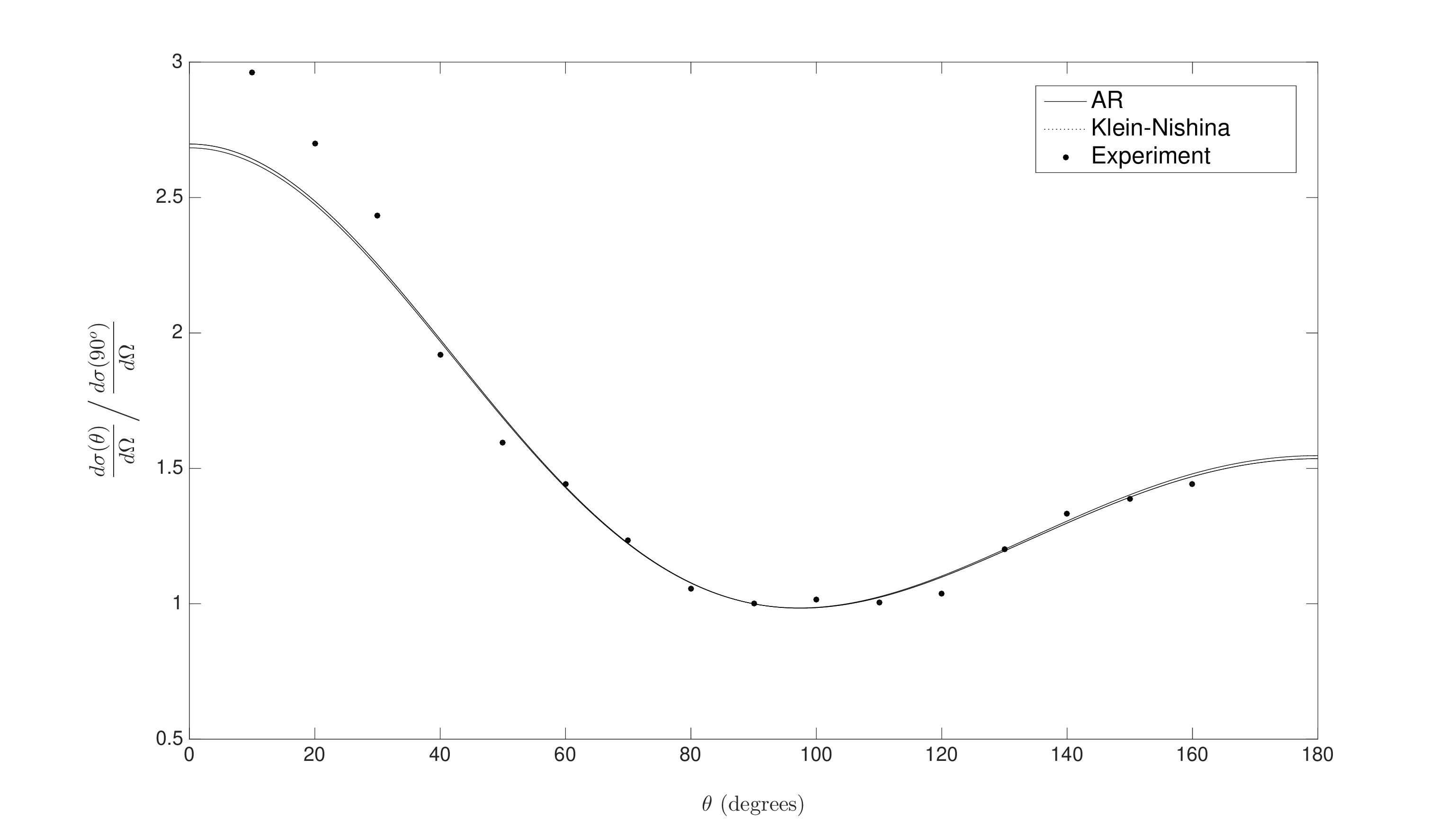} 
\caption{\it  Angular distribution of the differential 
cross section for Compton scattering expressed as a fraction of the value at $\theta=90^o$.  
The incident
photon energy is $\om\so = 0.173\  \me c\sq$.}
\label{figurew}
\end{figure}

For large $\om\so$ (incident photon energy) pair production is known to mask the Compton scattering \cite{muir}.    Experimental studies of the
Compton scattering have focused largely on the angular distribution of differential
scattering cross section for unpolarized incident photons.  The widely
referenced experimental data is due to Friedrich and Goldhaber
\cite[Page 703, Table 1]{frgo} for incident photon energy of $\om\so = 0.173 \, \me c\sq$.  They present intensity of scattered radiation at angle $\theta$ as a fraction of the intensity at $\th=90^o$.  In order to compare the prediction of autoregularization with that of Friedrich and Goldhaber's data, we have plotted the differential cross section at angle $\th$ as a fraction of the value at $\th=90^o$ for both autoregularization---Equation \ct{npdcr}---and the
Klein-Nishina formula
---Equation
\ct{hpdcr}---in Figure \ref{figurew}.
The intrinsic scale, $\tau$, of the scattering process is calculated in autoregularization numerically using 
\ct{tau} and \ct{Gamm}. 
Figure \ref{figurew} also shows
Goldhaber's experimental measurements, taken at $\th=10^o, 20^o,
\ldots, 160^o$.  \colb{The raw data underlying the graphs is presented in the 
footnote\footnote{\colb{In the following table the experimental measurements are labeled Exp.,
the predictions of autoregularization are labeled AR and the predictions of the Klein-Nishina
formula are labeled KN.  The angles are in degrees.  The entries in the table are expressed as 
fractions of the value at 90$^o$. Please see the discussion in the main text.}
\fs{10}{11}{
\bc
\colb{
\begin{tabular}{|l|c|c|c|c|c|c|c|c|}\hline
$\th$ (in $^o$) \j 10 \j 20 \j 30 \j 40 \j 50 \j 60 \j 70 \j 80   \\ \hline
Exp. \j 2.963 \j 2.699 \j 2.434 \j 1.919 \j 1.595 \j 1.443 \j 1.234 \j 1.055\\
AR\j 2.6428 \j 2.4869 \j 2.2537 \j 1.9765 \j 1.6918 \j 1.4324 \j 1.2229 \j 1.0775\\
KN\j 2.6291 \j 2.4747 \j 2.2435 \j 1.9686 \j 1.6861 \j 1.4285 \j 1.2206 \j 1.0764 \\ \hline
$\th$ (in $^o$)\j 90 \j 100 \j 110 \j 120 \j 130 \j 140 \j 150 \j 160   \\ \hline
Exp. \j 1 \j 1.016 \j 1.004 \j 1.037 \j 1.202 \j 1.332 \j 1.387 \j 1.442\\
AR\j 1 \j 0.9856 \j 1.0232 \j 1.0984 \j 1.1951 \j 1.298 \j 1.3932 \j 1.4696\\
KN\j 1 \j 0.9867  \j 1.0258 \j 1.1024 \j 1.2009 \j 1.3054 \j 1.4022 \j 1.4798\\ \hline
\end{tabular}
}
\ec}
}}. 

The RMS (Root Mean Square) error of
 the prediction of autoregularization  relative to the 
experimental data is 0.11195, which is about 4.02\% smaller than the RMS error
of 0.11664 for the prediction of the Klein-Nishina formula with respect to the same data.

\colb{In the above calculation of Compton scattering amplitude we made two 
assumptions. First, we assumed that the scattering 
occurs in vacuum,  devoid of any background field.  Secondly,
we assumed that the scattering photon and electron are free
particles in the asymptotic past and future.  
}

\colb{For several scattering processes   of interest 
in atomic, molecular and optical physics,  and astrophysics,  one or both of the above assumptions need to be relaxed.  For example,  
the photon scattering in the Delbruck process  
occurs
in the background of  a nuclear Coulomb field \cite{delb1,delb2,pamo}.  
Compton scattering that occurs in the  intense magnetic field 
of neutron stars is believed to play an important role in the 
interaction between matter and radiation in the vicinity of 
a neutron star \cite{mnpo}. 
Examples of  Compton scattering involving bound electrons  
include  
the scattering of a photon  
 off a bound electron in an atom \cite{kirc} or off a bound electron
in positronium \cite{kppr}.}

\colb{Autoregularization can be adapted to strong-field and bound-state
scattering problems, and may offer distinct advantages.  The primary
advantage is that the exponential suppression due to the
Gibbs factors makes the integrals hyperconvergent,  facilitating 
numerical calculations. For example, the forbidding computational
complexity of the Delbruck
scattering calculations has restricted theoretical calculations to
certain ranges of energy and/or scattering angles \cite{delb1,delb2,syss}; reduction
of the computational complexity by autoregularization might enable calculations to span broader 
ranges, including the 20-100 MeV range \cite{delb1,delb2}. }

\colb{A second advantage is that 
autoregularization naturally incorporates the effects of  changes in the energy density
and particle density in the scattering environment.  
For example, a Coulomb field  
increases the energy density of the background. A Coulomb field also 
increases  the vacuum polarization and hence the  virtual particle density \cite{uehl,wikr},
 which affects processes like   
 Delbruck 
scattering.  
}

\colb{ Autoregularization incorporates the 
impact of changes in background energy and particle densities  on a scattering process, as follows. 
The chemical potential of a particle depends on its number density\footnote{\colb{For example, the chemical potential of a system of monatomic ideal gas atoms of mass $m$ at absolute} \colb{ 
temperature $\tau$ is $\m \deq \tau \ln\brf n {n\uu Q}$, where $n$ is the density of atoms and $n\uu Q \deq \lrr{m\,\tau/(2\pi\, \hbar)\sq}\tu{3/2}$, the quantum concentration \cite{kikr}.}}. 
As the number density in the background changes  
the chemical potentials in relevant Gibbs factors are
modified in autoregularization to reflect the
change.  Similarly, the increased
energy density due to a background field introduces a new energy scale,
which is not present in vacuum. A  new energy scale, if sufficiently high,
modifies the 
temperature in the Gibbs factors of all particles\footnote{
 \colb{The Lorentz-invariant temperature of the process  is the taken to be
the maximum among the kinematic scales shown in \ct{tau}.  If the scale of the background field is larger than the $\tau$ defined in \ct{tau}, then the} \colb{ temperature needs
to be modified in the Gibbs factors.}}.  
Thus, in addition to including the direct impact of a background 
field in its Feynman diagrams, autoregularization also incorporates
the indirect impact that a background field has on scattering
through the changes it induces in the scattering environment.} 

\colb{
Autoregularization can be applied in strong-field and bound-state
calculations using the mode expansions in the Furry picture \cite{inmo,furr}. 
For example,  using the Furry picture, one can apply 
autoregularization to study Compton scattering of photons (and electrons)  off
bound electrons in a fully relativistic setting.  A particularly interesting  
application of autoregularization would be to calculate the ionization 
probability in the scattering of low-energy photons off bound 
electrons \cite{kirc}; in such calculations,
Gibbs factors would play a key role in  
describing the fluctuations of bound electrons and in suppressing
the infrared divergences  due to soft photons
\cite{kppr}.  
The $S$-matrix formalism and perturbation theory for 
bound-state QED are discussed in \cite{inmo,furr}.  Application of
autoregularization to selected strong-field and bound-state scattering
processes is the focus of our ongoing work.
}
 
 \sss{\colb{$e\tu + e\tu -$ pair annihilation}}\label{apa}
\color{black}{}
In the following discussion we use autoregularization to calculate the tree-level differential cross section 
 for pair annihilation  process
\bea 
e\tu - + e\tu + \ra \g + \g 
\leqn{Pa}
\eea 
We also compare the predictions of both autoregularization and the standard QED
with the experimental data   
obtained at  center-of-momentum energy of  206.671 GeV by 
ALEPH Collaboration et al \cite{alep}. 
Using the RMS error  as a goodness-of-fit  
 measure, 
the prediction of autoregularization  
is found to fit the experimental data  
better 
than the prediction of the standard 
QED.

\begin{figure}[hbt]. 
\centering
\includegraphics[height=1.5in]{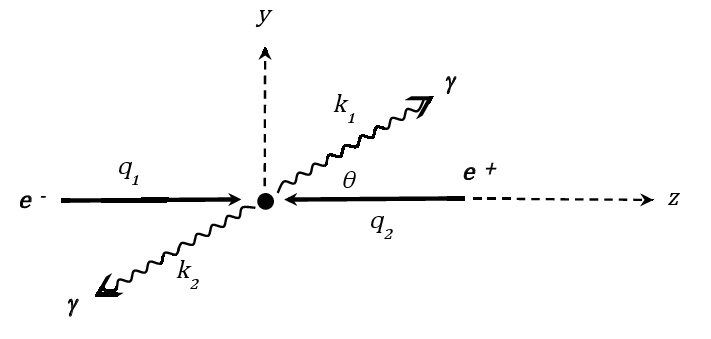}
\caption{\it Kinematics of $e\tu- \, e\tu +$ annihilation.}
\label{figurePairAnnKin}
\end{figure}
The kinematics of the scattering process
 are shown in Figure \ref{figurePairAnnKin}.  The
4-momenta of the incoming electron and positron are $q\so$ and $q\sw$,
and their $z$-spins $r\so$ and $r\sw$. 
The 4-momenta
of the outgoing photons are $k\so$ and $k\sw$ 
and their polarization states $s\so$ and $s\sw$ . 
We assume that the 3-momenta of the electron  
and positron  are oriented along the $z$ axis. We work in the center of momentum
frame and set $E\dfn q\so\uz \deq q\sw\uz $.  Since  the scattering is ultra-relativistic  
$ \lrb{\vec q\so} \deq \lrb{\vec q\sw}\, \approx \, E$.    Thus, 
\bea
q\uu{1,2} \deq (E, 0, 0, \pm E),  \quad  k\uu{1,2} \deq (E,0, \pm E\sth, \pm E\cth) \leqn{vs}
\eea

We work in  
Feynman 
gauge. The tree-level scattering amplitude for pair annihilation, shown in 
\ct{Pa}, is 
\beass
\cls \aea  
\lrs{\tpf \delf{q\so + q\sw - k\so - k\sw}}\,  i \, \clm
\eeass
where  
\beasm
\clm \aea 4\pi \a\, \xi \, \bar v(q\sw, r\sw) \, 
\clq(s\so, s\sw)
\, u(q\so, r\so), \qquad 
\xi  \deq \dsf{ g\uu e(q\so)\, g\uu e(q\sw)\, g\uu p(k\so) g\uu p(k\sw)}{Z\uu e\, Z\uu \g\, \cln}\nonumber\\[0.7\bls]
\clq 
(s\so, s\sw) \aea 
-
\lmu {s\so}{l\so}  \, \lmu {s\sw}{l\sw} \, f\so(\th) \lrs{\epsl {k\so} {l\so} (q\sl\so - k\sl\sw  +\me) \epsl {k\sw} {l\sw}} \leqn{qq}\\ \aba
  -
 \lmu {s\so}{l\so}  \, \lmu {s\sw}{l\sw} f\sw(\th) \lrs{\epsl {k\sw}{l\sw} (q\sl\so - k\sl\so + \me) \epsl {k\so}{l\so}}\nonumber
\eeasm
\beass
f\so(\th) \ada \dsf{2}{
\lrs{\exp\lrc{(c(\th)- \mu\uu e)/2E)}+1}\tu{1/2}  \, c\sq(\th)}; \quad c(\th) \deq \lrs{4E\sq \cos\sq\lrr{\dsf \th 2}+\me\sq}\tu{1/2}
\nonumber\\
f\sw(\th) \ada \dsf{\ds 2}
{\lrs{\exp\lrc{(s(\th)- \mu\uu e)/2E)}+1}\tu{1/2}\, s\sq(\th)}; \quad s(\th) \deq \lrs{4E\sq \sin\sq\lrr{\dsf \th 2}+\me\sq}\tu{1/2}\nonumber
\eeass
$r\so$ and $r\sw$ are the $z$-spins of the incoming electron and positron and
$\e\upm(\vk,s)$ denote the photon polarization vectors\footnote{
We take $\e\upm(\vk, 0) \dfn (1,0,0,0), \ \e\upm(\vk, 1) \dfn (0,\hat v), \ \e\upm(\vk, 0) \dfn (1,\hat w), \ \e\upm(\vk, 0) \dfn (0, \hat k)$ where $\hat k$ denotes the unit vector along $\vk$ and
$\hat v \times \hat w \deq \hat k$. 
}, $s = 0, 1, 2, 3$.  Sums over $l\so$ and $l\sw$ 
are implied.  $\sqrt{Z\uu e}$ and $\sqrt{Z\uu \g}$ are the wavefunction {\color{black} normalization} constants of the 
electron and photon fields.  $\cln$  
{\color{black} a normalizing factor}\footnote{
$\cln \deq \mtrxel{0\uu F}{T\lrs{\exp\lrc{-i\imp \clh'(x) \, \dfx}}}{0\uu f}$, where $\ket{0\uu F}$ is the vacuum of the 
non-interacting theory and $\clh'$ is the interaction Hamiltonian density in the interaction picture. $T$ is the
time-ordering operator.
}.  $\eta$ is the Lorentz metric. $\mu\uu e$ is defined in \ct{chp}. The outgoing photons are assumed to have
the polarization vectors $\e\upm(\vk\so, s\so)$
and $\e\upm(\vk\sw, s\sw)$ respectively
where $s\so, s\sw = 1,2$, since the external photons  are physical
transverse photons.
Averaging over the spins of the incoming fermions we get 
\bea 
\dsf 1 4 \sum_{r\so, r\sw} \lrb{\clm}\sq \aea  
4\pi\sq \, \a\sq\lrs{  |\xi|\sq  \, 
Tr\lrc{
\clq
(s\so, s\sw) \, (q\sl\so + \me) \, \ti
\clq
(s\so, s\sw) \, (q\sl\sw - \me)}}
\leqn{mirst}
\eea 
where 
\bea 
\ti 
\clq
(s\so, s\sw) 
\ada
-\lmu {s\so}{a\so}  \, \lmu {s\sw}{a\sw} \, f\so(\th) \lrs{\epsl {k\sw} {a\sw} (q\sl\so - k\sl\sw + \me) \epsl {k\so} {a\so}} \leqn{qti}\\ \aba
-\lmu {s\so}{a\so}  \, \lmu {s\sw}{a\sw} f\sw(\th) \lrs{\epsl {k\so}{a\so} (q\sl\so - k\sl\so + \me) \epsl {k\sw}{a\sw}}
\nonumber 
\eea 
Summing 
over the polarizations of the 
physical 
outgoing photons
we get, using \ct{mirst}, 
\bea
\lrb{\clm_{obs}}\sq \aea 
  \sum_{s\so, s\sw=1}^2
   4\pi\sq \, \a\sq\lrs{  |\xi|\sq  \, 
Tr\lrc{
\clq
(s\so, s\sw) \, (q\sl\so + \me) \, \ti
\clq
(s\so, s\sw) \, (q\sl\sw - \me)}} \hs{0.35}
\leqn{mirs}
\eea
The differential cross section for the $e\tu - e\tu +$ pair to
scatter into two photons in the center-of-momentum frame 
and ultra-relativistic
limit, 
is given by \cite{sred}
\bea 
\brf{d\sg}{d\Omega}\uu{CM} \aea 
\dsf{\lrb{\clm_{obs}}\sq}{64 \, \pi\sq\, s}, \qquad s \deq 4 E\sq
\leqn{dcs}
\eea 
$s$ is the Mandelstam variable representing the square of the center-of-momentum energy. 
Using \ct{mirs} and \ct{dcs} we get the differential cross section for pair annihilation predicted by 
autoregularization,
\bea
\brf{d\sg}{d\Omega}\uu{CM} \sps{AR} \aea 
\brf{\lrb{\xi}\sq \, \a\sq} 
{16 \,s}\sum_{s\so, s\sw=1}^2
Tr\lrc{
\clq
(s\so, s\sw) \, (q\sl\so + \me) \, \ti
\clq
(s\so, s\sw) \, (q\sl\sw - \me)}\hspace{0.25in} \leqn{dci}
\eea
From \ct{qq} and  \ct{qti} it follows that calculating \ct{dci} involves computing traces of the form
\beas
T(a\so, \ldots, a\sn) \ada Tr(a\sl\so \ldots a\sl_{n}), \qquad n= 4, 6, 8
\eeas
since the trace of an odd number of $\g$ matrices vanishes.  The required traces are computed
recursively using the relations
\beas
tr(a\sl\so a\sl\sw) \aea 4 a\so\d a\sw, \qquad a\so\d a\sw \dfn (a\so)\dnm (a\sw)\upm\\ 
tr(a\sl\so \ldots a\sl_{2n}) \aea \sum_{i=2}^{2n} (-1)^i\, (a\so \d a\si) 
 tr(a\sl\sw \ldots a\sl_{i-1} a\sl\si \ldots a\sl_{2n})
\eeas
The dot products needed to compute the traces are calculated numerically using the vectors shown in 
\ct{vs} and the polarization vectors shown below.
\beas
 \e\upm(\vk\so,0) \aea  
(1,0,0,0) \qquad \qquad   \quad \deq \e\upm(\vk\sw,0)\\
 \e\upm(\vk\so,1) \aea  
 (0, 1,0,0)\qquad \qquad  \quad \deq \e\upm(\vk\sw,1)\\
 \e\upm(\vk\so,2) \aea  
 (0, 0, \cth, - \sth)  \deq -\e\upm(\vk\sw,2)\\
 \e\upm(\vk\so,3) \aea  
 (0, 0, \sth, \cth) \quad   \deq -\e\upm(\vk\sw,3)
 \eeas
The differential cross section for pair annihilation in the center of momentum frame in the standard 
theory is given by \cite{pesc}
\beasm
\brf{d\sg}{d\Omega}\uu{CM}\sps{QED}  \deq \brf{2\pi\a\sq E}{s p}  
\lrs{\dsf{E\sq + p\sq\cos\sq(\th)}{\b} + \dsf{2\me\sq}{\b} - \dsf{2m\tu 4}{\b\sq}}, \quad \b \dfn \me\sq + p\sq \sin\sq(\th) 
\eeasm
where $p = \lrb{\vq\so}$.   

\begin{figure}[hbt]
\centering
\hs{-0.4}\includegraphics[width=6in]{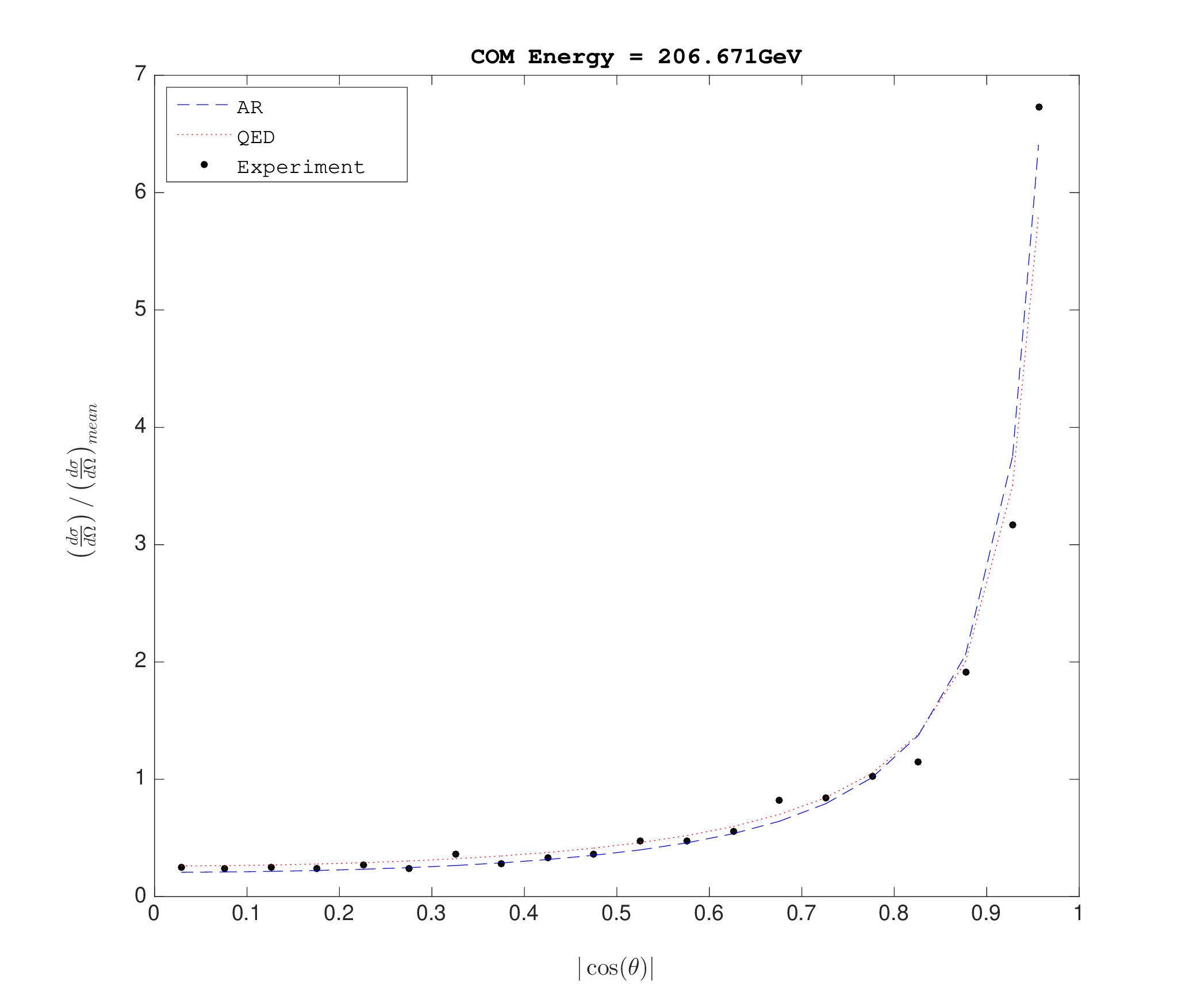}
\caption{Tree-level predictions of autoregularization and  the standard QED vs. experimental data for $e\tu+\, e\tu -\ra \g\, \g$.}
\label{206iqx}
\end{figure}

We compare the  the predictions of autoregularization and QED with the experimental data obtained at 
center-of-momentum energy of $\sqrt s = 206.671$ GeV by the ALEPH collaboration \cite{alep}. 
The data reported in \cite{alep} spans an angular range of $0.029 \leq \lrb{\cth} \leq 0.956$, in angular
steps of approximately $\cos\inv(0.05)$. 
To illustrate the angular dependence of the cross section we
plot the differential cross section as a fraction of the 
mean value over the angular range.
The graph is shown in Figure \ref{206iqx}.

 The discrepancy between theoretical predictions
 and experimental data is determined by computing the RMS (root mean
 square) error for the theoretical plots (AR or QED) in Figure \ref{206iqx} 
 relative to the plot of the experimental data. The ratio of the RMS errors is
 \beas
 \dsf{RMS_{error}(QED)}{RMS_{error}(AR)} \aea 1.3506
 \eeas
 showing that the angular dependence of the differential cross section predicted by autoregularization
  is in better agreement with experimental data than that predicted by the standard 
 QED. 
 Thus, we note that both in the low energy regime, in the context of Compton scattering (Figure \ref{figurew}),
as well as the ultra-relativistic regime, in the context of pair annihilation (Figure \ref{206iqx}), the 
predictions of autoregularization are in better agreement with the experimental data than the corresponding 
predictions of the standard QED. 

\colb{As with 
Compton scattering (Section \ref{acs}) 
the pair annihilation calculation presented above also assumed that the  annihilation occurs in vacuum, and that the incoming particles were free in the asymptotic past.   
Besides the simple setting we considered above, 
pair annihilation process is also being investigated 
in other settings.} 

\colb{Pair annihilation   in the background of the strong magnetic field of pulsars (and magnetars) \cite{daha, baru}, the strong  Coulomb field of heavy nuclei \cite{mzys} as well as the background field of intense lasers \cite{fedo} are the focus of active research.  
Annihilation of pairs in which one or both of the particles are bound
is also of current interest; examples include   
annihilation of bound electrons
in the inner shells of heavy atoms by free positrons \cite{mzys}, and positronium decay to photons \cite{bmms}.  
} 

\colb{An interesting phenomenon that occurs in a background field, but not in vacuum, is pair annihilation into a single photon \cite{mzys,fedo}.  
It is believed that in the annihilation  of an inner-shell bound electron by a free 
positron, the  1-photon annihilation dominates in heavy atoms and 
2-photon annihilation in light atoms.   A satisfactory theoretical
understanding  of the 2-photon annihilation amplitude in the vicinity of a 
heavy nucleus has been stymied, however, by the intractability of the
involved calculation \cite{mzys}.   
Autoregularization may provide a useful framework
for such calculations, as we describe below.  For remarks on  
 applying  autoregularization to strong-field and 
bound-state problems, please see Section \ref{acs}.
}

\colb{Besides the general computational intractability of strong-field
calculations, the calculation of 2-photon annihilation in a strong 
Coulomb field is beset with an additional challenge--the infrared 
divergence that arises due to soft photons \cite{mzys}.  The Gibbs 
factors exponentially suppress field fluctuations {\it both in the 
ultraviolet and infrared regimes} and would hence ensure
hyperconvergence of the integrals even when one of the outgoing
photons is soft.  Further, since the scattering amplitudes in 
autoregularization are finite at every order of perturbation theory 
no regularization is needed to isolate and remove infrared divergences
at a fixed order \cite{mzys}.}

\colb{Another context in which   
autoregularization would simplify calculations pertains to 
investigation of strong-field QED 
in intense lasers.  As the background field strength and/or the pulse length of the lasers  
increase, the contributions of  higher order processes become  important \cite{fedo}.
Again, the finiteness of  amplitudes at all orders in autoregularization, and 
the hyperconvergence of the integrals due to Gibbs factors  make \ipsr  a particularly convenient framework for calculations in strong fields
and/or long background field pulses.  As we remarked in Section \ref{acs},
application of autoregularization to selected strong-field and bound-state
processes, including pair annihilation in strong field, is the focus of our
ongoing work.}

\s{\colb{Vacuum energy density in the Standard Model}}\label{ved}
In this section we calculate the energy density of vacuum fluctuations
of the free fields in the Standard Model using autoregularization.  Previous theoretical 
estimates of the vacuum energy density using known regularization
schemes---such as cutoff regularization \cite{weincc} or dimensional
regularization with modified minimal subtraction renormalization \cite{mart}---exceed 
the observational upper bounds by several dozens of orders of magnitude, 
giving rise to the infamous cosmological constant problem.  

In regularizing the amplitude of a scattering process $\clp$  in Section \ref{psr}, we extracted an intrinsic `energy' scale
of the process from its kinematics   as described in \ct{tau}.  The vacuum fluctuations of the quantum 
fields is not a scattering process and has no 
asymptotic momentum eigenstates. However, the vacuum fluctuations of the free fields in the 
Standard Model (SM) do have
an intrinsic `energy' scale namely the scale of the vacuum energy density $\cle\uu{SM}$ they generate.
In natural units $\cle\uu{SM}$ can be written as 
\bea
\cle\uu{SM}\aea \tau\uu{SM}\tu 4
\leqn{6p1}
\eea
where $\tau\uu {SM}$ is the energy scale of $\cle\uu{SM}$.   Unlike the $\tau\uu\clp$ 
shown in \ct{tau}, which can be calculated {\it a priori}, the 
$\tau\uu {SM}$ shown in \ct{6p1}, cannot be calculated {\it a priori}.  However, we
can calculate $\tau\uu {SM}$ by solving the self-consistent Equation \ct{me} described below. 

Secondly, we used a center-of-momentum frame $\clf\uu\clp$ of a scattering process $\clp$
to define the $\xi$ function in \ct{2p2}.     
The property of $\clf\uu\clp$ that we used in the definition \ct{2p2} was that $\clf\uu\clp$
  was unambiguously specified (up to irrelevant rotations) and thus, every 
Lorentz observer could to transform to $\clf\uu\clp$ from his/her frame.   
In the calculation of $\cle\uu{SM}$ 
{\color{black} we  choose   $\clf\uu\clp$ to be 
the frame $\clf\uu{\clo}$ that is momentarily co-moving with the 
observer; such a 
$\clf\uu{\clo}$ is also defined unambiguously  (up to irrelevant rotations). 

Using $\tau\uu{SM}$ and the $\clf\uu{\clo}$, described above,  autoregularization,
described in Section \ref{psr}, can be applied without any other modification 
to regularize vacuum fluctuations of free quantum fields.  Specifically, the Gibbs
factors defined in \ct{2p5} are inserted into the mode expansions of
the free fields to regularize the vacuum fluctuations.  

The energy density of vacuum fluctuations of the free quantum fields
in the Standard Model, in natural units\footnote{$\hbar = c = 1$}, at any
specified scale $\tau$ is 
\bea
\cle\uu{SM}(\tau) \aea  \f12 \sum_{i} \,\ti d\si  \int 
\dvk \,  \, \lrs{ |\vk|\sq + m\si\sq  }^{1/2} g\si\sq (\vk, \tau)
\leqn{ed}
\eea
where the sum is over all of the free fields in the Standard Model; $\ti d\si, m\si$ and 
$g\si$ denote  the degeneracy factor,
mass and Gibbs factor of particle $i$, respectively.  
The
 magnitude of the degeneracy factor $\ti d\si$ is the number of 
creation operators per 3-momentum mode
in the mode-expansion of the free field and
its conjugate\footnote{
That is, $|\ti d\si| = d\si$, where $d\si$ is the degeneracy term in
the Gibbs factor \ct{2p5}. 
Thus, $|\ti d\si|$ is  
1 for scalar boson, 
2 for massless vector bosons
and self-conjugate (Majorana) fermions, 
3 for massive vector bosons 
and 4 for Dirac fermions.   
The massless gauge bosons are quantized in covariant gauge
and their contributions are subsequently corrected by including the contributions
of the gauge fixing terms and the associated Faddeev-Popov
ghosts.
}.  
The sign of $\ti d\si$ is positive for physical\footnote{Non-ghost.} bosons   and negative for fermions
and ghosts.  
 {\color{black} We do the calculation in the special frame $\clf\uu{\clo}$ that is momentarily comoving
 with the observer.  In frame $\clf\uu{\clo}$}  the Lorentz-invariant Gibbs factors, described in \ct{2p5}, are   
\beas
g\si \tu 4(\vk, \tau) \aea \left\{ 
\ba{ll}
d\uu f/ \lrr{\exp\lrc{ (1 /\tau) \lrr{\sqrt{|\vk|\sq + m\si\sq}-\mu\si} } + 1}    \j \mbox{ for fermions}\\
d\dn m / \lrr{\exp\lrc{(1 /\tau)\lrr{\sqrt{|\vk|\sq + m\si\sq}-\mu\si}} - 1}    \j \mbox{\rule{0pt}{14pt}  for massive  bosons}\\[0.2\bls]
d\sz /\lrr{\exp\lrc{|\vk|/\tau + \tau/|\vk|} - 1}    \j \mbox{\rule{0pt}{14pt} for massless bosons}
\ea
\right.
\eeas 
where, as specified in \ct{2p5}, $d\uu f, \, d\uu m$ and $d\sz$ are the magnitudes of the degeneracy factors 
for fermions, massive bosons and massless bosons, described above. $\mu\si$ is the chemical potential 
corresponding to the particle, described in \ct{chp}.       

As mentioned above, the intrinsic energy scale $\tau\uu{SM}$ of the vacuum fluctuations
is extracted from  $\cle\uu{SM}$   and is taken to be its characteristic energy scale.  
Thus we have 
\bea
\cle\uu{SM}(\tau\uu{SM}) \aea  \tau\uu{SM}\tu 4 \leqn{wow}
\eea
From \ct{ed} and \ct{wow} we obtain the  equation
\bea
\f12 \sum_{i} \,\ti d\si  \int 
\dvk \,  \, \lrs{ |\vk|\sq + m\si\sq  }^{1/2} g\si\sq (\vk, \tau\uu{SM}) \aea \tau\uu{SM}\tu 4
\leqn{me}
\eea
that is satisfied by the intrinsic scale $\tau\uu{SM}$ of the vacuum fluctuations of the
free fields in the SM. 

{\color{black} The left hand side of \ct{me} has three
types of integrals corresponding to fermions, massive bosons and massless bosons. Specifically, the 
integral  
for a fermion of mass $m\si$ and electric charge $q\si$,  with $d\si$ degrees of freedom   is  
\beas
 F(m\si, d\si, q\si) \aea  -\dsf {d\si} 2 \int d\cu k   \bsf{d\si \lrr{|\vk|\sq + m\si\sq}}
 {\eto{(\sqrt{|\vk|\sq + m\si\sq} - m\si \b\, q\si)/\tau\uu{SM}}+1}\tu{1/2}, 
 \qquad \b  \deq   \sqrt{\dsf{9\a}{16\pi}} 
 \eeas
 In terms of the dimensionless variable $x \dfn |\vk|/\tau\uu{SM}$ the integral can be written as
 \beasm
 F(m\si, d\si, q\si)  \aea
 -\lrr{2 \pi d\si \sqrt{d\si}} \tau\uu{SM}\tu 4\izi dx \, \bsf{x\tu 4 (x\sq + (m\si/\tau\uu{SM})\sq)}
 {\eto{(\sqrt{x\sq + (m\si/\tau\uu{SM})\sq} - (m\si/\tau\uu{SM})(\b\, q\si))}+1}\tu{1/2}\nonumber\nlasal
 \aea -\tau\uu{SM}\tu 4 \,  (2\pi \, d\si\, \sqrt{d\si})   \, I\uu +(m\si/\tau\uu{SM}, q\si)
 \leqn{fermi}
\eeasm
where 
\beas
I\uu{\pm}(a,b) \ada \izi dx \bsf{x\tu 4 (x\sq + a\sq)}{\eto{ \sqrt{x\sq + a\sq} - a \b b}\pm 1}\tu{1/2}
\eeas
Similarly the integral  corresponding to a massive boson of mass $m\si$,
$d\si$ degrees of freedom
and electric charge $q\si$ is
\bea
B_m(m\si, d\si, q\si) \aea \tau\uu{SM}\tu 4 \ (2\pi\,  d\si\sqrt{d\si})   \, I_-(m\si/\tau\uu{SM}, q\si) \leqn{massb}
\eea
Finally, the contribution of the photon, $B\sz$, comprising the contributions from the gauge-invariant term, the  gauge fixing term and  the ghost term in the Lagrangian, is
\bea
B\sz \aea  (3\sqrt 2 - 1) \lrs{2\pi \izi dx \ \dsf{x\cu}{\lrs{e^{x + 1/x}-1}\tu{1/2}}}\tau\uu{SM}\tu 4 \deq
1808.8740 \, \tau\uu{SM}\tu 4 \leqn{phot}
\eea
Using \ct{fermi}, \ct{massb} and \ct{phot}, Equation \ct{me} can be written as 
\beasm
\lrs{- \sum_i 2\pi \, d\si\tu{3/2} I_+(m\si/\tau\uu{SM},q\si) + \sum_j  2\pi\, d\sj\tu{3/2}   I_-(m\sj/\tau\uu{SM},q\sj) + 
1808.8740} \tau\uu{SM}\tu 4 \aea \tau\uu{SM}\tu 4 \nn \leqn{321}
\eeasm

On the left hand side of \ct{321} we  sum over all the elementary {\it free } fields in the SM\footnote{{\color{black}
The term Standard Model is a slight abuse of notation  since we assume neutrinos are massive.}
}.
In the electroweak sector   the 
 leptons, photon, W and Z bosons, and Higgs  can be regarded as free fields in the leading
approximation, with their interactions treated as perturbations.  
As we show below, the energy scale of the nontrivial solution of \ct{321}
is several orders of magnitude below $\Lambda_{QCD}$;  at that
low energy scale, the QCD degrees of freedom  
are not the quarks and gluons but rather the color-neutral 
hadrons\footnote{
{ \color{black}  
Since an effective description of the  dynamics
of color-neutral hadrons  at energy scales well below $\Lambda_{QCD}$ 
does not involve color degrees of freedom, the chemical potential in the
Gibbs factor of a hadron does not need to encompass the color charge.
}
}. 
Thus in \ct{321} the sum over $i$ spans the leptons and the 
known fermionic hadrons
 \cite{zyla}.
The sum over $j$ spans the known bosonic hadrons \cite{zyla}, $W\tu{\pm},$ $Z$ and Higgs bosons.

We observe that $\tau\uu{SM} \deq 0$ is a trivial solution of 
 \ct{321}.  The nontrivial solution of  \ct{321} satisfies
\beasm
- \sum_i 2\pi \, d\si\tu{3/2} I_+(m\si/\tau\uu{SM},q\si) + \sum_j  2\pi\, d\sj\tu{3/2}   I_-(m\sj/\tau\uu{SM},q\sj) +  
1808.8740\aea 1  \qquad\quad \leqn{feq}  
\eeasm

At present, we do not know if the masses of the three neutrinos, denoted $m\so, m\sw$ and $m\sr$, satisfy the normal hierarchy ($m\so < m\sw < m\sr$) or the inverted
hierarchy ($m\sr < m\so < m\sw$). We also do not know if the neutrinos are Dirac or 
Majorana fermions.    
We do know that the neutrino masses satisfy the following constraints.
\bea
m\sw\sq - m\so\sq \aea \scn{7.49}{-5} \, eV\sq\nn
|m\sr\sq - m\so\sq| \aea \scn{2.48}{-3} \, eV\sq \leqn{ncs}
\eea 

The upper bound on $\sum_r m\uu r = m\so + m\sw + m\sr$
derived from the $\L CDM+r+\sum m\uu\n$ model\footnote{{\color{black}Using the datasets called TTTEEE+BAO+PAN+BK14+$\tau 0p055.$}} \cite{chch}  is   
\bea
\sum_{\n=1}^3 m\uu\n < 0.110 \  eV
\leqn{nub}
\eea
We denote the mass of the lightest neutrino as $m^*\uu\n$.  At present there is no known lower bound on $m^*\uu\n$.   However, using \ct{ncs} and \ct{nub} we obtain an upper 
bound for $m^*\uu\n$,
\bea
m^*\uu\n <  u \deq \left\{ \ba{ll} 0.0262 \ \mbox{eV} \j \mbox{ normal hierarchy}\\ 
0.0082 \ \mbox{eV} \j \mbox{ inverted hierarchy}\ea \rtd
\leqn{nuub}
\eea
We choose $m\str\uu\n$ to saturate\footnote{
{\color{black} If  the actual
$m\str\uu\n$ is 
less than the value that saturates the upper bounds in \ct{nuub}, then the $\cle\uu{SM}$ values shown in Table I would be   lower.}
} the upper bounds shown in \ct{nuub}.   

The solution\footnote{{\color{black}
We can compare the contributions of the various fields to the left  hand side of \ct{feq}. In 
Appendix F we show that for $a> 1$ and $0\leq b\leq 2$, 
\beas
 I\uu\pm(a,b) \aea  \izi dx \bsf{x\tu 4 (x\sq + a\sq)}{\eto{ \sqrt{x\sq + a\sq} - a \b b}\pm 1}\tu{1/2}  < 
12288 \, a\cu \, \eto{-0.46 a}
\eeas
Further, we note that for $a>7$, $\log(12288 \, a\cu \, \eto{-0.46 a})$ decreases monotonically with $a$. 
At  $\tau< 10^{-8} \, \me$,  for an electron we have $a\dfn  \me/\tau  > 10^8$ and 
$b\deq 1$.  Thus for an electron  
$
\log\lrr{I\uu+(a, b) } < - 10^7 
$     
and
the magnitude of electron's (negative)
 contribution to the left hand side of \ct{feq} is less than 
$e\tu{-10^7}$, at $\tau< 10^{-8} \me$. A fermion/boson that is more massive than electron makes a 
contribution whose magnitude is even smaller than $e\tu{-10^7}$, which is 
negligible compared to the magnitude of a neutrino's contribution, which is
 $\sim  10^2$ and the contribution of 
$1808.8740$ by the photon. Thus, at $\tau < 10^{-8}\, \me$, the main contributions to the left hand side
of \ct{feq} come from the photon and the neutrinos.}} of Equation \ct{feq}, 
in the four different scenarios corresponding to neutrinos, are summarized
in the following Table I.
\bc
\bt{|c|c|c|}\hline
\multicolumn{3}{|c|}{TABLE I}\\ \hline
 \multicolumn{1}{|l}{Neutrinos are} \j \mc{1}{c|}{Dirac fermions} \j Majorana fermions\\ \cline{1-3}
 Normal hierarchy \j  \multicolumn{1}{|c|}{
 $\ba{rcl} \ \\ \tau\uu{SM} \aea \scn{5.6275}{-9} \ \me\\
 \cle\uu{SM} \aea 0.035904  \ \mbox{GeV/m$^3$}\\ \ \ea$}
 \j \multicolumn{1}{c|}{
 $\ba{rcl} \tau\uu{SM} \aea \scn{9.2094}{-9} \ \me\\
 \cle\uu{SM} \aea 0.25753  \ \mbox{GeV/m$^3$}\ea$}
 \\  \cline{1-3}
 Inverted hierarchy \j \multicolumn{1}{|c|}{
 $\ba{rcl} \ \\ \tau\uu{SM} \aea \scn{2.3659}{-9} \ \me\\
 \cle\uu{SM} \aea 0.0011216  \ \mbox{GeV/m$^3$}\\ \  \ea$}
 \j  \multicolumn{1}{c|}{
 $\ba{rcl} \tau\uu{SM} \aea \scn{8.1625}{-9} \ \me\\
 \cle\uu{SM} \aea 0.15892  \ \mbox{GeV/m$^3$}\ea$}
 \\  \cline{1-3}
\et
\ec

The energy scales shown in Table I,  $\tau\uu{SM} \sim 10^{-8}\, \me \sim \scn{5}{-12}$ GeV,
are several orders of magnitude smaller than $\Lambda_{QCD} \sim 10\tu{-1}$ GeV  \cite{pdgp}, which
justifies using the color-neutral hadrons, rather than quarks and gluons,  as 
the free fields in the QCD sector.

Based on the recently measured value of 
$H\sz \approx
67.4$ km.s$\inv$.Mpc$\inv$ of the Hubble parameter \cite{plan}, the  current estimate of the critical density is $\rho\uu c \deq 4.79$ GeV/m$\tu 3$.
From Table I, we see that regardless of whether the neutrinos are Dirac or Majorana fermions, and whether the neutrino masses satisfy the normal or inverted hierarchy 
autoregularization predicts that the tree-level energy density of vacuum fluctuations   is less than the current estimate of
the critical density.

\colb{While autoregularization  predicts that  the vacuum energy density of the free fields in the standard model, $\esm$, is less than the observed
critical density $\rhoc$, the known regularization methods
predict that $\esm$  exceeds $\rhoc$ by  several dozens of orders of magnitude  \cite{mart}.   
The staggering excess of the predicted $\esm$ over $\rhoc$, as well as the occurrence of divergent scattering amplitudes in the standard theory,  suggest that the standard theory may be an  {\it underconstrained} description of quantum fields.
That is, the behavior of quantum fields may   
actually be restricted by
some additional constraints that are missing in the standard theory.    
Autoregularization appears to add the  missing constraints to the  standard theory resulting in an  $\esm$ that is smaller than $\rhoc$ and   finite scattering amplitudes.  We elaborate below.}

\colb{Autoregularization is based on a new view that a quantum field can be regarded as   a 
system in thermal and diffusive equilibrium with a reservoir; please see Section \ref{psr} and Appendix \ref{agf}. The  fluctuations of a system that
is in thermal and diffusive equilibrium with a reservoir are described 
by the well-tested paradigm of Grand Canonical Distribution (GCD) in 
statistical mechanics.  Accordingly, as in GCD, autoregularization
{\it constrains} the field fluctuations with  Gibbs factors,
which  exponentially suppress  the high-energy modes, yielding an 
$\esm$ that is lower than $\rhoc$ as well as finite scattering amplitudes. 
}

\colb{Lacking the above statistical mechanical constraints in its framework, the standard theory allows every mode---regardless of its energy---to make an unweighted contribution to $\esm$ resulting in a divergent 
value of $\esm$.  The divergence in $\esm$, as well as the divergences in scattering amplitudes, can thus be attributed to the fact that   
constraints on field fluctuations---such as those in GCD---are missing    
in the standard theory.   That is, the standard theory provides
an {\it underconstrained} description of the 
behavior of quantum fields.  Known regularization methods,  
which have been successful
when applied to scattering amplitudes,  are unable to correct for the
underconstrained description of quantum fields in the case of $\esm$.
On the other hand, 
when the missing constraints on field fluctuations are incorporated into
the standard theory, as autoregularization does, the   
theoretically predicted value of $\esm$ becomes compatible
with the observed value of $\rhoc$ and  in addition the 
divergences in scattering amplitudes 
disappear as well.}

\colb{
In ongoing work we are computing higher-order corrections to the vacuum 
energy density.   
It seems 
plausible  that the magnitude of the higher-order corrections in autoregularization
will be less than that
of the leading contribution from free fields, owing to the additional suppression
due to the coupling constants and   extra Gibbs factors. 
On the other hand, as
can be easily verified for a 
simple scalar field, if the results are expressed in terms of the renormalized mass,
including the 1-loop correction does not modify the predicted value  of 
$\esm$   for the free field in the standard theory.  
Hence, as higher-order corrections are included we expect that 
autoregularization will yield consistently lower results for 
vacuum energy density than the standard theory.  
}
 
\s{Remarks}\label{remrks}
While the preliminary calculations show that the predictions of autoregularization are in good agreement with the 
experimental data at the tree-level and at 1-loop, fuller validation of autoregularization requires vetting its higher order
corrections   against  experimental data.  The electron's anomalous magnetic moment
provides a rich test bed for vetting autoregularization since it has been experimentally measured to high
accuracy \cite{hann,hhga} and, over the last several decades, it has been calculated theoretically 
up to tenth order  term in QED \cite{akni}, revealing a remarkable agreement between theory and
experiment.   
 
 \s{Acknowledgment}
 I thank Alan Guth, Jeffrey Goldstone and Samir Mathur for several helpful conversations.
 I thank the anonymous referees for their many meticulous comments, suggestions and questions
 which have improved the content and the presentation of the paper significantly. 
 
\titleformat{\section}{\large\bfseries}{\appendixname~\thesection :}{0.5em}{} 

\begin{appendices}
\s{\colb{Gibbs factor}}\label{agf} 
In this Appendix we present a heuristic derivation of 
\ct{2p5}  based on the interpretation of a quantum 
field as a system with a large number of interacting degrees of 
freedom.  
While the behavior of a system with a small number of degrees
of freedom is satisfactorily described by the laws of classical
and quantum mechanics, as is well known,
an effective description of the 
behavior of a system with a large number of degrees of freedom
necessitates the postulation of new laws, namely the laws of 
statistical mechanics. 
These new statistical laws
 ``{\it are of a different kind}'' and ``{\it cannot
in any way be reduced to purely mechanical laws}'' \cite{lali}.  

A quantum field has infinitely many degrees of freedom.  Hence the
behavior of quantum fields is constrained not only by the laws that 
govern the microscopic local interactions among their modes but 
also by the laws of statistical mechanics that emerge as a result of
the collective interactions of the infinitely many modes.

An interacting quantum field can be regarded as a system that is
in thermal and diffusive contact with the other quantum fields.
Specifically, let $\lrc{{\cal Q}\so, {\cal Q}\sw, \ldots}$ be the set of  
all quantum fields that coinhabit spacetime.  Then,
we  can regard an arbitrary quantum field, say $\cls=\lrc{{\cal Q}\so}$, 
as our system of interest and the complement
$\clr\uu {\cls}=\lrc{{\cal Q}\sw, {\cal Q}\sr, \ldots}$ as the reservoir
with which the system is in thermal and diffusive contact.  Being in 
thermal and diffusive contact, the system can exchange both energy 
and particles with the reservoir.  

 A system that is allowed to exchange energy and particles with a reservoir 
reaches thermal and diffusive equilibrium with the reservoir over a time 
scale, called  {\it relaxation time scale}, that is a characteristic of the 
system-reservoir complex.   We assume that   the relaxation time scale of a 
quantum field--- that is, the time scale over which a quantum field returns 
to thermal and diffusive equilibrium with its reservoir when perturbed---is 
significantly smaller than the time resolution of our measuring instruments.  
Stated differently, we assume that what we observe in a scattering process   
 is the equilibrium or near-equilibrium behavior of the participating fields. 
 
The behavior of a system that is in thermal and 
diffusive   equilibrium 
with a reservoir is described by the {\it Grand 
Canonical Distribution}  (GCD).   The GCD states that the probability 
 of a 
fluctuation away from equilibrium  is exponentially suppressed by the 
Gibbs factor.  Specifically, the probability  of finding a system in a state
with energy $E$ and with $N$ particles is given by 
\bea
GCD: \qquad P(E, N; \tau, \m) \aea Z \eto{- (E-\m N)/\tau}
\leqn{gcd}
\eea
where $Z$ is the probability of finding the system in the vacuum 
state ($E=N=0$).  $\m$ and $\tau$ are the 
chemical potential and temperature that characterize the diffusive
and thermal equilibrium between the system and its reservoir.  
$e^{-(E-\m N)/\tau}$ is called the Gibbs factor in the literature \cite{kikr}.

Consider a vacuum-to-vacuum fluctuation of the system in which
in three successive observations the system is found first in the vacuum
state, then in a definite single-particle state of 4-momentum
$k=(E,\vk)$, and finally again in
the vacuum state, as seen by an observer at rest in Lorentz
frame $\clf$. Denoting the vacuum and single-particle states as 
$\lrc{0}$ and $\lrc{k}$, the probability of observing the above fluctuation is 
\bea
P\uu{GCD}\lrr{\lrc{0}  \ra \lrc{k}\ra \lrc{0}}\aea Z\cu \eto{-(E-\m)/\tau} 
 \propto 
\eto{-(E-\m)/\tau}
\leqn{v2vgcd}
\eea

The amplitude of the above fluctuation can   be computed in 
quantum field theory (QFT) as 
\beass
{\cal A} (x,y;\vk) \dfn \bra{0 }\vp(y) \mid \vk\ \rangle \langle\  \vk\ \mid \vp(x) \ket{0 }
\eeass
where 
$x \dfn(t\so, \vx), \  y\dfn (t\sw, \vy)$ and$\ket{\vk}$ denotes the 1-particle state   mentioned
above.  For simplicity we have taken the quantum field $\vp$ to be a free scalar field of mass $m$.  
Using the mode expansion
\bea
\vp(x) \aea \lii k \lrc{
\opa(\vk) \eto{-i \uvk x} + \opad(\vk) \eto{i \uvk x}}, \qquad
\uvk\dfn (\om(\vk), \vk),  
\leqn{mephi}
\eea
where $E=\om(\vk) = \lrs{|\vk|\sq + m\sq}\tu{1/2}$, and the 
standard commutation relations for $\opa(\vk)$ and $\opad(\vk)$ we 
see that $\cla(x,y;\vk) \deq e\tu{i\uvk(x-y)}$.  Thus the probability
of the vacuum-to-vacuum 
fluctuation in QFT is given by 
\bea
P\uu{QFT}\lrr{\ket 0 \ra \ket{\vk} \ra \ket 0} \propto |\cla(x,y;\vk)|\sq 
\leqn{v2vqft}
\eea
which is independent of 
$E$.  

We can resolve the discrepancy between the prediction of GCD,
namely \ct{v2vgcd},  and the prediction
of quantum field theory that the vacuum-to-vacuum fluctuation is 
independent of $E$ by scaling the creation and annihilation operators
 in \ct{mephi}  to obtain the following modified 
mode expansion
\bea
 \vp(x) \aea \int \dsf{\dvk}{\tpc \tok} \d \lrs{g(\vk)} \d \lrc{
\opa(\uvk)\emiuvkx + \opad(\uvk) \eiuvkx}. \qquad \label{eqn10}
\eea
keeping the commutation relations among $\opa$ and $\opad$ 
unchanged.   Then 
\beas
P\uu{QFT}\lrr{\ket 0 \ra \ket{\vk} \ra \ket 0} \j \propto \j
g\tu 4(\vk)
\leqn{v2vqft2}
\eeas
which suggests that 
\bea
g(\vk) \sim \lrs{e\tu{-(E-\m)/\tau}}\tu{1/4}
\leqn{gfd}
\eea
With some abuse of notation, we call $g(\vk)$, shown in \ct{10}, the Gibbs factor.  The Gibbs factors 
 in \ct{2p5} are formulated  to be
({\it i\,})   consistent with \ct{gfd}  in the center-of-momentum (COM) frame\footnote{See \ct{2p4}.} for $(E-\m) \gg \tau$ and ({\it ii\,})   Lorentz-invariant.  Further,  the Gibbs factors  in \ct{2p5}
also ensure that $P\uu{QFT}$, in \ct{v2vqft2}, resembles Fermi-Dirac or Bose-Einstein distributions in the 
COM frame, depending on whether the particle is a 
 fermion or boson. 
 
In 
finite temperature 
field theory it is known that the  `{\it
chemical potential}' of an electron plays a role analogous to the $A\uz$ 
component of the dynamical gauge (photon) field  coupled  to the electron
\cite{lavu,schm,turk}. Therefore, we take the chemical potential of 
an electron {\color{black} to be\footnote{{\color{black} Setting $\m\uu e \deq   \lrb{\langle A\uz \rangle_{\lambda_e}}$,
 equates the chemical potential  of the electron (particle) 
and that of positron (antiparticle).  The  heuristic argument 
underlying \ct{mooi} 
is elaborated below.
The chemical potential $\m$ of a system is   
 \bea
 \m \ada - \tau  \brf{\pl \sg}{\pl N} _{V, E} \leqn{chpo}
 \eea
 where $\tau$ is the fundamental temperature,
 $\sg$ the entropy, $N$ the number of particles,
 $V$   the volume and   $E$ the internal energy of the system.
Consider a system $S$ that comprises a single particle 
and a different system $\bar S$ that comprises  a single antiparticle.  Assume that $S$
and $\bar S$  have equal $V$ and $E$, and that they are in thermal equilibrium
with separate reservoirs $R$ and $\bar R$ at temperature $\tau$.   If we add an additional particle to $S$
and an additional antiparticle to $\bar S$, keeping $V$ and $E$ unchanged,
the resultant changes in entropy of $S$ and $\bar S$ are equal. 
 Hence, we take the 
 chemical potential of a particle  to be equal to that of its antiparticle.  
 The  Gibbs factors \ct{2p5},  formulated using the above
 heuristic argument,  predict that the $g$-factors of electron and positron are
 equal,  
which is consistent with the 
 experimental measurement \cite{dsde} that showed the ratio of the $g$-factors
  of the electron and positron, $g(e\tu-)/g(e\tu +) = 1 + (0.5\pm 2.1)\times 10^{-12}$. 
  }}} 
\bea
\mu\uu e
\deq {\color{black} \lrb{\langle A\uz \rangle_{\lambda_e}}}, 
\leqn{mooi}
\eea
where the average 
 is taken over 
the characteristic Lorentz-invariant length scale of the electron, namely its
Compton wavelength $\l\uu e \deq {\hbar}/{\me c}$. Thus 
\bea
\mu\uu e \deq \brf 3 2 
\sqrt{\dsf{\a}{4\pi}}.
\leqn{mue}
\eea 
The chemical potential of an arbitrary particle is obtained by adapting
 the above argument to the particle of interest.
  
 The energy scale $\tau$
 in the Gibbs factor is taken to be the intrinsic scale of the process or phenomenon of   interest, guided by the heuristic that the equilibrium
 temperature between a system \cls---comprising fields participating in a 
 scattering process or phenomenon---and its reservoir $\clr\uu\cls$ varies as  the
 intrinsic scale of the process or phenomenon. 
\s{\colb{Electron's gyromagnetic ratio}}\label{aamm}
In this Appendix we derive the two factors $\chi\so$ and $\chi\sr$ shown in 
\ct{chi1} and \ct{chit}. 
At $O(\de)$ the scattering amplitude shown in \ct{for1} 
can be expanded in frame $\clf$ in powers of  $ e$ as
\beasm
{\cal A} (p,s,q,s') \aea \dsf{(-i)(i)}{Z\d \cln\d \gef{p} \gef{q}}\nl 1
\aba \int \dfx \ \dfy \ \eto{-ip\d x+iq\d y} \  \bar u(p,s) \ (i\psl\dn y -
\me)
\bra 0
\nn
\aba
(-ie) \int d^4 z\ T\lrc{\psi(y) \ \bar\psi(x)\lrs{\bar\psi(z)\asl\dnc(z) \psi(z) } }+\nn
\aba  \dsf{3\d (-ie)\cu}{3!} \int d^4 z \ d^4 v \ d^4 w\
T\left\{\psi(y) \ \bar\psi(x)\lrs{\bar\psi(z)\asl\dnc(z) \psi(z)} \lrs{\bar\psi(v)\asl (v) \psi(v)}
\right.\nn
\aba \left.
\lrs{\bar\psi(w)\asl (w) \psi(w)}\right\}
\vac
(i\overleftarrow\psl\dn x +
\me) \  u(q,s') + O(e^4)  \nl 1
\ada {\cal A}\so \d e +   {\cal A}\sr   \d
e\cu + O(e^4) 
\label{eqn102}
\eeasm
The factor 3  in the $O(e \cu)$ term accounts for the three ways in which the $A\dnc$ can be
chosen from the product $(-i \int {\cal H}\dni(z) \df z)(-i\int {\cal H}\dni(v)\df v)(-i\int
{\cal H}\dni(w) \df w)$.  Hereafter, we will abbreviate $\gef p$
and $\gpf p$ to $g\uu e(p)$ and $g\uu p (p)$. 

Since $p$ and $q$ are on mass shell,   using Gordon's identity
\beass
\bar u(q,s') \gmu u(p,s) \aea \bu(q,s') \lrc{\dsf{(p+q)\upm}{2\me} +
\dsf{2i \sigma\unm(p-q)\dnn}{2\me}} \ju
\eeass
where $\sigma\unm :=\brf i 4 [\gnu,\gmu]$ and
noting that $\mathfrak g\uu{tree} = 2$,  we have
\beasm
\cla \so 
\aea \lrs{\mathfrak g\uu{tree} 
\bsf{(i) (-i) g\uu e(p) g\uu e(q)}{Z \cln (2\me)} \bu(q,s') \, \sigma\unm (p-q)\dnn \, \cam\,  \ju} \nonumber\nlb
\ \nonumber\nlb 
+ \bsf{(-i)g\uu e(p) g\uu e(q) }{  \ Z \cln (2\me)}
\bu \ (p+q)\upm\  \cam\  \ju
  \label{eqn4}
\eeasm 
From \ct{3p2} and \ct{4}
we have
\bea
\chi\so  \aea
\dsf{  g\uu e(p) g\uu e(q)}{Z \cln(2\me)} \label{eqn101}
\eea 
From \ct{102}  we have
\beasm
{\cal A} \dn 3  
\aea
-\chi\so \d (2\me)
\int
\  \ \dsf{d\tu 4 k}{\tpf} 
 \lrs{\gue{p+k}\gue{q+k}\gup{k}}\sq \nonumber\nn[0.5\baselineskip]
\aba
\bsf{
\bar u(q,s') \lrc{\g\tu\a (q\sl + k\sl + \me) \gmu (p\sl + k\sl + \me)  \g\dna} u(p,s)}{
\lrr{(p+k)\sq - \me\sq + i\e}\lrr{(q+k)\sq - \me\sq + i\e}
\lrr{k\sq   + i\e}}
\ (A_c)\dnm(p-q) 
\label{eqn103}
\eeasm
Setting \h{a = q+k} and \h{b=p+k}
and recalling that
\beas
\gau \gad = 4; \
\gau \gmu\gad = -2 \gmu; \
\gau \gru \gmu \gad = 4 \gu\rho\m; \
\gau \gru\gmu\gsu\gad = -2 \gsu\gmu\gru
\eeas
the numerator within the brackets in \ct{103} can be rewritten as
\beas
\bu(q,s') \lrc{\gau (a\sl + \me ) \gmu (b\sl + \me) \gad} \ju \aea
(-2\me \sq)
\lrs{\bu(q,s') \ \gmu \ \ju}   \nlb
+ (4 \me ) \lrs{\bu(q,s') \ \lrc{a\tu\m + b\tu\m}\ \ju}
\nlb
- 2  \lrs{ \bu(q,s')   \lrr{b\,\sl \ \gmu\ a\sl} \ju
}
\eeas
The middle term does not contribute to the anomalous magnetic moment
because it contracts \h{\cam} with $(a+b)\upm$.
Consider
the third term.
\beas
-2 \lrs{\bu(q,s') \lrc{\gsu\gmu\gru} \ju \,b\dn\sg\, a\dn\rho}\hs{3.5}
\eeas

\vs{-0.55}

\beas
\aea  2 \lrs{\bu(q,s') \lrc{\gsu\gru\gmu-2\gsu \ \gu\m\rho} \ju \,b\dn\sg\, a\dn\rho} \\
 \hs 2 \aea 2 \lrs{\bu(q,s') \lrc{\lrr{\gu\sg\rho -2 i \spn\sg\rho}\gmu
 -2\gsu \ \gu\m\rho} \ju \,b\dn\sg\, a\dn\rho}
\eeas
If $\de=0$, that is the classical background field is turned off, then $|\, \vq - \vcp \ | = 0$.
Therefore, in the weak-field limit $\de \ra 0$, $|\ \vq - \vcp \ | \ra 0$; that is,
$| \ \vq - \vcp \ |$ can be made as small as we please by taking $\de$ to be
sufficiently small.  Therefore, in the
weak-field limit
$
\spn\sg\rho b\dn\sg a\dn\sg \deq  \spn\sg\rho(p+k)\dn\sg (q+k)\dn\sg \rightarrow  0
$
and
\beas
- 2  \lrs{ \bu(q,s')   \lrr{ b\sl  \ \gmu\ a\sl} \ju }\aea
2 (a\upa b\dna) \lrs{\bu(q,s') \,\gmu\, \ju} - 4 \,\bu(q,s')\  b\sl \ \ju \ a\upm
\eeas
Putting it all together and using Gordon's identity again we get
\beass
\bu(q,s') \lrc{\gau (a\sl + \me ) \gmu (b\sl + \me) \gad} \ju \cam\hspace{3in}
\\[-1\bls]
\eeass

\vspace{-0.35in}

\beasm
 \aea
\mathfrak g\uu{tree}\d (a\upa b\dna - \me\sq) \brf{2i}{2\me} \lrs{\bu(q,s') \  \spn\n\m(p-q)\dnn\  \cam \ \ju } \nn
 \aba + \bu(q,s') \lrs{2(a\upa b\dna -\me\sq) \dsf{(p+q)\upm}{2\me}
 + 4\me (a+b)\upm - 4 b\sl \ a\upm} \ju \cam
\label{eqn105}
\eeasm
The second term contracts $(p+q)\upm$, $(a+b)\upm$ and $a\upm$ with $\cam$
and hence does not contribute to $\xi\sr(e,m)$.

From  
\ct{3p2},  
\ct{103} and
\ct{105} we have,
\beasm
\chi\sr 
\aea
-(2i) \d \chi\so 
\int
\dsf{d\tu 4 k}{\tpf} 
\bsf{\lrs{\gue{p+k}\gue{q+k}\gup{k}}\sq \lrc{(p+k)\upa (q+k)\dna - \me\sq}}{
\lrr{(p+k)\sq - \me\sq + i\e}\lrr{(q+k)\sq - \me\sq + i\e}
\lrr{k\sq  + i\e}}
\label{eqn106}
\eeasm

\s{\colb{Lamb shift}}\label{als}
In  
\colb{Section} \ref{ls} we summarized the calculation of the Lamb shift
based on autoregularization. In this Appendix, we present the details of
the calculation underlying the summary presented in  
\colb{Section} \ref{ls}.  
The following discussion is 
adapted from  Dyson's calculation \cite{dyso2}. 
We use the natural units $c=\hbar = \e\sz = 1$.

The total Hamiltonian for an
atomic electron interacting with the Maxwell field is given by
\beas
H \aea H\dn A + H\dn M + \l H\dn{int}
\eeas
$H\dn A$ is the Hamiltonian of an orbiting electron\footnote{
In Hydrogen atom.}, whose dynamics is
described by nonrelativistic quantum mechanics. $H\dn M$ is the Hamiltonian
of the free Maxwell field; $H\dn M$ is regarded as a quantum field and is 
regularized using autoregularization. 
$$H\dn{int}= \int \, \dvx\, j\upm(\vx,t) A\dnm(\vx,t)$$ is the Hamiltonian 
describing the interaction between
an electron and the Maxwell field; $j\upm(\vx,t)$ is the  electron current
that couples to the Maxwell field $A\upm(\vx,t)$.
$\l$ is a dimensionless parameter that is used for book-keeping. $\l$ will be set to 1 eventually.

We denote the time-independent orthonormal family of energy eigenstates of
$H\dn A$  as     $\ket 1, \ket 2, \ldots, $ $\ket k, \ldots$
with eigenvalues $E\so, E\sw, \ldots , E\sk, \ldots$.  We seek to
compute the correction to energy level $\ket{n}$; thus, the label $n$
is to be distinguished from all other labels $k\neq n$ in
the following discussion.  Labels like $k$ and $n$ represent
triples of principal, azimuthal and magnetic  quantum numbers.

In the interaction picture, let
$\kpi$ denote the state of the electron interacting
with the electromagnetic field  
with the initial condition $\ket{\psi_{_I}(0)} = \ket n$.  In the
interaction picture\footnote{
Evolution of operators is determined by the unperturbed
Hamiltonian $H\sz := H\uu A + H\uu M$, and the evolution
of states is determined by the interaction Hamiltonian
operator $\l H\tu I\uu{int}$ in the interaction picture.
} we can expand $\kpi$ as
\bea
\kpi \aea a\so(t) \ket 1 + a\sw(t) \ket 2 + \ldots a\sn(t) \ket n + \ldots
\label{eqn202}
\eea
Clearly, $a\sk(t) = \dotp{k}{\psi_{_I}(t)}, \ k = 1, 2, \ldots$.  
The initial condition is  $a\sn(0) = 1$,
$a\sk(0) = 0$ for $\ k\neq n$.
Define
 \beas
 \ket{\Psi\dni (t)} \ada \kpi \otimes \vacm, \qquad \ket K \ \dfn \ \ket k \otimes \vacm, \qquad k = 1,\, 2,\, \ldots
 \eeas
 where $\vacm$ is the vacuum of the free electromagnetic field.
 We note that
 \beas
\dotp K {\Psi\dni (t)} \aea \dotp k {\psi\dni(t)} \d \dotp{0\dn M}{0\dn M} \  =\  a\sk(t), \qquad k=1,\, 2,\, \ldots
 \eeas

The evolution of $\bkpi$  in the interaction picture is given by
the equation
\bea
i\plt \bkpi \aea \l\, \hii(t) \bkpi \label{eqn204}
\eea
At $O(\l\uz)$, that is in the absence of the interaction term, $\bkpi$ and
hence $\kpi$ is
time-independent, $a\sk(t) = 0$ for $k\neq n$ and $a\sn(t) = 1$.
 
 At $O(\l)$, for $k=1,\, 2,\, \ldots$, using \ct{202} and \ct{204} we have
 \beasm
 i\,\dot  a\sk(t) \aea \l \langle \, K\, |\hii(t)\bkpi \ = \ \l\d \mtrxel k {j\upm(\vx,t) } {\psi\dni (t)}
 \mtrxel{\zm}{A\dnm(\vx,t)}{\zm}  \nl 1
 \aea  0 
  \label{eqn205}
 \eeasm
 since the vacuum expectation value of $A\dnm(\vx,t)$ vanishes.  Thus, at $O(\l)$ we
 can consistently set $a\sn(t) = 1$ and $a\sk(t)=0$ for $k\neq n$.  And at $O(\l)$ the solution of \ct{204} can be
 written as
 \bea
 \bkpi \aea \lrc{1 - i\int_{0}^t \l\, \hii(\tau) \, d\tau} \d a\sn(t) \d \ket N;  
 \label{eqn206}
 \eea
 The time-evolution of $a\sn(t)$ at $O(\l\sq)$ can be obtained by inserting
 \ct{206} into \ct{205}.
 \bea
 i\, \dot a\sn(t) \aea -i \,\l\sq\, \mtrxel{N} {\,\hii(t)  \int_{0}^t  \, \hii(\tau) \, d\tau }
 {N} \d a\sn(t)
 \eea
 or
 \beasm
 i\d \dsf{\dot a\sn(t)}{a\sn(t)} \aea - i \,\l\sq\, \int \, \dvx \, \dvy \, \int_0^t \,
 d\tau \,  \mtrxel n {j\upm(\vx,t)
 j\upn(\vy,\tau) } n  \ \d \  \mtrxel{\zm} {A\dnm(\vx,t) A\dnn(\vy,\tau)} {\zm}  \nn
 \label{eqn209}
 \eeasm 
In Coulomb gauge, we expand the Maxwell field as
\beasm
 A\dnm(x; \clf) \aea \int \dsf{d\vk}{\tpc\ 2 |\vk|} \,  \gup{\uvk}\sum_{s=1}^2 \,
 \e\dnm(\vk,s) \lrc{\hat{a}(\vk,s) \eto{-i\,\uvk\, x} + \hat{a}\hc(\vk,s) \eto{i\,\uvk\, x}} \quad \label{eqn2142}
 \eeasm
 where $\uvk = (|\vk\,|, \vk)$.  $\clf$ is taken to be the momentarily comoving reference
 frame of the electron.  $g\uu p(\uvk; \clf, \clp)$  
 is abbreviated to $\gup{\uvk}$.  Hereafter, we also abbreviate $A\dnm(x;\clf)$ 
 to $A\dnm(x)$.
The $\e\dnm(\vk,1)$ and $\e\dnm(\vk,2)$ are chosen to be two polarization
 vectors such that $\e\upm(\vk,s) = (0, \hat \e(\vk,s))$, for $s=1,2$, and
 ($\hat \e(\vk,1), \hat \e(\vk,2), \hat k$) form a right-handed orthonormal
  triad, where
 $\hat k$ is the unit vector along $\vk$.   
 Further, using the Coulomb gauge commutation relations\footnote{
 \label{fn:CoulombGauge}
In Coulomb gauge, the free ffield is expanded as
\beass
 A\upm(x ) \aea \int \dsf{d\vk \d \gup{\uvk}}{ (2\pi)\cu (2\,|\, \vk\ |)}\d
\lrc{\sum_{s=1}^2 \e\upm(\uvk,s)\lrs{\cuks\  \eto{-i\uvk x}
+ \cduks\  \eto{i\uvk x}} }
\eeass
and the commutation relations between the creation and annihilation
operators are postulated to be
\beasm
\ba{l}
\ [\cuks, \cdukpsp]\deq \tpc\d (2\, |\,\vk\,|) \d \de(\vk-\vk\ ')\d \de_{s,s'}, \qquad  \\ \
[\cuks, \cukpsp] \deq [\cduks, \cdukpsp]
\deq  0, \qquad \qquad \quad\ s,\, s' = 1,2
\ea \label{eqn1413}
\eeasm
 }  
 we obtain
 \beass
 \mtrxel{\zm} {A\dnm(\vx,t ) A\dnn(\vy,\tau )} {\zm}\hs 4
 \eeass

\vs{-0.45}

 \beasm
\aea \int\dsf{\dvk \d  \gup{\uvk}\sq  }{\tpc  \d     |\vk|} \d
 \sum_{s =1}^2 \e\dnm(\vk,s) \e\dnn(\vk,s ) \eto{-i(|\vk|\, (t-\tau)+ \vk\d (\vx - \vy))}
 \label{eqn207}
 \eeasm
 Recalling that   in the interaction picture
$j\upm(\vx,t) = e^{iH\dn A t} j\upm(\vx,0) e^{-iH\dn At}$ and
inserting $1 = \sum_r \ket r \, \bra r$ between $j\upm(\vx,t)$ and $j\upn(\vy,\tau)$
we obtain
\bea
\mtrxel n {j\upm(\vx,t) j\upn(\vy, \tau)} n \aea \sum_r  \eto{-i(E\dn r-E\sn)(t-\tau)}
\lrab{j\upm(\vx,0)}_{nr} \, \lrab{j\upn(\vy,0)}_{rn} \leqn{upakruthi}
\eea
where $\lrab{j\upm(\vx,0)}_{nr} := \mtrxel n {j\upm(\vx,0)} r$.   Noting that the
electron current is a Hermitian operator we have
\bea
\lrab{j\upm(\vx,0)}_{nr} \aea
\lrab{j\upm(\vx,0)}_{rn}^*\leqn{nalavim}
\eea
With a slight abuse of notation we define $\lrab{j\upm(\vk)}_{nr} := 
\int \eto{-i\vk\d\vx} \lrab{j\upm(\vx,0)}_{nr} \ \dvx$.
Using \ct{nalavim} we have
\bea
\lrab{j\upm(-\vk)}_{rn} \aea j\upm(\vk)_{nr}^*
\leqn{ghana}
\eea
Then, noting that $\e\sz(\vk,s) = 0$,  $\e\dnm(\vk,s)$ is real, for $\mu=1,2,3$,
and using \ct{ghana} we have
\beas
\int \dvx \dvy \ \eto{-i\vk\d(\vx-\vy)} \lrs{\lrab{j\upm(\vx,0)}_{nr} \e\dnm(\vk,s) }
\lrs{\lrab{j\upn(\vy,0)}_{rn} \  \e\dnn(\vk,s)}  \hspace{3in}
\eeas

\vs{-0.6}

\bea
\aea \lrab{j\upm(\vk)}_{nr} \e\dnm(\vk,s) \lrab{j\upn(-\vk)}_{rn} \e\dnn(\vk,s) \deq
 | j\upm(\vk) \ \e\dnm(\vk,s) |\, \sq
 \leqn{vana}
\eea
Inserting \ct{207}, \ct{upakruthi} and \ct{vana} into \ct{209} we have
\beafns
i\d \dsf{\dot a\sn(t)}{a\sn(t)}
=  -  i \,\l\sq \int_0^t d\tau \int \dsf{\dvk\d \gup{\uvk}\sq  }{\tpc \  2|\vk|}
\sum_r  \sum_{s=1}^2  \left|\lrab{j\tu\mu(\vk)}_{nr} \ \e\dnm(\vk,s)\right|\sq
\eto{-i(E\dn r -E\sn + |\vk|)(t-\tau)}\nonumber
\eeafns
Using the identity \cite[Equation 258]{dyso2},
 $\ds \int_0^\infty \eto{-ia\tau} \  d\tau = \pi \de(a) + \dsf 1{ia}$,  we have for $t\gg 0$,
 \bea
 \dsf{\dot a\sn(t)}{a\sn(t)}
  \aea - \dsf{ \l\sq}{16\pi\cu} \int \dsf{\dvk }{|\, \vk|} \
 \gup{\uvk}\sq  \d \sum_r  \sum_{s=1}^2 \left|\lrab{j\tu\mu(\vk)}_{nr}  \ \e\dnm(\vk,s)\right|\sq \nn
 \aba \hspace{0.5in}\lrs{
 \pi\de(E\sm -E\sn + |\vk|) + \dsf 1{i(E\sm -E\sn + |\vk|)}}
 \label{eqn210}
 \eea
 Recalling the initial condition $a\sn(0) = 1$,
 we conclude
 that the solution of \ct{210} is (setting $\l =1$),
 \beasm
a\sn(t)
\aea \eto{-i(\Delta E\sn - i\Gamma\sn/2)t}, \qquad \mbox{ where }\nn
\Delta E\sn
 \aea - \dsf{ 1}{16\pi\cu} \int \dsf{\dvk }{|\,\vk|} \
\gup{\uvk}\sq  \d \sum_r  \dsf{\sum_{s=1}^2 \left|\lrab{j\tu\mu(\vk)}_{nr}  \ \e\dnm(\vk,s)\right|\sq}{E\dn r -E\sn + |\vk|}
   \nn
\Gamma\sn
 \aea\dsf{  \l\sq}{8\pi\sq} \int \dsf{\dvk }{|\vk|} \
 \gup{\uvk}\sq   \sum_r \sum_{s=1}^2 \left|\lrab{j\tu\mu(\vk)}_{nr}  \ \e\dnm(\vk,s)\right|\sq \d \de(E\dn r -E\sn + |\vk|)\label{eqn213}
 \eeasm

 Following Dyson we  
use the dipole approximation, in which $\vk\d\vr \approx 0$,
to set \h{\vec j(\vk)  = e\, \vec v
= \brf e {\me} \vec p} \ where  $\vec v$ and
$\vec p$
are the  velocity and
momentum
operators.
Recalling that $\e\sz(\vk,s) = 0$,
we have
\beasm
\sum_{s=1}^2 \left|\lrab{j\tu\mu(\vk)}_{nr}  \ \e\dnm(\vk,s)\right|\sq \aea
\brf e {\me} \sq \sum_{s=1}^2 \left|\lrab{\vec p}_{nr}  \d   \hat\e\, (\vk,s)\right|\sq
\label{eqn214}
\eeasm
As noted earlier, the choice of $\hat \e(\vk,1)$  determines $\hat \e(\vk,2)$. 
In the expansion \ct{214} we chose a single $\hat\e(\vk,1)$ vector on the unit 
circle in the plane perpendicular to $\vk$ in the ($k\uo, k\sq, k\cu$) space. The
$A\dnm(x)$ acting on $\vacm$ produces for each $\vk$ two photons that have a 
net polarization of $\hat\e(\vk,1) + \hat\e(\vk,2)$ and a phase  $\eto{i\uk(x)}$.
The net polarization vector could have been chosen to be any vector on a circle of
radius $\sqrt 2$ in the plane perpendicular to $\vk$ in ($k\uo, k\sq, k\cu$) space.
Thus summing over all possible net polarizations in \ct{213} 
involves 
multiplying the right hand side of \ct{213} with $2\pi$. Using \ct{214} and summing 
over all net polarizations we obtain
\bea
\Delta E\sn
\aea - \dsf {e\sq}{8\pi\sq \, m\sq} \int \dsf{\dvk }{|\,\vk|} \
 \gup{\uvk}\sq  \d \sum_r  \dsf{ \sum_{s=1}^2 \left|\lrab{\vec p}_{nr}  \d   
 \hat\e\, (\vk,s)\right|\sq }{E\dn r -E\sn + |\,\vk|}
\label{eqn215}
\eea
We note that
\beas
\sum_{s=1}^2 \left|\lrab{\vec p}_{nr}  \d   \hat\e\, (\vk,s)\right|\sq
\aea \left| \lrab{\vec p}\dn{nr} \right|\sq (1-\cos\sq\th)
\eeas
where $\th$ is the angle between $k\cu$ and $\lrab{\vec p}_{nr}$.

Therefore setting $\kappa =|\vk|$, and
\beas
\ti g(\kappa) \ada \lrs{\dsf 2 {\eto{\kappa/\tau_p + \tau_p/\kappa}-1}}^{1/4} \deq \gup{\uvk}; \qquad \tau\uu\clp \deq \me  
\eeas
\ct{215} can be rewritten as
\bea
\Delta E\sn \aea - \dsf {e\sq}{8\pi\sq\, \me\sq}
\int_0^\infty d\kappa \d \kappa \d \ti g(\kappa)\sq \d \sum_r \dsf{\left| \lrab{\vec p}\dn{nr} \right|\sq}{E\dn r - E\sn + \kappa}
\int_0^\pi \sin \th \ (1-\cos\sq\th) \ d\th \int_0^{2\pi} d\varphi\nn
\aea -\dsf {e\sq} {3\pi\, \me\sq} \int_0^\infty d\kappa \d \kappa\d
\ti g(\kappa)\sq
\d \sum_r \dsf{\left| \lrab{\vec p}\dn{nr} \right|\sq}{E\dn r - E\sn + \kappa} \label{eqn216}
\eea
The  
interaction  with the  
electromagnetic field  
shifts not only the energy of an electron in a Hydrogen orbital,  
as shown in \ct{216}, but also the self-energy of a free electron. 
The shift in the self-energy for a free electron  is 
\bea
\Delta E\sps{free} \aea  
-\dsf 1 {\me} \lrs{\dsf {2 e\sq\ } {3\pi\, } \int_0^\infty \ti g(\kappa) \sq \d d\kappa
} \dsf{(\vec P\sn)\sq }{2\me}   \qquad \leqn{frel}
\eea 
In order to interpret \ct{frel} we look at the shift in the kinetic energy of a
free electron of momentum $\vec P$ when a mass correction $\De m$ is added to its bare mass $m$,
to account for the interactions between the electron and the  electromagnetic field,
to obtain the physical mass $\me\deq m+\De m$.  At $O(\De m)$, we have
\bea
\Delta E\sps{free} 
\aea  - \dsf 1 \me \d \De m \d \dsf{(\vec P)\sq}{2\me} \qquad \leqn{eshift}
\eea
Comparing \ct{frel} and \ct{eshift} we conclude that the energy shift given in
\ct{frel} is the change in the kinetic energy of the free electron resulting from
the addition of electromagnetic mass $\De m \deq \dsf{2 e\sq}{3\pi}\izi \ti g(\kappa)\sq \, d\kappa$
to the bare mass of the electron to obtain the physical mass $\me$.   For a bound electron, in
state $\ket n$,
we obtain the shift in kinetic energy $\Delta E\uu{\De m}^{(n)}$ due to the addition of the electromagnetic mass
by replacing $(\vec P\sn)\sq $ in \ct{eshift} with $\bra n  \vec p \ \sq \ket n=\langle  \vec p\  \sq\rangle\uu{nn}$,  the
expected value of the $\vec p\ \sq$ operator.
\beass
\Delta E\uu{\De m}^{(n)} \aea - \dsf 1 \me \lrs{\dsf {2 e\sq\ } {3\pi\, } \int_0^\infty d\kappa
\d \ti g(\kappa) \sq} \dsf{\langle  \vec p\  \sq\rangle\uu{nn}}{2\me} \deq - \lrs{\dsf { e\sq\ } {3\pi\, \me\sq} \int_0^\infty d\kappa
\d \ti g(\kappa) \sq} \sum_r |\langle \vec p\, \rangle\dn{nr}|\sq
\eeass
The shift $\Delta E\uu{\De m}^{(n)}$ is already included in the energy $E\sn$ of $\ket n$,
that we compute using the Bohr model of the Hydrogen atom in which we use not the
bare mass of the electron  but the physical mass $\me$. The Lamb shift we measure is
the change in energy relative to $E\sn$.  Therefore, from the total shift computed in \ct{216} we
need to subtract the shift in kinetic energy $\Delta E\uu{\De m}^{(n)}$ due to mass
correction  to obtain the theoretically predicted Lamb shift.

Subtracting  $\Delta E\uu{\De m}^{(n)}$ from the $\Delta E\sn$ in \ct{216} and inserting the factors $\hbar, c$ and
$\e\sz$ (including in $\tau\uu\clp = \me c\sq$) to obtain an expression in SI  units, we get
the prediction for the observed energy shift
\beasm
\Delta E\sn\sps{predicted} \aea \Delta E\sn - \Delta E\sn\sps {free} =
\dsf {e\sq} {3\pi\, \e\sz\, \me\sq\, c\sq} \int_0^\infty d\kappa  \d
\ti g (\hbar\, c \,\kappa)\sq
\d \sum_r\dsf{\left| \lrab{\vec p}\dn{nr} \right|\sq\d (E\dn r-E\sn)}{E\dn r - E\sn + \hbar\,  c\,  \kappa} \nn
\aea \dsf {e\sq} {3\pi\, \e\sz\, \me\sq\, c\cu\, \hbar}
\sum_r  \left| \lrab{\vec p}\dn{nr} \right|\sq\d (E\dn r-E\sn) \lrs{\int_0^\infty dz  \d
\dsf{ \ti g(z)\sq}{E\dn r - E\sn + z}}
\label{eqn218}
\eeasm
where $z = \hbar \, c\, \kappa$.

For
$^2S\dn{1/2}$ and $^2P\dn{1/2}$
levels the principal quantum number is 2.  To avoid conflict with the symbol
$n$ used above, we'll use $n\uu{principal}$ to denote the principal quantum number of
an energy level in the Hydrogen atom.  As mentioned before, the symbols $r$ and $n$, which
are used to denote the orbitals,  actually stand for triples comprising the principal,
azimuthal and magnetic quantum numbers.
Using the energy eigenfunctions of the Hydrogen atom
one can verify that
\beass
\mtrxel{\psi\dn{2,\,0,\,0}} \nabla {\psi\dn{1,\,0,\,0}} =
\mtrxel{\psi\dn{2,\,1,\,0}} \nabla {\psi\dn{1,\,0,\,0}}=
\mtrxel{\psi\dn{2,\,1,\,1}} \nabla {\psi\dn{1,\,0,\,0}}=
\mtrxel{\psi\dn{2,\,1,\,-1}} \nabla {\psi\dn{1,\,0,\,0}}=0
\eeass
Further at $n\uu{principal}=2$, $E\dn{(2,\, l'\, ,\, m')} - E\dn{(2\, ,\, l\, ,\, m)} = 0$.  So
the sum in \ct{218} starts at the principal quantum number $n\uu{principal}=3$
and $\left| \lrab{\vec p}\dn{nr} \right|\sq(E\dn r - E\sn) > 0$ for all $r$ in the sum.
Therefore, we conclude that $\Delta E\sn\sps{predicted} \geq 0$.
Further, the integral in \ct{218} decreases
monotonically with increasing $E\dn r$.  Therefore, we can bound the integral using
the two limits $n\uu {principal} = 3$ and $n\uu{principal} \ra \infty$.
Denoting $E_{\lrs{n\uu{principal} = j}}$ as $E_{[j]}$ and noting  that
$E_{[\infty]} = 0$, we have for $j = 3, 4, 5, \ldots $,
\beass
I\dn{min} \dfn \izi dz \d \dsf{\ti g(z)\sq}{0 - E\dn{[2]} + z}
< \izi dz \d \dsf{\ti g(z)\sq}{E\dn{[j]} - E\dn{[2]} + z}
< \izi dz \d \dsf{\ti g(z)\sq}{E\dn{[3]} - E\dn{[2]} + z} \dfn I\dn{max}
\eeass
Hence, from \ct{218} we have
\beasm
 \dsf {e\sq \ I\uu{min}} {3\pi\, \e\sz\, \me\sq\, c\cu\, \hbar}
\sum_r  \left| \lrab{\vec p}\dn{nr} \right|\sq (E\dn r-E\sn) <
\Delta E\sn\tu{(predicted)} <   \dsf {e\sq\ I\uu{max}} {3\pi\, \e\sz\, \me\sq\, c\cu\, \hbar}
\sum_r  \left| \lrab{\vec p}\dn{nr} \right|\sq (E\dn r-E\sn)\nl 1
\leqn{yay}
\eeasm
Using
\ct{2p5}
and $E\dn{[j]} = -\dsf{E\so}{j\sq}$,  where $E\so = \a\sq m\dn e c\sq /2$, 
and noting that $\tau\uu\clp = \me c\sq$ as noted after \ct{lsp} we have
\beasm
I(j) \aea \izi \ dz\  \bsf{ 2 }{
 \eto{z/(m\dn e c\sq) + (m\dn e c\sq)/z} - 1 }^{1/2}
\bsf 1 {E\so \lrr{\dsf 1 4 - \dsf 1 {j\sq}} + z}\qquad j=3,4,5,\ldots \nn
\aea
\izi \ dq\  \bsf{ 2 }{
 \eto{q + 1/q} - 1 }^{1/2}
\bsf 1 { \lrc{\dsf{\alpha\sq}2 \lrr{\dsf 1 4 - \dsf 1 {j\sq}} + q}}, \qquad
q \dfn \dsf{z}{\me c\sq}\label{eqn219}
\eeasm 
 
Numerical evalution of \ct{219} yields
\bea
I\dn{min} = I(\infty) = 1.244289, \qquad I\dn{max} = I(3) =  1.244294
\label{eqn220}
\eea
Noting that the nonrelativistic Hamiltonian of the Hydrogen atom
$H\dn A =  \dsf{\vec p \d\vec p}{2m} - \dsf{e\sq}{4\pi\e\sz |\,\vec x\,|}$  we have
\bea
\sum_r |\lrab{\vec p}\dn{nr}|\,\sq \, (E\dn r- E\sn)
\aea \mtrxel n {\vec p \d H\vec p - H\vec p \d \vec p} n
\deq \dsf{e\sq \hbar\sq}{2\e\sz} |\psi\sn(0)|\sq \label{eqn221}
\eea
Inserting \ct{221} into \ct{yay} we  have
\bea
 \dsf {e\sq I\uu{min} } {3\pi\, \e\sz\, \me\sq\, c\cu\, \hbar}
\bsf{e\sq \hbar\sq}{2\e\sz} |\psi\sn(0)|\sq\ < \ \Delta E\sn\tu{(predicted)}\ <   
\ \dsf {e\sq\ I\uu{max}} {3\pi\, \e\sz\, \me\sq\, c\cu\, \hbar}
 \bsf{e\sq \hbar\sq}{2\e\sz} |\psi\sn(0)|\sq
 \leqn{mve}
\eea

We note that $|\psi_{(2\, ,\, 1\, ,\, m)} (0)|\sq =0$, where $m= -1, 0, 1$.
Therefore, from \ct{218} and \ct{221} we conclude that there is no
energy shift in the $^2P_{1/2}$ level.

On the other hand,
 $|\psi_{2\, ,\, 0\, ,\, 0}(0) |\sq= \dsf{1}{8\pi \, a\sz\cu}$,
 where $a\sz = \dsf{\hbar}{\a m\dn e c}$.
From
\ct{220} and \ct{mve}
we have
\beass
1.244289 \bsf{e\tu 4 \hbar}{48 \pi^2 \e\sz\sq \me\sq c\cu a\sz\cu}
\ < \ E\sw\sps{predicted}
\ < \  1.244294
\bsf{e\tu 4 \hbar}{48 \pi^2 \e\sz\sq \me\sq c\cu a\sz\cu}
\eeass
Dividing throughout by $h$ we get the Lamb shift
predicted by the
above calculation to be
\bea
1060.476 \,MHz \ <\  \Delta \nu\sps{predicted}_{_{LS}} \ <\  1060.480\, MHz
\eea

\s{\colb{Photon propagator}}\label{afpp}
The following calculation, adapted from \cite{huan}, derives the full photon propagator 
shown in \ct{phprop}.  
Expanding the free Maxwell field in Lorentz frame $\clf$, as
\beass
A\upm(x;\clf,\clp) \aea \int \dsf{\dvk\ \gpfv k}{\tpc \tok}  \sump \l
  \e\upm(\uvk,\l) \lrc{\opa(\uvk,\l) \emiuvkx + \opad(\uvk,\l) \eiuvkx} \qquad \leqn{2401}
\eeass
where $\uvk \dfn (|\vk|, \vk)$,
and using the standard commutation relations of the creation and annihilation
operators, one obtains the propagator of the free
Maxwell field   
\beas
M\umn(x,y;\clf)
\aea \msrf {k}  \ti M\umn(k; \clf,\clp)  \eto{-ik(x-y)},
\eeas
where
\beas
\ti M \umn(k;\clf,\clp) \ada i\ \bsf{(-\eta\umn) \lrs{\gpf {k}}\sq}{(k)\sq + i\e}
\eeas
Hereafter we abbreviate $M\umn(k;\clf,\clp)$ to $M\umn(k)$.

As is well known the  full photon propagator can be written, in
momentum space, as
\beasm
\ti M\umn\uu{full}(k)
\aea \ti M\umn(k) +   \ti M\tu{\m\a}(k)  \Pi\dn{\a\b}(k)   \ti M\tu{\b\n}(k) +   
\ti M\tu{\m\a}(k)  \Pi\dn{\a\b}(k)  \ti M\tu{\b\rho}(k)\  \Pi\dn{\rho\sg}(k)
\, \ti M\tu{\sg\n}(k) + \ldots
\nl 1 \leqn{fppr}
\eeasm
Due to conservation of electron current $\Pi\uu{\a\b}(k)$ satisfies
$k\upa \Pi\uu{\a\b}(k) = k\upb \Pi\uu{\a\b}(k) = 0$.  Therefore,
 we can take $\Pi\dn{\a\b}(k)$ to be
\beas
\Pi\uu{\a\b}(k) \aea i \, e \sq\, (k\sq \eta\uu{\a\b} - k\dna k\dnb) \,\chi(k), 
\qquad \chi(k) \dfn \dsf{\Pi\upd\m\m(k)
}{3\, i\, e \sq\, k\sq}
\leqn{dochi}
\eeas
$e$ is the coupling constant in the unrenormalized Lagrangian. $\eta$ is the Lorentzian metric tensor. 

Defining
\beas
P\upd  \m \n (k)\ada \de\upd \m \n - \dsf{k\upm k\dnn}{k\sq},
\eeas
and noting that $P\upd \m\a(k)  P\upd\a\n(k) = P\upd \m \n(k)$,  we have
\bea
 \prod_{j=1}^n \Pi\upd {\mu\sj}{\a\sj} (k) \ti M\upd {\a\sj}{\nu\sj} (k) \aea \l^n(k) 
 \ P\upd{\mu\so}{\nu\sn}, \qquad \l(k) \dfn e \sq \, \gpf k\sq \, \chi(k),
 \quad \leqn{wwic}
\eea
Inserting \ct{wwic} into \ct{fppr} we have
\beas
\ti M\umn\uu{full} (k)
\aea \ti M\upd\m\a (k) \lrc{\dsf{\eta\tu{\a\n}}{1-\l} - \dsf\l{1-\l}\  \brf{k\upa k\upn}{k\sq}}
\eeas
The second term (containing $k\upa k\upn$)
vanishes when contracted with the conserved electron current $j\dnn$.
Therefore,
\beas
\ti M\umn\uu{full}(k) \aea \dsf {\ti M\umn(k)}{1- e\sz\sq \d \gpf k\sq \d \chi(k)} 
\eeas 
 {\color{black}
 \s{Upper bound on $I\uu{\pm}$}\label{uboi}
We show that for  $a > 1$, $0 \leq b \leq 2$, and $\b \dfn \sqrt{\dsf{9\a}{16\pi}}$ 
 \bea
 I\uu{\pm}(a,b) \ada \izi   \bsf{x\tu 4 (x\sq + a\sq) }{e\tu{\sqrt{x\sq + a\sq}- a \b b} \pm 1}\tu{1/2}  dx
 \ < \ 12288 \, a\cu \, e^{-0.46 a} \leqn{iie}
 \eea
{\bf Proof:} Set  $\th \dfn b \b$ and note that $\th < 0.0724$.  Changing variables from $x$ to $y = x\sq + a\sq$ we have
 \beas
 I\uu{\pm}(a,b) \aea \dsf 1 2 \int_{a\sq}^\infty \sqrt{y-a\sq} \bsf{y}{e^{\sqrt y - a\th}\pm 1}\tu{1/2} dy
 \eeas
 For $a>1$, we have $\dsf 1 2 e^{a(1-\th)} > 1$.  Since $y \geq a\sq$ in the integrand, we have  
 \beas
 e^{\sqrt y - a\th } \pm 1 \j \geq \j  e^{\sqrt y - a\th } - 1 \ > \ e^{\sqrt y - a\th } - \brf 1 2 e^{\sqrt y - a\th }  \deq \brf 1 2 e^{\sqrt y - a\th }\\
 \aea  \brf 1 2 \lrs { e^{(1/2) \sqrt y} }\lrs{e^{(1/2)\sqrt y - a\th }} \ \geq \ \brf 1 2 \lrs { e^{(1/2) \sqrt y} }\lrs{e^{a \lrr{\f12  -  \th}}}
 \eeas
 Therefore,
 \beas
 I\uu\pm(a,b) \j \leq \j \dsf 1 2 \int_{a\sq}^\infty \dsf{y}{\lrs{\brf 1 2 \lrs { e^{(1/2) \sqrt y} }\lrs{e^{a \lrr{\f12  -  \th} }}}^{1/2}}
 \ < \ \brf 1 {\sqrt 2} e^{-a (0.25-\th/2)}\int_{a\sq}^\infty y \ e^{-(1/4) \sqrt y} \ dy \\
 \aea \brf 1 {\sqrt 2} e^{-a(0.5-\th/2)} (8 a\cu + 96 a\sq + 768 a + 3072)\\
 \j \leq \j    (4)(3072\,  a\cu) \ e^{-(0.46) a}  \deq 12288 \ a\cu \ e^{-(0.46) a}.
 \eeas}
\end{appendices}
\colb{

}

\begin{thebibliography}{ab}
\bibitem{hatf} 
Hatfield  B, {\it Quantum Field Theory of Point Particles and Strings},
Frontiers in Physics, Addison-Wesley, 1992. 
\bibitem{huan} 
Huang K, {\it Quantum Field Theory: from Operators to Path Integrals},
second, revised edition, Wiley-VCH Verlag GmbH \& Co. KGaA, Weinheim, 2010. 
\bibitem{itzu} 
Itzykson C and Zuber J-B,
{\it Quantum Field Theory}, Dover Publications, 2006. 
\bibitem{mand} 
Mandl F, {\it Statistical Physics}, John Wiley and Sons, 1999. 
\bibitem{pesc} 
Peskin ME and Schroeder DV,
{\it An Introduction to Quantum Field Theory},
Addison-Wesley, 1995.
\bibitem{ryde} 
Ryder L, {\it Quantum Field Theory}, Cambridge University Press, 1996.
\bibitem{sred} 
Srednicki M, {\it Quantum Field Theory}, Cambridge University Press, 2007.
\bibitem{wein} 
Weinberg S, {\it The Quantum Theory of Fields, Volume I: Foundations}, Cambridge University
Press, 2005.
\bibitem{zee} 
Zee A, {\it Quantum Field Theory in a Nutshell}, Princeton University Press, 2010.
\bibitem{frgo} 
Friedrich W and Goldhaber G, Zeits. f. Physik. 44, 700, 1927. 
\bibitem{alep} 
The ALEPH Collaboration {\it et al.}, 
Phys. Rept. 532, 119, 2013.
\bibitem{schw} 
Schwinger J,  Phys. Rev. 73, 416L, 1948. 
\bibitem{akni} 
Aoyama T, Kinoshita T and Nio M,   
Phys. Rev. D 97, 036001, 
2018.
\bibitem{hann} 
Hanneke D, Fogwell S, Gabrielse G,
Phys. Rev. Lett.
100,  120801, 2008.
\bibitem{hhga} 
Hanneke D, Hoogerheide SF and Gabrielse G,
Phys. Rev. A 83, 052122, 2011.
\bibitem{gibb}
Gibbs JW, {\it Elementary Principles in Statistical Mechanics}, Charles Scribner's Sons, New York, 1902.
\bibitem{kikr} 
Kittel C and Kroemer H, {\it Thermal Physics}, 2$^{nd}$ edition,
W.H. Freeman, 1980.
\bibitem{bethe} 
Bethe HA, Phys. Rev.
 72 (4),  339, 1947.
\bibitem{lamb} 
Lamb WE and Retherford RC,
Phys. Rev.
72 (3), 241, 1947. 
\bibitem{hind} 
Hindmarsh WR, {\it Atomic Spectra}, Pergamon Press Limited, 1967.
\bibitem{anne} 
Andrews DA and Newton G, Phys. Rev. Lett. 37 (19)  1254, 1976.
\bibitem{eide} 
Eides MI, Phys. Rep. 342,  63, 2001. 
\bibitem{kiye} 
Kinoshita T and Yennie DR, {\it High Precision Tests of Quantum Electrodynamics -- an Overview}, 
in {\it Quantum Electrodynamics}, ed. Kinoshita T, World Scientific, 1990.
\bibitem{krla} 
Kroll NM and Lamb WE,
Phys. Rev. 75,  388, 1949.
\bibitem{lambnl} 
Lamb WE,
Nobel Lecture, December 12, 1955.
\bibitem{grif} 
Griffiths DJ, {\it Introduction to Quantum Mechanics}, Prentice Hall, 1994. 
\bibitem{bbfe} 
Baranger M, Bethe HA and Feynman RP, Phys. Rev. 92 (2),  482, 1953.
\bibitem{tpla}  
Taylor BN, Parker WH and Langenberg DN, Rev. Mod. Phys.
41, 375, 1969. 
\bibitem{tdla} 
Triebwasser S, Dayhoff ES and Lamb Jr. WE, Phys. Rev. 89, 98, 1953.
\bibitem{rosh} 
Robiscow RT and Shyn TW, Phys. Rev. Lett. 24, 559, 1970.
\bibitem{psya} 
Pal'chikov VG, Sokolov YL and Yakovlev VP, JETP Lett. 38,  418, 1983.
\bibitem{soya} 
Sokolov YU and Yahovlev VP, Sov. Phys.-JETP 56,  7, 1982.
\bibitem{naun} 
Newton G, Andrews DA and Unsworth PJ, Phil. Trans. Roy. Soc.
London 290, 373, 1979. 
\bibitem{lupi} 
Lundeen SR and Pipkin FM,
Phys. Rev. Lett.
 46 (4),  232, 1981. 
\bibitem{vhdr} 
van Wijngaarden A, Holuj F and Drake GWF,
Can. J. Phys. 76,  95, 1998.
\bibitem{hapi} 
Hagley EW and Pipkin FM, Phys. Rev. Lett.
72,  1172, 1994. 
\bibitem{klni} 
Klein O and Nishina Y, Zeits. f. Physik. 52, 853, 1928. 
\bibitem{muir} 
Muirhead H, {\it The Physics of Elementary Particles}, Pergamon Press, 1965.
\bibitem{delb1}
Schumacher M, Rad. Phys. Chem. 56, 101, 1999.
\bibitem{delb2}
Milstein AI and Schumacher M, Phys. Rep. 243, 183, 1994. 
\bibitem{pamo}
Papatzacos P and Mork K, Phys. Rep. 21(2), 81, 1975. 
\bibitem{mnpo}
Mushtukov AA, Nagirner DI and Poutanen J, 
Phys. Rev. D 93, 105003, 2016.
\bibitem{kirc}
Kirchner M \etal, Nat. Phys. 16, 756, 2020. 
\bibitem{kppr}
Kaliman Z, Pisk K and Pratt RH, Phys. Rev. A 83, 053406, 2011.
\bibitem{syss}
Sommerfeldt J \etal, Phys. Rev. Lett. 131, 061601, 2023.
\bibitem{uehl}
Uehling EA, Phys. Rev. 48, 55, 1935; 
Serber R, Phys. Rev. 48, 49, 1935;
Gyulassy M,
Phys. Rev. Lett. 32 (24), 1393, 1974;
Gyulassy M, 
Phys. Rev. Lett. 33 (15), 921, 1974;  
Gyulassy M, Nucl. Phys. A 244, 497, 1975;
Soff G and Mohr PJ, Phys. Rev. A 38 (10), 
5066, 1988;  Neghabian AR, Phys. Rev. A 27, 2311, 1983.
\bibitem{wikr}
Wichmann EH and Kroll NM, Phys. Rev. 101 (2), 843, 1956.
 \bibitem{inmo}
 Indelicato P and Mohr P, {\it Introduction to Bound-state Quantum
 Electrodynamics}, 1--110 in Liu W (ed.) Handbook of Relativistic Quantum Chemistry, Springer, Berlin, Heidelberg, 2016.
  \bibitem{furr}
 Furry  WH, Phys. Rev. 81 (1), 115, 1951.
\bibitem{daha}
Zhu TH and Ruderman M, Astrophys. J. 478, 701, 1997; Daugherty JK and Harding AK, AIP Conf. Proc. 101, 387, 1983.
\bibitem{baru}
Bander M and Rubinstein HR, Astroparticle Phys 1, 277, 1993. 
\bibitem{mzys}
Mandrykina ZA {\it et al.}, Phys. Rev. A, 105, 062806, 2022.
\bibitem{fedo}
Fedotov A \etal, Phys. Rep. 1010, 1, 2023. 
\bibitem{bmms}
Bass SD \etal,  Rev. Mod. Phys. 95, 021002, 2023;
Karshenboim SG, Int. J. Mod. Phys. A 19, 3879, 2004.
\bibitem{weincc} 
Weinberg S,  
Rev. Mod. Phys.
61, 1, 1989.
\bibitem{mart} 
Martin J, Comp. Ren. Phys. 13 (6-7), 566, 2012.
\bibitem{zyla} 
Zyla PA {\it et al.} (Particle Data Group), Prog. Theor. Exp. Phys. 2020, 083C01, 2020.
\bibitem{chch} 
Choudhury SR and Choubey S, JCAP 9, 017, 2018. 
\bibitem{pdgp} 
Bethke S, Dissertori G and Salam GP, 
https://pdg.lbl.gov/2014/reviews/rpp2014-rev-qcd.pdf,
retrieved on March 20, 2022.
\bibitem{plan} 
Planck Collaboration,  Astron. Astrophys. 641, A6, 2020.
\bibitem{lali} 
Landau LD and Lifshitz EM, {\it The Classical Theory of Fields},
Vol. 2, Course of Theoretical Physics, Pergamon Press, 1975. 
\bibitem{lavu} 
Laine M and Vuorinen A, {\it Basics of Thermal Field Theory, a Tutorial
on Perturbative Computations},  Lecture Notes in Physics 925, Springer
2016.
\bibitem{schm} 
Schmitt A, {\it Basics of Quantum Field Theory at Finite Temperature and 
Chemical Potential}, 
Lecture Notes in Physics 811, 123, Springer, 2010. 
\bibitem{turk} 
Turko L, Phys. Lett. 104B, 2, 153, 1981.
\bibitem{dsde} 
Van Dyck RS, Schwinberg PB, Dehmelt HG, Phys. Rev. Lett. 59, 26, 1987. 
\bibitem{dyso2} 
Dyson FJ, {\it Advanced Quantum Mechanics}, second edition, World Scientific Publishing Co. Pvt. Ltd., 2011. 
\end{thebibliography}
\end{document}